\documentclass[12pt,a4paper]{article}
\usepackage[utf8]{inputenc}
\usepackage[english]{babel}
\usepackage{amsmath}
\usepackage{amsfonts}
\usepackage{amssymb}
\usepackage{graphicx}
\usepackage{cancel}
\usepackage{placeins}
\usepackage{soul}
\usepackage[usenames,dvipsnames]{color}
\usepackage{slashed, epsf, latexsym}
\usepackage{epsfig}
\usepackage{graphics}
\usepackage{jheppub}

\begin{document}

	\preprint{}
	\title{Stress Tensor for Large-$D$ membrane at subleading orders}
	\author{Parthajit Biswas}
	\affiliation{School of Physical Sciences, National Institute of Science Education and Research, HBNI, Jatni - 752050, India}
	\emailAdd{parthajit.biswas@niser.ac.in}
	
	\abstract{In this note, we have extended the result of \citep{radiation} to calculate the membrane stress tensor up to ${\cal O}\left(\frac{1}{D}\right)$ localized on the codimension-one membrane world volume propagating in asymptotically flat/AdS/dS space-time. We have shown that the subleading order membrane equation follows from the conservation equation of this stress tensor.}

\maketitle


\section{Introduction}\label{sec:intro}
It has recently been shown that in a large number of space-time dimensions($D$) there is one-to-one corresondence\cite{membrane,arbBack,yogesh1,secondorder} between dynamics of a black hole and a co-dimension one membrane propagating in asymptotic spacetime of the black hole.\footnote{See \cite{Emparan:2013moa,Emparan:2013xia, Emparan:2014cia,EmparanCoupling,Prester:2013gxa,Emparan:2014jca,QNM:Emparan} for initial developments on black holes in large-$D$ limit. See\cite{Chmembrane,yogesh2,Suzuki:2015iha,Suzuki:2015axa,Emparan:2015gva,Tanabe:2015hda,Tanabe:2015isb,Chen:2015fuf,EmparanHydro,Sadhu:2016ynd,Herzog:2016hob,Tanabe:2016opw,Rozali:2016yhw,Chen:2016fuy,Chen:2017wpf,Chen:2017hwm,Rozali:2017bll,Chen:2017rxa,Herzog:2017my,Kar:2019kyz,Dandekar:2019hyc} for works related to large $D$ expansion.} In \cite{radiation} the author has constructed a stress tensor up to ${\cal O}(1)$ defined on the membrane world volume propagating in asymptotically flat space-time and demonstrated that the leading order membrane equation follows from the conservation equation of the stress tensor. In this note,  we have extended the result of \cite{radiation} to calculate the stress tensor up to ${\cal O}\left(\frac{1}{D}\right)$ defined on the membrane world volume but propagating in asymptotically flat/dS/AdS space-time and have demonstrated that the subleading order membrane equations \cite{secondorder} are just the conservation equation of this stress tensor.

The computation of \cite{radiation} essentially captures the leading non-trivial part of the dual membrane stress tensor. But for several reasons, it seems that determining the stress tensor at next order is going to be quite useful. 

For example, in \cite{Dandekar:2017aiv} the authors have written an improved stress tensor that would work even for finite $D$ at least for stationary metrics. However, in the same paper, they have also reported that their finite $D$ improvement, once compared against the stress tensor for gravity systems with a dual hydrodynamic description in derivative expansion, does not match up to the relevant orders.    Needless to say,  in their construction the leading and the first subleading  terms of the membrane stress tensor  played a very important role and the subsubleading terms of the stress tensor will add very useful  new data to the whole programme  of finite $D$ improvement.


\subsection{Final Result:}
In this section, we shall write our final result - the expression of the membrane stress tensor up to corrections of order ${\cal O}\left(1\over D\right)^2$. Conservation of this stress tensor would result in the membrane equation derived in \citep{secondorder} - the equation that governs the dynamics of the membrane.

In our case the membrane, which is a codimension-1 hypersurface,  is embedded in AdS/ dS space. More precisely, the metric of the embedding space satisfies the following equation
 $$ R_{AB}-\left(\frac{R-\Lambda}{2}\right)G_{AB}=0 $$
Where, dimension($D$) dependence of $\Lambda$ is parametrized as follows $$\Lambda=[(D-1)(D-2)]\lambda,~~~\lambda\sim {\cal O}(1)$$

The membrane is characterized by its shape (encoded in its extrinsic curvature ${\cal K}_{\mu\nu}$) and a velocity field ($u_\mu$), unit normalized with respect to the induced metric of the membrane. The membrane stress tensor, that we report below, is a symmetric two-indexed tensor, constructed out of this velocity field, extrinsic curvature and its derivatives.\\
For convenience, we shall decompose the stress tensor in the following way
\begin{equation}\label{eq:STRESS_TENSOR}
\boxed{8\pi T_{\mu\nu}={\cal S}_1~u_\mu u_\nu+{\cal S}_2~g_{\mu\nu}^{(ind)}+{\cal V}_\mu ~u_\nu+{\cal V}_\nu ~u_\mu+{\cal W}_{\mu\nu}}
\end{equation}
Where,
\begin{equation}
\begin{split}
{\cal S}_1&=\frac{\cal K}{2}+\frac{1}{2}\bigg(\frac{\bar{\nabla}^2 {\cal K}}{{\cal K}^2}-\lambda~\frac{D-1}{\cal K}-\frac{1}{K}{\cal K}_{\alpha\beta}{\cal K}^{\alpha\beta}\bigg)+\frac{1}{\cal K}\bigg[-u\cdot {\cal K}\cdot {\cal K}\cdot u-13\bigg(\frac{u\cdot\nabla {\cal K}}{\cal K}\bigg)^2\\
&+2~u^\alpha {\cal K}_{\alpha\beta}\bigg(\frac{\bar{\nabla}^\beta {\cal K}}{\cal K}\bigg)+14~\bigg(\frac{u\cdot\nabla {\cal K}}{\cal K}\bigg)(u\cdot {\cal K}\cdot u)-\frac{\cal K}{D}\bigg(\frac{u\cdot\nabla {\cal K}}{\cal K}\bigg)+\frac{\cal K}{D}(u\cdot {\cal K}\cdot u)+\frac{1}{{\cal K}^3}\bar{\nabla}^2\left(\bar{\nabla}^2 {\cal K}\right)\\
&-4~(u\cdot {\cal K}\cdot u)^2-8~\lambda \frac{D}{\cal K}\left(\frac{u\cdot\nabla {\cal K}}{\cal K}\right)+4~\lambda\frac{D}{\cal K}~(u\cdot {\cal K}\cdot u)-2~\bigg(\frac{\bar{\nabla}_\alpha {\cal K}}{\cal K}\bigg)\bigg(\frac{\bar{\nabla}^\alpha {\cal K}}{\cal K}\bigg)+\lambda-\lambda^2~\frac{D^2}{{\cal K}^2}\bigg]\\
&+\frac{1}{\cal K}\left(2~\text{Zeta}[3]-1\right)\bigg[-\frac{\cal K}{D}\bigg(\frac{(u\cdot\nabla){\cal K}}{\cal K}-u\cdot {\cal K}\cdot u\bigg)-\lambda-u\cdot {\cal K}\cdot {\cal K}\cdot u+2\left(\frac{\nabla_\alpha {\cal K}}{\cal K}\right)u^\beta {\cal K}^\alpha_\beta\\
&-\bigg(\frac{u\cdot\nabla {\cal K}}{\cal K}\bigg)^2+2\left(\frac{u\cdot\nabla {\cal K}}{\cal K}\right)(u\cdot {\cal K}\cdot u)-\bigg(\frac{\bar{\nabla}^\alpha {\cal K}}{\cal K}\bigg)\bigg(\frac{\bar{\nabla}_\alpha {\cal K}}{\cal K}\bigg)-(u\cdot {\cal K}\cdot u)^2\bigg]\nonumber
\end{split}
\end{equation}
\begin{equation}
\begin{split}
{\cal S}_2&=-\frac{1}{2}\left(u\cdot {\cal K}\cdot u\right)-\frac{1}{2{\cal K}}~{\cal K}^{\alpha\beta}{\cal K}_{\alpha\beta}-\frac{1}{\cal K}\bigg(\frac{u\cdot\nabla {\cal K}}{{\cal K}}-\frac{1}{2}\left(u\cdot{\cal K}\cdot u\right)-\frac{\cal K}{2~D}\bigg)\left(u\cdot {\cal K}\cdot u\right)+\frac{\lambda}{K}\\
&+\frac{1}{\cal K}~{\cal K}^{\alpha\beta}(\nabla_\alpha u_\beta)-\frac{2}{\cal K}~u_\alpha {\cal K}^{\alpha\beta}\bigg(\frac{1}{2}\frac{\bar{\nabla}_\beta{\cal K}}{\cal K}-\frac{\bar{\nabla}^2 u_\beta}{\cal K}\bigg)
\end{split}
\end{equation}
\begin{equation}
\begin{split}
{\cal V}_\mu&=\frac{1}{2}\bigg(\frac{\bar{\nabla}_\mu {\cal K}}{\cal K}\bigg)-\bigg(\frac{\bar{\nabla}^2 u_\mu}{\cal K}\bigg)+\frac{1}{\cal K}{\cal K}^\alpha_\mu {\cal K}_{\alpha\beta}u^\beta-\frac{1}{{\cal K}^3}\bar{\nabla}^2\left(\bar{\nabla}^2 u_\mu\right)+\frac{1}{\cal K}\bar{\nabla}_\mu\bigg(\frac{u\cdot\nabla {\cal K}}{\cal K}\bigg)\\
&+\frac{1}{\cal K}\bigg(\frac{\bar{\nabla}^2 u_\mu}{\cal K}\bigg)\bigg(-2~(u\cdot {\cal K}\cdot u)+4~\frac{u\cdot\nabla {\cal K}}{\cal K}+2~\lambda\frac{D}{\cal K}-\frac{\cal K}{D}\bigg)+\frac{1}{2~{\cal K}}\bigg(\frac{\bar{\nabla}_\mu {\cal K}}{\cal K}\bigg)(u\cdot {\cal K}\cdot u)
\end{split}
\end{equation}
\begin{equation}
\begin{split}
{\cal W}_{\mu\nu}&=\frac{1}{2}{\cal K}_{\mu\nu}-\frac{1}{2}\left(\bar{\nabla}_\mu u_\nu+\bar{\nabla}_\nu u_\mu\right)-\frac{1}{\cal K}~{\cal K}_{\mu\nu}\left(u\cdot {\cal K}\cdot u\right)+\frac{1}{2~{\cal K}}\left(\bar{\nabla}_\mu u_\nu+\bar{\nabla}_\nu u_\mu\right)\left(u\cdot {\cal K}\cdot u\right)\\
&+\frac{1}{2~{\cal K}}\bigg[\bar{\nabla}_\mu\bigg(\frac{\bar{\nabla}^2 u_\nu}{\cal K}\bigg)+\bar{\nabla}_\nu\bigg(\frac{\bar{\nabla}^2 u_\mu}{\cal K}\bigg)+\bar{\nabla}_\mu\left(u^\alpha {\cal K}_{\alpha\nu}\right)+\bar{\nabla}_\nu\left(u^\alpha {\cal K}_{\alpha\mu}\right)-2~\bar{\nabla}_\mu\bigg(\frac{\bar{\nabla}_\nu {\cal K}}{\cal K}\bigg)\bigg]\\
&-\frac{1}{{\cal K}}\left(\bar{\nabla}^\alpha u_\mu\right)\left(\bar{\nabla}_\alpha u_\nu\right)-\frac{1}{{\cal K}}\bigg(\frac{\bar{\nabla}^2 u_\mu}{\cal K}\bigg)\bigg(\frac{\bar{\nabla}^2 u_\nu}{\cal K}\bigg)\\
\end{split}
\end{equation}
Here, $g_{\mu\nu}^{(ind)}$ is the induced metric on the membrane, $\bar{\nabla}_\mu$ is the covariant derivative with respect to $g_{\mu\nu}^{(ind)}$. Membrane velocity ($u_\mu$) can also be viewed as a vector field($u_A$) in the full background space-time.  $u_\mu$ is related to $u_A$ through the following equation
\begin{equation}
u_\mu=\left(\frac{\partial X^A}{\partial y^\mu}\right)u_A
\end{equation}
Where, $X^A$ are the coordinates in the full space-time and $y^\mu$ are the coordinates on the membrane world volume.\\
The extrinsic curvature of the membrane ${\cal K}_{\mu\nu}$ is defined as follows
\begin{equation}
{\cal K}_{\mu\nu}=\left(\frac{\partial X^A}{\partial y^\mu}\right)\left(\frac{\partial X^B}{\partial y^\nu}\right)K_{AB}, ~~~\text{Where,}~~ K_{AB}=\Pi_A^C\nabla_C n_B
\end{equation} 
Here, $n_A$ is the normal to the membrane and $\Pi_{AB}$ is the projector orthogonal to the membrane defined as $\Pi_{AB}=g_{AB}-n_A n_B$.

\subsection{Strategy:}
The two key principles that fix this stress tensor are the following
\begin{itemize}
\item Conservation of the stress tensor should reproduce the membrane equation up to the relevant order.
\item This stress tensor should be the source of the gravitational radiation, generated from the massive fluctuating membrane.
\end{itemize}
In fact, it is the second principle that finally determines the algorithm to be used to derive the stress tensor. The algorithm is such that the first principle is automatically ensured and we have used it in the end  as a consistency check for our long calculation.

Below, we shall just write down the steps to be used so that the final construction is consistent with the second principle. However,  we shall not write the justification for any of these steps as they are explained in detail in \citep{radiation} and explanation is completely independent of the order in terms of (1/D) expansion.
\begin{itemize}
\item Step-1: Codimension one membrane is given by a single scalar equation $\psi = 1$.
Define $\psi>1$  region as ` outside of the membrane'  and $\psi<1$ as ` inside of the membrane'. `Outside region' is the one that  extends towards asymptotic infinity and contains the gravitational radiation. 

\item Step-2:
Next, we would like to  write a space-time metric for both outside and inside region, with the following properties.
\begin{enumerate}
\item The metric  would solve Einstein equation (in presence of cosmological constant) linearized around  pure AdS/dS metric.

\item The metric would fall off as $\psi^{-D}$ in the outside region and would be regular in the inside region.

\item The metric should be continuous across the membrane though its first normal derivative need not be.

\end{enumerate}

It turns out that in ${1\over D}$ expansion, the above two conditions uniquely fix the metric on both sides,   in terms of the induced metric on the membrane, which we read off from the large-$D$ metric determined in \cite{secondorder}.

\item Step-3: Once we have determined the metric on both sides, the discontinuity of its normal derivative across the membrane is also fixed unambiguously. The conserved stress tensor associated  with the membrane is  computed from this discontinuity. More precisely,  it is the difference between the two Brown York  stress tensors on the membrane evaluated with respect to the inside and outside metric.
\begin{equation}
T_{AB}=T_{AB}^{(in)}-T_{AB}^{(out)}
\end{equation}
Here,
\begin{equation}
8\pi T_{AB}^{(in)}=K_{AB}^{(in)}-K^{(in)}\mathfrak{p}_{AB}^{(in)}~~{\text{and,}}~~ 8\pi T_{AB}^{(out)}=K_{AB}^{(out)}-K^{(out)}\mathfrak{p}_{AB}^{(in)}
\end{equation}
are respectively the Brown York stress tensors of internal solution and external solution evaluated on the membrane. $K_{AB}^{(in)}$ and $\mathfrak{p}_{AB}^{(in)}$ are respectively extrinsic curvature and projector on to the membrane viewed as a submanifold of the background space-time perturbed by the internal solution. Similarly, $K_{AB}^{(out)}$ and $\mathfrak{p}_{AB}^{(out)}$ are respectively extrinsic curvature and projector on to the membrane viewed as a submanifold of the background space-time perturbed by the external solution. $T_{AB}^{(out)}$ and  $T_{AB}^{(in)}$ both satisfies $n^AT_{AB}^{(out/in)}=0$. So, $T_{AB}$ can equally well be regarded as a tensor $T_{\mu\nu}$ that lives on the membrane world volume.
\end{itemize}

Calculationally, this is very lengthy. In the main text, we have just written the final results, most of the lengthy derivations are in the appendices. The organization of this note is as follows: In section \ref{sec:outside} we have linearized the Large -$D$ solution known up to subleading order and have changed the gauge and subsidiary condition (as discussed just below eq.\eqref{eq:2.1}). In section \ref{sec:inside} we have constructed a linearized solution of Einstein's equation in the inside region of the membrane. In section \ref{sec:stress} we have calculated the membrane stress tensor and in the section  \ref{sec:conservation} we have shown that the subleading order membrane equation follows from the conservation of this stress tensor.

\section{Linearized Solution : Outside($\psi>1$)}\label{sec:outside}
In this section, we shall work out the metric in the outside region. \
However, what   we are finally interested in is just the difference between  Brown York stress tensor across the membrane. To compute it, we need to know the metric only very near the membrane. The large $D$ solution as described in \cite{secondorder} already determined the metric in this near membrane region even at non-linear order. For our purpose, we shall simply read off the `outside metric' from  \cite{secondorder}. In fact, we have to pick out only the part that is enough to solve the linearized equations. In other words, we need only that part of the metric which could be recast as
\begin{equation}\label{eq:2.1}
G^{(out)}_{AB}=g_{AB}+\psi^{-D}\mathfrak{h}_{AB}=g_{AB}+\psi^{-D}\sum_{m=0}^\infty (\psi-1)^m h^{(m)}_{AB}
\end{equation}
In the first subsection, we have described the large-$D$ solution and read off the piece needed.

The main calculation of this section involves a change of gauge and `subsidiary conditions' (conventions that fix how the basic fields would evolve away from the membrane, see \cite{arbBack} for more details). In the next two subsections,  we have described the new set of conventions, that are more useful for our purpose and performed the required changes on the metric,  read off in the first subsection. Needless to say, all steps are worked out in an expansion in ${1\over D}$.

%
%
\subsection{Large-$D$ Metric upto sub-subleading order : Linearized}
In this subsection we will just quote the solution of Einstein's equation up to second subleading order in $\frac{1}{D}$ expansion as derived in \cite{secondorder} and we will linearize the solution in $\psi^{-D}$. The solution is given by
\begin{equation}\label{schemesol}
{\cal G}_{AB} = g_{AB} + \psi^{-D}O_A O_B  +\left(\frac{1}{D}\right)^2 G^{(2)}_{AB}  + \cdots
\end{equation}
Here, $g_{AB}$ is the background metric and $O_A=n_A-u_A$.
\begin{equation}\label{eq:parameter1}
\begin{split}
G^{(2)}_{AB}& =\bigg[O_A O_B\left(f_1(R)~\mathfrak{s}_1+f_2(R)~\mathfrak{s}_2 \right) +  t(R)~\mathfrak{t}_{AB} 
+  v(R)~\big(~ \mathfrak{v}_A O_B + \mathfrak{v}_B O_A\big)\bigg]\\
&\text{where}~~ R\equiv D(\psi-1),~~P_{AB} = g_{AB} - n_A n_B +u_A u_B\\
&\text{and,}~~n^A~\mathfrak{v}_A = u^A~\mathfrak{v}_A=0,~~n^A~\mathfrak{t}_{AB} =u^A~\mathfrak{t}_{AB}=0,~~g^{AB}~\mathfrak{t}_{AB}=0
\end{split}
\end{equation}
Where,
\begin{equation}\label{funcn}
\begin{split}
&t(R)=-~2\left(\frac{D}{K}\right)^2\int_R^{\infty}\frac{y~dy}{e^y-1}\\
&v(R)=2\left(\frac{D}{K}\right)^3\bigg[\int_R^{\infty}e^{-x}dx\int_0^x\frac{y~e^y}{e^y-1}dy~-~e^{-R}\int_0^{\infty}e^{-x}dx\int_0^x\frac{y~e^y}{e^y-1}dy\bigg]\\
&f_1(R)=-2\left(\frac{D}{K}\right)^2\int_R^{\infty}x~e^{-x}dx+2~e^{-R}\left(\frac{D}{K}\right)^2\int_0^{\infty}x~e^{-x}dx\\ 
&f_2(R)=\left(\frac{D}{K}\right)\Bigg[\int_R^{\infty}e^{-x}dx\int_0^x\frac{v(y)}{1-e^{-y}}dy-e^{-R}\int_0^{\infty}e^{-x}dx\int_0^x\frac{v(y)}{1-e^{-y}}dy\Bigg]\\
&~~~~~~~~-\left(\frac{D}{K}\right)^4\Bigg[\int_R^{\infty}e^{-x} dx\int_0^x\frac{y^2~ e^{-y}}{1-e^{-y}}dy-e^{-R}\int_0^{\infty}e^{-x} dx\int_0^x\frac{y^2 ~e^{-y}}{1-e^{-y}}dy\Bigg]
\end{split}
\end{equation}
And,
\begin{equation}\label{structure}
\begin{split}
\mathfrak{t}_{AB}&=P^C_A P^{D}_B\bigg[ \bar{R}_{FCDE}O^E O^F+\frac{K}{D}\bigg(K_{CD}-\frac{\nabla_C u_{D}+\nabla_{D}u_C}{2}\bigg)\\
&- P^{EF}(K_{EC}-\nabla_E u_C)(K_{FD}-\nabla_F u_{D})\bigg]\\[6pt]
\mathfrak{v}_{A}&={P}^B_{A}\bigg[\frac{K}{D}\left(n^D u^E O^F \bar{R}_{FBDE}\right)+\frac{K^2}{2D^2}\left(\frac{{\nabla}_B {K}}{K}+(u\cdot{\nabla})u_B-2~u^D {K}_{D B}\right)\\
&-{P}^{F D} \left(\frac{{\nabla}_F {K}}{D}-\frac{K}{D} (u^E {K}_{E F})\right)\left({K}_{D B}-\nabla_D u_B\right)\bigg]\\[6pt]
\mathfrak{s}_1&=u^E u^F n^D n^C\bar{R}_{CEFD}+\left(\frac{u\cdot{\nabla}K}{K}\right)^2+\frac{\hat{\nabla}_A {K}}{K}\bigg[4~u^B {K}^A_B-2\left[(u\cdot{\nabla})u^A\right]-\frac{\hat{\nabla}^A {K}}{K}\bigg]\\
&-(\hat{\nabla}_A u_B)(\hat{\nabla}^A u^B)-(u\cdot {K}\cdot u)^2-\big[(u\cdot\hat{\nabla})u_A\big]\big[(u\cdot\hat{\nabla})u^A\big]+2\left[(u\cdot{\nabla})u^A\right](u^B {K}_{BA})\\
&-3~(u\cdot {K}\cdot {K}\cdot u)-\frac{K}{D}\left(\frac{u\cdot{\nabla}{K}}{K}-u\cdot {K}\cdot u\right)\\[6pt]
\mathfrak{s}_2&=\frac{K^2}{D^2}\bigg[-\frac{{K}}{D}\left(\frac{u\cdot{\nabla}{K}}{K}-u\cdot {K}\cdot u\right)- 2~\lambda- (u\cdot {K}\cdot {K}\cdot u)+2 \left(\frac{{\nabla}_A{K}}{K}\right)u^B {K}^A_B-\bigg(\frac{u\cdot{\nabla}{K}}{K}\bigg)^2\\
&+2\left(\frac{u\cdot{\nabla}{K}}{K}\right)(u\cdot {K}\cdot u)-\bigg(\frac{\hat{\nabla}^D K}{K}\bigg)\bigg(\frac{\hat{\nabla}_D K}{K}\bigg)-(u\cdot {K}\cdot u)^2+n^B n^D u^E u^F\bar{R}_{FBDE}\bigg]
\end{split}
\end{equation}
Here, $\bar{R}_{ABCD}$ is the Riemann tensor of the background metric $g_{AB}$ and $\hat{\nabla}$ is defined through the following equation - for a generic $n$-index tensor $W_{A_1 A_2\cdots A_n}$
\begin{equation}\label{eq:hat_def}
\hat{\nabla}_A W_{A_1 A_2\cdots A_n}=\Pi_A^C \Pi_{A_1}^{C_1}\Pi_{A_2}^{C_2}\cdots\Pi_{A_n}^{C_n}\nabla_C W_{C_1 C_2\cdots C_n}
\end{equation}

We want the sub-subleading order metric in linearized order in $\psi^{-D}$. So, we need to calculate the above integration \eqref{funcn} in linearized order in $\psi^{-D}$. The answers are the following. See \ref{app:integration} for details.
\begin{equation}\label{Int:linear}
\begin{split}
t(R)&=-2\left(\frac{D}{K}\right)^2 e^{-R}\left[R+1\right]
+{\cal O}\left(e^{-2R}\right)\\
v(R)&=2\left(\frac{D}{K}\right)^3\left(1+R+\frac{R^2}{2}\right)e^{-R}+{\cal O}\left(e^{-2R}\right)\\
f_1(R)&=-2\left(\frac{D}{K}\right)^2R~e^{-R}+{\cal O}\left(e^{-2R}\right)\\
f_2(R)&=2\left(\frac{D}{K}\right)^4e^{-R}\left(2~\text{Zeta}[3]-1\right)+{\cal O}(e^{-2R})
\end{split}
\end{equation}
Using \eqref{Int:linear}, we can write the full metric $G_{AB}$ as 
\begin{equation}
\begin{split}
G_{AB}&=g_{AB}+\psi^{-D}O_A O_B+\psi^{-D}\frac{1}{D^2}\bigg[-2\left(\frac{D}{K}\right)^2(R+1)~\mathfrak{t}_{AB}-2\left(\frac{D}{K}\right)^2R~\mathfrak{s}_1~O_AO_B\\
&+2\left(\frac{D}{K}\right)^4\left(2~\text{Zeta}[3]-1\right)\mathfrak{s}_2~O_AO_B+2\left(\frac{D}{K}\right)^3\left(1+R+\frac{R^2}{2}\right)\big(~ \mathfrak{v}_A O_B + \mathfrak{v}_B O_A\big)\bigg]\\[8pt]
&=g_{AB}+\psi^{-D}\bigg[O_A O_B+\frac{1}{K^2}\bigg\{2\frac{D^2}{K^2}\left(2~\text{Zeta}[3]-1\right)\mathfrak{s}_2 O_AO_B-2~\mathfrak{t}_{AB}+2\frac{D}{K}\big(\mathfrak{v}_A O_B + \mathfrak{v}_B O_A\big)\bigg\}\bigg]\\
&+R~\psi^{-D}\frac{1}{K^2}\left[-2~\mathfrak{t}_{AB}-2~\mathfrak{s}_1~ O_A O_B+2\left(\frac{D}{K}\right)\big(~ \mathfrak{v}_A O_B + \mathfrak{v}_B O_A\big)\right]\\
&+R^2~\psi^{-D}\frac{1}{K^2}\left(\frac{D}{K}\right)\big(~ \mathfrak{v}_A O_B + \mathfrak{v}_B O_A\big)+{\cal O}\left(\frac{1}{D^3} ,\psi^{-2D}\right)
\end{split}
\end{equation}
Now, if we write $G_{AB}$ as
\begin{equation}\label{eq:GAB}
\begin{split}
G_{AB}=g_{AB}+\psi^{-D}M_{AB}=g_{AB}+\psi^{-D}\sum_{n=0}^\infty(\psi-1)^nM^{(n)}_{AB}
\end{split}
\end{equation}
We will get
\begin{equation}\label{eq:M0}
\begin{split}
M^{(0)}_{AB}&=O_A O_B+\frac{2}{K^2}\Big[-\mathfrak{t}_{AB}+\Big(\frac{D}{K}\Big)^2\left(2~\text{Zeta}[3]-1\right)\mathfrak{s}_2O_AO_B+\frac{D}{K}\big(\mathfrak{v}_A O_B + \mathfrak{v}_B O_A\big)\Big]+{\cal O}\Big(\frac{1}{D}\Big)^3\\
M^{(1)}_{AB}&=-\frac{2D}{K^2}\Big[\mathfrak{t}_{AB}+\mathfrak{s}_1~ O_A O_B-\frac{D}{K}\big(~ \mathfrak{v}_A O_B + \mathfrak{v}_B O_A\big)\Big]+{\cal O}\Big(\frac{1}{D}\Big)^2\\
M^{(2)}_{AB}&=\Big(\frac{D}{K}\Big)^3\big(~ \mathfrak{v}_A O_B + \mathfrak{v}_B O_A\big)+{\cal O}\Big(\frac{1}{D}\Big)
\end{split}
\end{equation}

\subsection{Change of Gauge Condition}\label{subsec:gauge}
Large-$D$ solution \cite{secondorder} has been derived in the gauge condition $O^A h_{AB}=0$. But, it turns out that in the calculation of the stress tensor it is more convenient to use the gauge condition $n^A h_{AB}=0$. In this subsection, we will implement this gauge transformation.
\begin{equation}
G_{AB}=g_{AB}+\psi^{-D}M_{AB}
\end{equation}
We do the following infinitesimal coordinate transformation 
\begin{equation}
\begin{split}
x^A&\rightarrow x^{\prime A}=x^A - \psi^{-D}\xi^A(x^A)\\
\end{split}
\end{equation}
Under the above coordinate transformation, metric transforms as follows
\begin{equation}
G^{\prime}_{AB}(x^\prime)=G_{AB}(x^\prime)+\nabla_A^\prime\left[\psi^{-D}\xi_B(x^\prime)\right]+\nabla_B^\prime\left[\psi^{-D}\xi_A(x^\prime)\right]
\end{equation}
%
Now, using \eqref{eq:GAB}, we get
\begin{equation}\label{Mprime}
\boxed{M^\prime_{AB}=M_{AB}+\psi^D\nabla_A\left[\psi^{-D}\xi_B\right]+\psi^D\nabla_B\left[\psi^{-D}\xi_A\right]}
\end{equation}
We choose the coordinate transformation in a way such that $n^A M^\prime_{AB}=0$. Now using the expansion $\xi_A=\sum_{n=0}^\infty(\psi-1)^n \xi_A^{(n)}$ we get
\begin{equation}\label{eq:Gauge}
-n^A\sum_{m=0}^\infty (\psi-1)^m M^{(m)}_{AB}=\psi^D(n\cdot\nabla)\left[\psi^{-D}\sum_{m=0}^\infty(\psi-1)^m\xi^{(m)}_B\right]+\psi^D n^A\nabla_B\left[\psi^{-D}\sum_{m=0}^\infty(\psi-1)^m\xi^{(m)}_A\right]
\end{equation}
Now, using the following decomposition
\begin{equation}\label{eq:Xi_decomposition}
\begin{split}
\xi_B^{(0)}&=\xi_B^{(0,0)}+\frac{1}{D}\xi_B^{(0,1)}+\frac{1}{D^2}\xi_B^{(0,2)}+\frac{1}{D^3}\xi_B^{(0,3)}+{\cal O}\left(\frac{1}{D}\right)^4\\
\xi_B^{(1)}&=\xi_B^{(1,0)}+\frac{1}{D}\xi_B^{(1,1)}+\frac{1}{D^2}\xi_B^{(1,2)}+{\cal O}\left(\frac{1}{D}\right)^3\\
\xi_B^{(2)}&=\xi_B^{(2,0)}+\frac{1}{D}\xi_B^{(2,1)}+{\cal O}\left(\frac{1}{D}\right)^2
\end{split}
\end{equation}
From \eqref{eq:Gauge}, we can determine $\xi_A^{(m,n)}$ order by order in $\frac{1}{D}$ expansion in terms of $M^{(n)}_{AB}$. See \ref{Appendix:Gauge} for details. Different components of $\xi_B^{(2)}$ become
\begin{equation}\label{eq:xi2}
\begin{split}
\xi_B^{(2,0)}&=0\\
\xi_B^{(2,1)}&=\frac{1}{N}\left[n^A M_{AB}^{(2)}+n^A M_{AB}^{(1)}-\frac{n_B}{2}\left(n\cdot M^{(2)}\cdot n+n\cdot M^{(1)}\cdot n\right)\right]\\
\end{split}
\end{equation}
Different components of $\xi_B^{(1)}$ become
\begin{equation}\label{eq:xi1}
\begin{split}
\xi_B^{(1,0)}&=0\\
\xi_B^{(1,1)}&=\frac{1}{N}\left[n^A M_{AB}^{(1)}+n^A M_{AB}^{(0)}-\frac{n_B}{2}\left(n\cdot M^{(1)}\cdot n+n\cdot M^{(0)}\cdot n\right)\right]\\
\xi_B^{(1,2)}&=\frac{1}{N}\left[(n\cdot\nabla)\xi_B^{(1,1)}+(n\cdot\nabla)\xi_B^{(0,1)}\right]+\frac{1}{N}\left[n^A\nabla_B\xi_A^{(1,1)}+n^A\nabla_B\xi_A^{(0,1)}\right]\\
&+2~\xi_B^{(2,1)}+\xi_B^{(1,1)}-\frac{n_B}{N}\left[n^A(n\cdot\nabla)\xi_A^{(1,1)}+n^A(n\cdot\nabla)\xi_A^{(0,1)}\right]\\
\end{split}
\end{equation}
Different components of $\xi_B^{(0)}$ become
\begin{equation}\label{eq:xi0}
\begin{split}
\xi_B^{(0,0)}&=0\\
\xi_B^{(0,1)}&=\frac{1}{N}\left[n^A M_{AB}^{(0)}-\frac{n_B}{2}\left(n\cdot M^{(0)}\cdot n\right)\right]\\
\xi_B^{(0,2)}&=\frac{1}{N}\left[(n\cdot\nabla)\xi_B^{(0,1)}+n^A\nabla_B\xi_A^{(0,1)}\right]+\xi_B^{(1,1)}-\frac{n_B}{N}\left[n^A(n\cdot\nabla)\xi_A^{(0,1)}\right]\\
\xi_B^{(0,3)}&=\frac{1}{N}\left[(n\cdot\nabla)\xi_B^{(0,2)}+n^A\nabla_B\xi_A^{(0,2)}\right]+\xi_B^{(1,2)}-\frac{n_B}{N}\left[n^A(n\cdot\nabla)\xi_A^{(0,2)}\right]
\end{split}
\end{equation}
Using \eqref{eq:xi2}, \eqref{eq:xi1} and \eqref{eq:xi0} we can calculate $M^\prime_{AB}$ from \eqref{Mprime}.
We expect the final answer to be fully projected and that is what we get. See \ref{Appendix:Gauge} for details.
\begin{equation}\label{MPRIME}
\begin{split}
M^\prime_{AB}&=\Pi^C_A\Pi^{C'}_B\bigg[M^{(0)}_{CC'}+(\psi-1)M^{(1)}_{CC'}+(\psi-1)^2M^{(2)}_{CC'}\bigg]+\hat{\nabla}_A \xi_B^{(0)}+\hat{\nabla}_B \xi_A^{(0)}\\
&+(\psi-1)\left(\hat{\nabla}_A\xi_B^{(1)}+\hat{\nabla}_B\xi_A^{(1)}\right)+{\cal O}\left(\frac{1}{D}\right)^3
\end{split}
\end{equation}
%
Using \eqref{eq:M0}, \eqref{eq:ter1} and \eqref{eq:ter3} we can finally write $M^\prime_{AB}$ as
\begin{equation}
\boxed{M^\prime_{AB}=(\psi-1)^m {M^\prime}^{(m)}_{AB}}
\end{equation}
Where,
\begin{equation}
\begin{split}
{M^\prime}^{(0)}_{AB}&=u_A u_B+\frac{1}{\psi K}\Big[u_A\frac{\hat{\nabla}_B K}{K}+u_B\frac{\hat{\nabla}_A K}{K}+K_{AB}-\hat{\nabla}_B u_A-\hat{\nabla}_A u_B\Big]\\
&+\frac{2}{K^2}\Big[-\mathfrak{t}_{AB}+\frac{D^2}{K^2}\left(2~\text{Zeta}[3]-1\right)\mathfrak{s}_2~u_Au_B-\frac{D}{K}\big(\mathfrak{v}_A u_B + \mathfrak{v}_B u_A\big)\Big]\\
&+\frac{1}{K^2}\Big[-\frac{(n\cdot\nabla)K}{K}\Big(4~u_A\frac{\hat{\nabla}_B K}{K}+K_{AB}-2~\hat{\nabla}_B u_A\Big)+2~u_A\hat{\nabla}_B\Big(\frac{n\cdot\nabla K}{K}\Big)\\
&+\hat{\nabla}_B\left\{u^E K_{AE}-(n\cdot\nabla)u_A\right\}-2\frac{\hat{\nabla}_B K}{K}\left\{u^E K_{AE}-\Pi_A^C(n\cdot\nabla)u_C\right\}\Big]\\
&+\frac{1}{K^2}\Big[-\frac{n\cdot\nabla K}{K}\Big(4~u_B\frac{\hat{\nabla}_A K}{K}+K_{AB}-2~\hat{\nabla}_A u_B\Big)+2~u_B\hat{\nabla}_A\Big(\frac{n\cdot\nabla K}{K}\Big)\\
&+\hat{\nabla}_A\left\{u^E K_{BE}-(n\cdot\nabla)u_B\right\}-2\frac{\hat{\nabla}_A K}{K}\left\{u^E K_{BE}-\Pi_B^C(n\cdot\nabla)u_C\right\}\Big]+{\cal O}\Big(\frac{1}{D}\Big)^3
\end{split}
\end{equation}
\vspace{-0.1in}
\begin{equation}
\begin{split}
{M^\prime}^{(1)}_{AB}&=-\frac{2D}{K^2}\Big[\mathfrak{t}_{AB}+\mathfrak{s}_1~ u_A u_B+\frac{D}{K}\big(~ \mathfrak{v}_A u_B + \mathfrak{v}_B u_A\big)\Big]\\
&~~~~+\frac{1}{K}\Big[u_A\frac{\hat{\nabla}_B K}{K}+u_B\frac{\hat{\nabla}_A K}{K}+K_{AB}-\hat{\nabla}_B u_A-\hat{\nabla}_A u_B\Big]+{\cal O}\Big(\frac{1}{D}\Big)^2
\end{split}
\end{equation}

\subsection{Change of Subsidiary Condition}\label{subsec:subsidiary}
${M^\prime}^{(m)}_{AB}$ can not yet be identified with ${h}^{(m)}_{AB}$ - we have used in the calculation of the stress tensor. Because, we have imposed the condition $\Pi^{C}_{A}\Pi^{C'}_{B}(n.\nabla){h}^{(m)}_{CC'}=0$ on ${h}^{(m)}_{CC'}$. We will expand 
${M^\prime}^{(m)}_{AB}$ in a power series expansion in $(\psi-1)$ and will determine different coefficients by satisfying $\Pi^{C}_{A}\Pi^{C'}_{B}(n.\nabla){h}^{(m)}_{CC'}=0$.\\
We define $h^{(0)}_{AB}$ in the following way such that $\Pi^{C}_{A}\Pi^{C'}_{B}(n.\nabla)h^{(0)}_{AB}=0$
\begin{equation}\label{eq:h0AB_def}
h^{(0)}_{AB}={M^\prime}^{(0)}_{AB}-(\psi-1)C^{(0)}_{AB}-(\psi-1)^2E^{(0)}_{AB}+{\cal O}(\psi-1)^3
\end{equation}
Acting on the above equation by $\Pi^{C}_{A}\Pi^{C'}_{B}(n.\nabla)$ and them equating the coefficient of $(\psi-1)^0$ we get
\begin{equation}\label{eq:C0AB_def}
C^{(0)}_{AB}=\frac{1}{N}\Pi^{C}_{A}\Pi^{C'}_{B}(n.\nabla){M^\prime}^{(0)}_{CC'}
\end{equation}
Equating the coefficient of $(\psi-1)$ we get
\begin{equation}\label{eq:E0AB_def}
E^{(0)}_{CC'}=-\frac{1}{2N}\Pi^{A}_{C}\Pi^{B}_{C'}(n.\nabla)C^{(0)}_{AB}
\end{equation}
The final form form of $h^{(0)}_{AB}$ on $\psi=1$ takes the following form. See  \ref{app:outside} for details
\begin{equation}\label{eq:h0AB}
\boxed{h^{(0)}_{AB}={\cal S}^{(0)}~u_A u_B+u_A{\cal H}^{(0)}_B+u_B{\cal H}^{(0)}_A+{\cal W}^{(0)}_{AB}}
\end{equation}
Where,
\begin{equation}
\begin{split}
{\cal S}^{(0)}&=1-\frac{2}{K^2}\bigg[u\cdot K\cdot K\cdot u-3\bigg(\frac{(u\cdot\nabla)K}{K}\bigg)^2-2~u^B K_{BD}\bigg(\frac{\hat{\nabla}^D K}{K}\bigg)+2~u\cdot K\cdot u~\bigg(\frac{(u\cdot\nabla)K}{K}\bigg)\\
&+\frac{K}{D}\bigg(\frac{(u\cdot\nabla)K}{K}\bigg)-\frac{K}{D}(u\cdot K\cdot u)\bigg]\\
&+\frac{2}{K^2}\left(2~\text{Zeta}[3]-1\right)\bigg[-\frac{K}{D}\bigg(\frac{(u\cdot\nabla)K}{K}-u\cdot K\cdot u\bigg)-\lambda-u\cdot K\cdot K\cdot u+2\left(\frac{\nabla_A K}{K}\right)u^B K^A_B\\
&-\bigg(\frac{(u\cdot\nabla)K}{K}\bigg)^2+2~\frac{(u\cdot\nabla)K}{K}(u\cdot K\cdot u)-\bigg(\frac{\hat{\nabla}^D K}{K}\bigg)\bigg(\frac{\hat{\nabla}_D K}{K}\bigg)-(u\cdot K\cdot u)^2\bigg]
\end{split}
\end{equation}
\begin{equation}
\begin{split}
{\cal H}^{(0)}_A&=\frac{1}{K}\bigg(\frac{\hat{\nabla}_A K}{K}\bigg)+\frac{2}{K^2}\hat{\nabla}_A\bigg(\frac{\hat{\nabla}^2 K}{K^2}\bigg)+\frac{2}{K^2}K^F_A\bigg(\frac{\hat{\nabla}_F K}{K}\bigg)-\frac{2}{K^2}\left(\hat{\nabla}^F u_A\right)\bigg(\frac{\hat{\nabla}_F K}{K}\bigg)\\
&+\frac{2}{K^2}\bigg(\frac{\hat{\nabla}^2 u_A}{K}\bigg)\bigg[u\cdot K\cdot u-2~\frac{(u\cdot\nabla)K}{K}\bigg]+\frac{2}{K^2}\bigg(\frac{\hat{\nabla}_A K}{K}\bigg)\bigg[u\cdot K\cdot u-2~\frac{(u\cdot\nabla)K}{K}+\lambda~\frac{D}{K}+\frac{K}{2D}\bigg]
\end{split}
\end{equation}
\begin{equation}
\begin{split}
{\cal W}^{(0)}_{AB}&=\frac{1}{K}\left[K_{AB}-\hat{\nabla}_A u_B-\hat{\nabla}_B u_A\right]\\
&-\frac{2}{K^2}~K_{AB}\left[\frac{(u\cdot\nabla)K}{K}-u\cdot K\cdot u\right]-\frac{2}{K^2}\left(\hat{\nabla}_A u_B+\hat{\nabla}_B u_A\right)\left[\frac{K}{2D}-2~\frac{(u\cdot\nabla)K}{K}+u\cdot K\cdot u\right]\\
&+\frac{2}{K^2}K^F_A K_{FB}-\frac{2}{K^2}\left(K^F_A\hat{\nabla}_F u_B+K^F_B\hat{\nabla}_F u_A\right)+\frac{2}{K^2}\big(\hat{\nabla}^F u_A\big)\big(\hat{\nabla}_F u_B\big)+\frac{2}{K^2}\bigg(\frac{\hat{\nabla}^2 u_A}{K}\bigg)\bigg(\frac{\hat{\nabla}^2 u_B}{K}\bigg)\\
&+\frac{2}{K^2}\bigg(\frac{\hat{\nabla}_A K}{K}\bigg)\bigg(\frac{\hat{\nabla}_B K}{K}\bigg)-\frac{2}{K^2}\bigg[\bigg(\frac{\hat{\nabla}_A K}{K}\bigg)u^EK_{EB}+\bigg(\frac{\hat{\nabla}_B K}{K}\bigg)u^EK_{EA}\bigg]\\
&+\frac{1}{K^2}\left[\hat{\nabla}_A\left(u^E K_{EB}\right)+\hat{\nabla}_B\left(u^E K_{EA}\right)\right]-\frac{1}{K^2}\bigg[\hat{\nabla}_A\bigg(\frac{\hat{\nabla}^2 u_B}{K}\bigg)+\hat{\nabla}_B\bigg(\frac{\hat{\nabla}^2 u_A}{K}\bigg)\bigg]
\end{split}
\end{equation}
Now, $h^{(1)}_{AB}$ on the surface $\psi=1$ becomes
\begin{equation}\label{eq:h1AB_def}
h^{(1)}_{AB}={M^\prime}^{(1)}_{AB}+C^{(0)}_{AB}
\end{equation}
The final form form of $h^{(1)}_{AB}$ on $\psi=1$ takes the following form. See  \ref{app:outside} for details
\begin{equation}\label{eq:h1AB}
\boxed{h^{(1)}_{AB}={\cal S}^{(1)}~u_A u_B+u_A{\cal H}^{(1)}_B+u_B{\cal H}^{(1)}_A+{\cal W}^{(1)}_{AB}}
\end{equation}
Where
\begin{equation}
{\cal S}^{(1)}=-2~\lambda \left(\frac{D}{K^2}\right)
\end{equation}
\begin{equation}
\begin{split}
{\cal H}^{(1)}_A&=\frac{D}{K}\bigg(\frac{\hat{\nabla}^2 u_A}{K}\bigg)+\frac{D}{K^2}\bigg(\frac{\hat{\nabla}_A K}{K}\bigg)\bigg[-5~\frac{(u\cdot\nabla)K}{K}+2~u\cdot K\cdot u-\lambda~\frac{D}{K}\bigg]\\
&+\frac{D}{K^2}\bigg(\frac{\hat{\nabla}^2 u_A}{K}\bigg)\bigg[-12~\frac{(u\cdot\nabla)K}{K}+6~u\cdot K\cdot u-2~\lambda~\frac{D}{K}+2~\frac{K}{D}\bigg]\\
&+\frac{D}{K^2}\bigg[-u^B K_{BD}K^D_A+\frac{1}{K^2}\hat{\nabla}^2\left(\hat{\nabla}^2 u_A\right)-3\bigg(\frac{\hat{\nabla}^B K}{K}\bigg)\hat{\nabla}_B u_A+\frac{1}{K^2}\hat{\nabla}_A\left(\hat{\nabla}^2 K\right)+K^D_A\bigg(\frac{\hat{\nabla}_D K}{K}\bigg)\bigg]
\end{split}
\end{equation}
\begin{equation}
\begin{split}
{\cal W}^{(1)}_{AB}&=\frac{D}{K^2}\left[u\cdot K\cdot u-\frac{K}{D}\right]K_{AB}+\frac{D}{K^2}\bigg[\frac{\hat{\nabla}^2 K}{K^2}-\lambda~\frac{D}{K}\bigg]\left(\hat{\nabla}_A u_B+\hat{\nabla}_B u_A\right)+\frac{D}{K^2}~K^F_A K_{FB}\\
&-\frac{D}{K^2}~\lambda~ \Pi_{AB}-\frac{D}{K^2}\left(K^F_A~\hat{\nabla}_F u_B+K^F_B~\hat{\nabla}_F u_A\right)+2~\frac{D}{K^2}\left(\hat{\nabla}^F u_A\right)\left(\hat{\nabla}_F u_B\right)\\
&+2~\frac{D}{K^2}\bigg(\frac{\hat{\nabla}^2 u_A}{K}\bigg)\bigg(\frac{\hat{\nabla}^2 u_B}{K}\bigg)+\frac{D}{K^2}\frac{1}{K}\hat{\nabla}_A\left(\hat{\nabla}_B K\right)-\frac{D}{K^2}\bigg[\bigg(\frac{\hat{\nabla}_A K}{K}\bigg)u^E K_{EB}+\bigg(\frac{\hat{\nabla}_B K}{K}\bigg)u^E K_{EA}\bigg]\\
&-\frac{D}{K^2}\frac{1}{K}\left[\hat{\nabla}_A\left(\hat{\nabla}^2 u_B\right)+\hat{\nabla}_B\left(\hat{\nabla}^2 u_A\right)\right]+\frac{D}{K^2}\bigg[\bigg(\frac{\hat{\nabla}_A K}{K}\bigg)\bigg(\frac{\hat{\nabla}^2 u_B}{K}\bigg)+\bigg(\frac{\hat{\nabla}_B K}{K}\bigg)\bigg(\frac{\hat{\nabla}^2 u_A}{K}\bigg)\bigg]
\end{split}
\end{equation}
%
%
%
%
%
%
So, finally, we have brought the large-D solution in the following form 
\begin{equation}
G^{(out)}_{AB}=g_{AB}+{\psi}^{-D}\mathfrak{h}_{AB}=g_{AB}+{\psi}^{-D}\sum_{m=0}^\infty (\psi-1)^m h^{(m)}_{AB}
\end{equation}
Where, $h^{(m)}_{AB}$ satisfies $~n^A h^{(m)}_{AB}=0~$ and $~\Pi^A_C\Pi^B_{C'}(n\cdot\nabla)h^{(m)}_{AB}=0~$

\section{Linearized Solution : Inside($\psi<1$)}\label{sec:inside}
In this section, we shall construct the `inside solution' i.e, the metric for region $\psi<1$. As we have mentioned before, we want this metric to be regular throughout the `inside region' in order to make sure that the membrane is the sole source of the gravitational radiation in this system. 

Note that the solution presented in \cite{secondorder} continued to be a solution even when $\psi<1$. However, this solution diverges at the location of the black hole, the point where $\psi$ approaches zero and  also it does not have any discontinuity across the event horizon - the location of the membrane. Therefore, unlike the `outside solution'  we have to construct the inside solution from scratch maintaining the regularity and the fact that on the membrane  it reduces to the same induced metric as the one read off from the `outside solution'.\\\\
We shall write the inside metric in the following form
\begin{equation}\label{metric}
G_{AB}^{(in)}=g_{AB}+\tilde{\mathfrak{h}}_{AB}=g_{AB}+\sum_{m=0}^\infty (\psi-1)^{m}\tilde{h}^{(m)}_{AB}
\end{equation}
Where, $g_{AB}$ is background metric. $\tilde{h}^{(m)}_{AB}$ satisfies the gauge condition $n^A\tilde{h}^{(m)}_{AB}=0$. At linearized order, Christoffel symbol for \eqref{metric} is given by
\begin{equation}
\Gamma^A_{BC}=\bar{\Gamma}^A_{BC}+\underbrace{\frac{1}{2}g^{AC'}\left[\nabla_C\tilde{\mathfrak{h}}_{BC'}+\nabla_B\tilde{\mathfrak{h}}_{CC'}-\nabla_{C'}\tilde{\mathfrak{h}}_{BC}\right]}_{\delta\Gamma^A_{BC}}+{\cal O}\left(\tilde{\mathfrak{h}}\right)^2
\end{equation}
Where, $\bar{\Gamma}^A_{BC}$ is Christoffel symbol of $g_{AB}$ and $\nabla_A$ is covariant derivative with respect to $g_{AB}$. Now, Ricci tensor is given by
\begin{equation}
R_{AB}^{(in)}=\bar{R}_{AB}+\nabla_D\left[\delta\Gamma^D_{AB}\right]-\nabla_B\left[\delta\Gamma^D_{AD}\right]
\end{equation}
Where, $\bar{R}_{AB}$ is Ricci tensor for $g_{AB}$.
\begin{equation}
\begin{split}
\delta\Gamma^A_{BA}&=\frac{1}{2}g^{AC'}\left[\nabla_A \tilde{\mathfrak{h}}_{BC'}+\nabla_B \tilde{\mathfrak{h}}_{AC'}-\nabla_A\tilde{\mathfrak{h}}_{BC'}\right]=\frac{1}{2}\nabla_B \tilde{\mathfrak{h}}
\end{split}
\end{equation}
Where, $\tilde{\mathfrak{h}}=g^{AC'}\tilde{\mathfrak{h}}_{AC'}$.
So, Ricci tensor for inside region$ (\psi<1)$
\begin{equation}
R_{AB}^{(in)}=\bar{R}_{AB}+\frac{1}{2}\nabla_D\nabla_A \tilde{\mathfrak{h}}_B^D+\frac{1}{2}\nabla_D\nabla_B\tilde{\mathfrak{h}}_A^D-\frac{1}{2}\nabla^2 \tilde{\mathfrak{h}}_{AB}-\frac{1}{2}\nabla_B\nabla_A\tilde{\mathfrak{h}}
\end{equation}
Einstein equation in the inside region
\begin{equation}\label{Einstein}
\begin{split}
&R_{AB}^{\text{(in)}}-(D-1)\lambda~G_{AB}^{\text{(in)}}=0\\
\Rightarrow~&\frac{1}{2}\nabla_D\nabla_A \tilde{\mathfrak{h}}_B^D+\frac{1}{2}\nabla_D\nabla_B\tilde{\mathfrak{h}}_A^D-\frac{1}{2}\nabla^2 \tilde{\mathfrak{h}}_{AB}-\frac{1}{2}\nabla_B\nabla_A\tilde{\mathfrak{h}}-(D-1)\lambda\tilde{\mathfrak{h}}_{AB}=0\\
\end{split}
\end{equation}
Projecting the above equation perpendicular to $n_A$ and $n_B$ we get
\begin{equation}\label{eq:Einstein_in}
\begin{split}
&\Pi^A_C\Pi^B_{C'}\bigg[\nabla_A\nabla_E\tilde{\mathfrak{h}}^E_B+\nabla_B\nabla_E\tilde{\mathfrak{h}}^E_A-\nabla^2 \tilde{\mathfrak{h}}_{AB}-\nabla_B\nabla_A\tilde{\mathfrak{h}}+2\bar{R}_{EABC}\tilde{\mathfrak{h}}^{EC}\\
&+\bar{R}_{AC}\tilde{\mathfrak{h}}^C_B+\bar{R}_{BC}\tilde{\mathfrak{h}}^C_A-2(D-1)\lambda\tilde{\mathfrak{h}}_{AB}\bigg]=0
\end{split}
\end{equation}
Using the following decomposition for $\tilde{h}^{(1)}_{AB}$
\begin{equation}\label{eq:h1t_def}
\tilde{h}^{(1)}_{AB}=\tilde{h}^{(1,1)}_{AB}+\frac{1}{D}\tilde{h}^{(1,2)}_{AB}
\end{equation}
We can solve for $\tilde{h}^{(1,1)}_{AB}, \tilde{h}^{(1,2)}_{AB}, \tilde{h}^{(2)}_{AB}$ by solving \eqref{eq:Einstein_in} order by order in $\frac{1}{D}$ expansion. The final form form of $\tilde{h}^{(1,1)}_{AB}$ on $\psi=1$ takes the following form. See  \ref{app:inside} for details
\begin{equation}\label{eq:htilde11}
\boxed{\tilde{h}^{(1,1)}_{CC'}=\tilde{\cal S}^{(1,1)}~u_C u_{C'}+u_C\tilde{\cal H}^{(1,1)}_{C'}+u_{C'}\tilde{\cal H}^{(1,1)}_C+\tilde{\cal W}^{(1,1)}_{CC'}}
\end{equation}
where,
\begin{equation}
\tilde{\cal S}^{(1,1)}={\cal O}\left(\frac{1}{D}\right)^2
\end{equation}
\begin{equation}
\begin{split}
\tilde{\cal H}^{(1,1)}_C&=-\frac{D}{K}\bigg(\frac{\hat{\nabla}^2 u_C}{K}\bigg)+\frac{D}{K^2}\bigg[\frac{\hat{\nabla}^2 K}{K^2}-\lambda\frac{D}{K}-\frac{K}{D}\bigg]\bigg(\frac{\hat{\nabla}^2 u_C}{K}\bigg)-\frac{D}{K^4}\hat{\nabla}_C\left(\hat{\nabla}^2K\right)-\frac{D}{K^2}K^D_C\bigg(\frac{\hat{\nabla}_D K}{K}\bigg)\\
&~~~~+\frac{D}{K^2}K^F_C K_{FD}u^D+\frac{D}{K^2}\bigg(\frac{\hat{\nabla}^F K}{K}\bigg)\left(\hat{\nabla}_F u_C\right)+\frac{D}{K^2}\bigg(\frac{\hat{\nabla}_C K}{K}\bigg)\bigg[2~\frac{\hat{\nabla}^2 K}{K^2}+\frac{u\cdot\nabla K}{K}-\lambda\frac{D}{K}\bigg]
\end{split}
\end{equation}
\begin{equation}
\begin{split}
\tilde{\cal W}^{(1,1)}_{CC'}&=-2\frac{D}{K^2}\left(\hat{\nabla}^D u_C\right)\left(\hat{\nabla}_D u_{C'}\right)-2\frac{D}{K^2}(u\cdot K\cdot u)K_{CC'}+\lambda \frac{D}{K^2}\Pi_{CC'}\\
&-\frac{D}{K^2}\bigg[\frac{\hat{\nabla}^2 K}{K^2}-\lambda \frac{D}{K}\bigg]\left(\hat{\nabla}_C u_{C'}+\hat{\nabla}_{C'} u_C\right)-\frac{D}{K^2}\bigg[\bigg(\frac{\hat{\nabla}^2 u_C}{K}\bigg)\bigg(\frac{\hat{\nabla}_{C'}K}{K}\bigg)+\bigg(\frac{\hat{\nabla}^2 u_{C'}}{K}\bigg)\bigg(\frac{\hat{\nabla}_C K}{K}\bigg)\bigg]\\
&+\frac{D}{K^2}\bigg[\bigg(\frac{\hat{\nabla}_C K}{K}\bigg)u^F K_{FC'}+\bigg(\frac{\hat{\nabla}_{C'} K}{K}\bigg)u^F K_{FC}\bigg]-\frac{D}{K^2}K^E_C K_{EC'}-\frac{D}{K^2}\frac{1}{K}\hat{\nabla}_C\left(\hat{\nabla}_{C'}K\right)\\
&+\frac{D}{K^2}\left[K^D_C\left(\hat{\nabla}_D u_{C'}\right)+K^D_{C'}\left(\hat{\nabla}_D u_C\right)\right]+\frac{D}{K^2}\frac{1}{K}\left[\hat{\nabla}_C\left(\hat{\nabla}^2 u_{C'}\right)+\hat{\nabla}_{C'}\left(\hat{\nabla}^2 u_C\right)\right]
\end{split}
\end{equation}\\
The final form form of $\tilde{h}^{(1,2)}_{AB}$ on $\psi=1$ takes the following form. See  \ref{app:inside} for details
\begin{equation}\label{eq:htilde12}
\boxed{\tilde{h}^{(1,2)}_{CC'}=\tilde{\cal S}^{(1,2)}~u_C u_{C'}+u_C\tilde{\cal H}^{(1,2)}_{C'}+u_{C'}\tilde{\cal H}^{(1,2)}_C+\tilde{\cal W}^{(1,2)}_{CC'}}
\end{equation}
Where,
\begin{equation}
\tilde{\cal S}^{(1,2)}=2~\lambda\left(\frac{D}{K}\right)^2
\end{equation}
\begin{equation}
\begin{split}
\tilde{\cal H}^{(1,2)}_C&=\frac{D}{K}\bigg[-1+\frac{D}{K}\bigg(\frac{\hat{\nabla}^2 K}{K^2}\bigg)+\lambda~\frac{D^2}{K^2}\bigg]\bigg(\frac{\hat{\nabla}^2 u_C}{K}\bigg)\\
&+\frac{D^2}{K^2}\bigg[-\bigg(\frac{\hat{\nabla}^2\hat{\nabla}^2 u_C}{K^2}\bigg)-2\bigg(\frac{\hat{\nabla}^E K}{K}\bigg)\left(\hat{\nabla}_E u_C\right)+2\hat{\nabla}_C\bigg(\frac{(u\cdot\nabla)K}{K}\bigg)\bigg]
\end{split}
\end{equation}
\begin{equation}
\begin{split}
\tilde{\cal W}^{(1,2)}_{CC'}&=-2~\frac{D^2}{K^2}\bigg(\frac{\hat{\nabla}^2 u_C}{K}\bigg)\bigg(\frac{\hat{\nabla}^2 u_{C'}}{K}\bigg)+2~\frac{D^2}{K^2}\left(\hat{\nabla}_C u_{C'}+\hat{\nabla}_{C'} u_C\right)\frac{(u\cdot\nabla)K}{K}
\end{split}
\end{equation}
The final form form of $\tilde{h}^{(2)}_{AB}$ on $\psi=1$ takes the following form. See  \ref{app:inside} for details
\begin{equation}\label{eq:htilde2}
\boxed{\tilde{h}^{(2)}_{CC'}=\tilde{\cal S}^{(2)}~u_C u_{C'}+u_C\tilde{\cal H}^{(2)}_{C'}+u_{C'}\tilde{\cal H}^{(2)}_C+\tilde{\cal W}^{(2)}_{CC'}}
\end{equation}
Where,
\begin{equation}
\tilde{\cal S}^{(2)}={\cal O}\left(\frac{1}{D}\right)
\end{equation}
\begin{equation}
\tilde{\cal H}^{(2)}_C=\frac{D}{K}\bigg[-\frac{1}{2}-2~\frac{D}{K}\bigg(\frac{\hat{\nabla}^2 K}{K^2}\bigg)+\lambda~\frac{D^2}{K^2}\bigg]\bigg(\frac{\hat{\nabla}^2 u_C}{K}\bigg)+\frac{D^2}{2K^2}\bigg[\frac{\hat{\nabla}^2\hat{\nabla}^2 u_C}{K^2}-2\bigg(\frac{\hat{\nabla}^E K}{K}\bigg)\left(\hat{\nabla}_E u_C\right)\bigg]
\end{equation}
\begin{equation}
\tilde{\cal W}^{(2)}_{CC'}=\frac{D^2}{K^2}\bigg(\frac{\hat{\nabla}^2 u_C}{K}\bigg)\bigg(\frac{\hat{\nabla}^2 u_{C'}}{K}\bigg)
\end{equation}\\
Adding \eqref{eq:htilde11} and \eqref{eq:htilde12} we get
\begin{equation}\label{eq:htilde1}
\boxed{\tilde{h}^{(1)}_{CC'}=\tilde{\cal S}^{(1)}~u_C u_{C'}+u_C\tilde{\cal H}^{(1)}_{C'}+u_{C'}\tilde{\cal H}^{(1)}_C+\tilde{\cal W}^{(1)}_{CC'}}
\end{equation}
Where,
\begin{equation}
\tilde{\cal S}^{(1)}=2~\lambda~\frac{D}{K^2}
\end{equation}
\begin{equation}
\begin{split}
\tilde{\cal H}^{(1)}_C&=-\frac{D}{K}\bigg(\frac{\hat{\nabla}^2 u_C}{K}\bigg)+\frac{D}{K^2}\bigg[2~\frac{\hat{\nabla}^2 K}{K^2}-2~\frac{K}{D}\bigg]\bigg(\frac{\hat{\nabla}^2 u_C}{K}\bigg)-\frac{D}{K^4}\hat{\nabla}_C\left(\hat{\nabla}^2K\right)-\frac{D}{K^2}K^D_C\bigg(\frac{\hat{\nabla}_D K}{K}\bigg)\\
&~~~~+\frac{D}{K^2}K^F_C K_{FD}u^D-\frac{D}{K^2}\bigg(\frac{\hat{\nabla}^F K}{K}\bigg)\left(\hat{\nabla}_F u_C\right)+\frac{D}{K^2}\bigg(\frac{\hat{\nabla}_C K}{K}\bigg)\bigg[2~\frac{\hat{\nabla}^2 K}{K^2}+\frac{u\cdot\nabla K}{K}-\lambda\frac{D}{K}\bigg]\\
&~~~~-\frac{D}{K^2}\bigg(\frac{\hat{\nabla}^2\hat{\nabla}^2 u_C}{K^2}\bigg)+2~\frac{D}{K^2}\hat{\nabla}_C\bigg(\frac{(u\cdot\nabla)K}{K}\bigg)
\end{split}
\end{equation}
\begin{equation}
\begin{split}
\tilde{\cal W}^{(1)}_{CC'}&=-2\frac{D}{K^2}\left(\hat{\nabla}^D u_C\right)\left(\hat{\nabla}_D u_{C'}\right)-2\frac{D}{K^2}(u\cdot K\cdot u)K_{CC'}+\lambda \frac{D}{K^2}\Pi_{CC'}\\
&-\frac{D}{K^2}\bigg[\frac{\hat{\nabla}^2 K}{K^2}-\lambda \frac{D}{K}-2~\frac{(u\cdot\nabla)K}{K}\bigg]\left(\hat{\nabla}_C u_{C'}+\hat{\nabla}_{C'} u_C\right)\\
&-\frac{D}{K^2}\bigg[2~\bigg(\frac{\hat{\nabla}^2 u_C}{K}\bigg)\bigg(\frac{\hat{\nabla}^2 u_{C'}}{K}\bigg)+\bigg(\frac{\hat{\nabla}^2 u_C}{K}\bigg)\bigg(\frac{\hat{\nabla}_{C'}K}{K}\bigg)+\bigg(\frac{\hat{\nabla}^2 u_{C'}}{K}\bigg)\bigg(\frac{\hat{\nabla}_C K}{K}\bigg)\bigg]\\
&+\frac{D}{K^2}\bigg[\bigg(\frac{\hat{\nabla}_C K}{K}\bigg)u^F K_{FC'}+\bigg(\frac{\hat{\nabla}_{C'} K}{K}\bigg)u^F K_{FC}\bigg]-\frac{D}{K^2}K^E_C K_{EC'}-\frac{D}{K^2}\frac{1}{K}\hat{\nabla}_C\left(\hat{\nabla}_{C'}K\right)\\
&+\frac{D}{K^2}\left[K^D_C\left(\hat{\nabla}_D u_{C'}\right)+K^D_{C'}\left(\hat{\nabla}_D u_C\right)\right]+\frac{D}{K^2}\frac{1}{K}\left[\hat{\nabla}_C\left(\hat{\nabla}^2 u_{C'}\right)+\hat{\nabla}_{C'}\left(\hat{\nabla}^2 u_C\right)\right]\\
\end{split}
\end{equation}

\section{Stress Tensor}\label{sec:stress}
In this section, we will derive the expression for membrane stress tensor. The membrane stress tensor is given by the discontinuity of the Brown-York stress tensor across the membrane.\footnote{See subsection 3.3 of \cite{radiation} for detailed discussion on this}
\subsection{Outside($\psi>1$) Stress Tensor}
The outside stress tensor is given by
\begin{equation}\label{eq:TAB:OUT}
8\pi T_{AB}^{(out)}=K^{(out)}_{AB}-K^{(out)}\mathfrak{p}_{AB}^{(out)}\bigg{|}_{\psi=1}
\end{equation}
\vspace{-0.1in}
\begin{equation}
\text{Where, }\mathfrak{p}_{AB}^{(out)}=G_{AB}^{(out)}-n_A^{(out)}n_B^{(out)};~G_{AB}^{(out)}=g_{AB}+\psi^{-D}\mathfrak{h}_{AB};~n_A^{(out)}=\frac{\partial_A \psi}{\sqrt{G^{AB}_{(out)}\partial_A \psi~\partial_B \psi}}\nonumber
\end{equation}
\begin{equation}\label{eq:KABout_def}
K^{(out)}_{AB}=\left[\mathfrak{p}^{(out)}\right]^C_A \left[\mathfrak{p}^{(out)}\right]^{C'}_B \left(\tilde{\nabla}_C n_{C'}^{(out)}\right)_{\psi=1}
\end{equation}
\begin{equation}
\text{Where,~} \mathfrak{p}_{AB}^{(out)}=G_{AB}^{(out)}-n_A^{(out)}n_B^{(out)} \text{ and,}~ \tilde{\nabla} \text{ is covariant derivative with respect to } G_{AB}^{(out)}
\end{equation}
The final expression for $K^{(out)}_{AB}$ and $K^{(out)}$ are the followings. See \ref{app:stress} for details.
\begin{equation}\label{eq:KAB:OUT}
\begin{split}
K_{AB}^{(out)}&=K_{AB}-\frac{ND}{2}h^{(0)}_{AB}+\frac{N}{2}h^{(1)}_{AB}+\frac{1}{2}\left(h^{(0)}_{BD}K^D_A+h^{(0)}_{AD}K^D_B\right)\\
K^{(out)}&=K-\frac{ND}{2}h^{(0)}+\frac{N}{2}h^{(1)}
\end{split}
\end{equation}
Putting the expression for $K_{AB}^{\text{(out)}}$ and $K^{\text{(out)}}$ from \eqref{eq:KAB:OUT} in \eqref{eq:TAB:OUT} we get the final expression of $ T_{AB}^{(out)}$.
\begin{equation}\label{eq:TABout_expr}
\begin{split}
8\pi T_{AB}^{(out)}&=K_{AB}-\frac{ND}{2}h^{(0)}_{AB}+\frac{N}{2}h^{(1)}_{AB}+\frac{1}{2}\left(h^{(0)}_{BD}K^D_A+h^{(0)}_{AD}K^D_B\right)\\
&-\left(\Pi_{AB}+h^{(0)}_{AB}\right)\left(K-\frac{ND}{2}h^{(0)}+\frac{N}{2}h^{(1)}\right)
\end{split}
\end{equation}

\subsection{Inside($\psi<1$) Stress Tensor}
The inside stress tensor is given by
\begin{equation}\label{eq:TAB:IN}
8\pi T_{AB}^{(in)}=K^{(in)}_{AB}-K^{(in)}\mathfrak{p}_{AB}^{(in)}\bigg{|}_{\psi=1}
\end{equation}
\begin{equation}
\text{Where, }\mathfrak{p}_{AB}^{(in)}=G_{AB}^{(in)}-n_A^{(in)}n_B^{(in)};~G_{AB}^{(in)}=g_{AB}+\tilde{\mathfrak{h}}_{AB};~n_A^{(in)}=\frac{\partial_A \psi}{\sqrt{G^{AB}_{(in)}\partial_A \psi~\partial_B \psi}}
\end{equation}
Now,
\begin{equation}\label{eq:KABin_def}
K^{(in)}_{AB}=\left[\mathfrak{p}^{(in)}\right]^C_A \left[\mathfrak{p}^{(in)}\right]^{C'}_B \left(\dot{\nabla}_C n_{C'}^{(in)}\right)_{\psi=1}
\end{equation}
\begin{equation}
\text{Where,~} \mathfrak{p}_{AB}^{(in)}=G_{AB}^{(in)}-n_A^{(in)}n_B^{(in)} \text{ and,}~ \dot{\nabla} \text{ is covariant derivative with respect to } G_{AB}^{(in)}
\end{equation}
The final expression for $K^{(in)}_{AB}$ and $K^{(in)}$ are the followings. See \ref{app:stress} for details.
\begin{equation}\label{eq:KAB:IN}
\begin{split}
K^{\text{(in)}}_{AB}&=K_{AB}+\frac{1}{2}\left(\tilde{h}^{(0)}_{BF}K^F_A+\tilde{h}^{(0)}_{AF}K^F_{B}+N\tilde{h}^{(1)}_{AB}\right)\\
K^{\text{(in)}}&=K+\frac{N}{2}\tilde{h}^{(1)}
\end{split}
\end{equation}
Putting the expression for $K_{AB}^{(in)}$ and $K^{(in)}$ from \eqref{eq:KAB:IN} in \eqref{eq:TAB:IN} and using the fact that $\tilde{h}^{(0)}_{AB}=h^{(0)}_{AB}$ we get the final expression of $ T_{AB}^{(in)}$.
\begin{equation}
8\pi T_{AB}^{\text{(in)}}=K_{AB}+\frac{1}{2}\left({h}^{(0)}_{BF}K^F_A+{h}^{(0)}_{AF}K^F_{B}+N\tilde{h}^{(1)}_{AB}\right)-\left(\Pi_{AB}+{h}^{(0)}_{AB}\right)\left(K+\frac{N}{2}\tilde{h}^{(1)}\right)
\end{equation}

\subsection{Membrane Stress Tensor}
Membrane stress tensor is given by
\begin{equation}\label{Stresstensor}
\begin{split}
8\pi T_{AB}&=8\pi\left[ T_{AB}^{\text{(in)}}-T_{AB}^{\text{(out)}}\right]\\
&=\frac{ND}{2}\left[h^{(0)}_{AB}-\Pi_{AB} h^{(0)}\right]-\frac{N}{2}\left[h^{(1)}_{AB}-\tilde{h}^{(1)}_{AB}-\Pi_{AB}\left(h^{(1)}-\tilde{h}^{(1)}\right)\right]+{\cal O}\left(h\right)^2
\end{split}
\end{equation}
We can simplify the calculation of stress tensor by using a trick. We define
\begin{equation}\label{eq:TABNT}
8\pi T_{AB}^{(NT)}=\frac{ND}{2}h^{(0)}_{AB}-\frac{N}{2}\left[h^{(1)}_{AB}-\tilde{h}^{(1)}_{AB}\right]
\end{equation}
Then from \eqref{Stresstensor} we can very easily see that $T_{AB}-T_{AB}^{(NT)}\propto \Pi_{AB}$. Let's call this proportionality factor $\Delta$. With this notation membrane stress tensor becomes
\begin{equation}
8\pi T_{AB}=8\pi \left[T_{AB}^{(NT)}+\Delta \Pi_{AB}\right]
\end{equation}
Now, from the condition $K^{AB}T_{AB}=0$ we get
\begin{equation}
\begin{split}
8\pi~\Delta&=-\frac{1}{K}~8\pi\left(K^{AB}T^{(NT)}_{AB}\right)\\
\end{split}
\end{equation}
Using \eqref{eq:h0AB}, \eqref{eq:h1AB}, \eqref{eq:htilde1} and identity \eqref{eq:identity3} in \eqref{eq:TABNT} and after some simplification we get the final form of $T_{AB}^{(NT)}$ as
\begin{equation}
\begin{split}
8\pi T_{AB}^{(NT)}={\cal S}_1~u_A u_B+{\cal V}_A ~u_B+{\cal V}_B ~u_A+\widetilde{\cal W}_{AB}
\end{split}
\end{equation}
\vspace{-0.3in}
\\Where,
\begin{equation}\label{eq:S1}
\begin{split}
{\cal S}_1&=\frac{K}{2}+\frac{1}{2}\bigg(\frac{\hat{\nabla}^2 K}{K^2}-\lambda~\frac{D-1}{K}-\frac{1}{K}K_{AB}K^{AB}\bigg)\\
&+\frac{1}{K}\bigg[-u\cdot K\cdot K\cdot u-13\bigg(\frac{u\cdot\nabla K}{K}\bigg)^2+2~u^B K_{BD}\bigg(\frac{\hat{\nabla}^D K}{K}\bigg)+14~\bigg(\frac{u\cdot\nabla K}{K}\bigg)(u\cdot K\cdot u)\\
&-\frac{K}{D}\bigg(\frac{u\cdot\nabla K}{K}\bigg)+\frac{K}{D}(u\cdot K\cdot u)+\frac{1}{K^3}\hat{\nabla}^2\left(\hat{\nabla}^2 K\right)-4~(u\cdot K\cdot u)^2-8~\lambda \frac{D}{K}\left(\frac{u\cdot\nabla K}{K}\right)\\
&+4~\lambda\frac{D}{K}~(u\cdot K\cdot u)-2~\bigg(\frac{\hat{\nabla}_B K}{K}\bigg)\bigg(\frac{\hat{\nabla}^B K}{K}\bigg)+\lambda-\lambda^2~\frac{D^2}{K^2}\bigg]\\
&+\frac{1}{K}\left(2~\text{Zeta}[3]-1\right)\bigg[-\frac{K}{D}\bigg(\frac{(u\cdot\nabla)K}{K}-u\cdot K\cdot u\bigg)-\lambda-u\cdot K\cdot K\cdot u+2\left(\frac{\nabla_A K}{K}\right)u^B K^A_B\\
&-\bigg(\frac{u\cdot\nabla K}{K}\bigg)^2+2\left(\frac{u\cdot\nabla K}{K}\right)(u\cdot K\cdot u)-\bigg(\frac{\hat{\nabla}^D K}{K}\bigg)\bigg(\frac{\hat{\nabla}_D K}{K}\bigg)-(u\cdot K\cdot u)^2\bigg]
\end{split}
\end{equation}
\begin{equation}\label{eq:VA}
\begin{split}
{\cal V}_A&=\frac{1}{2}\bigg(\frac{\hat{\nabla}_A K}{K}\bigg)-\bigg(\frac{\hat{\nabla}^2 u_A}{K}\bigg)+\frac{1}{K}K^F_A K_{FD}u^D-\frac{1}{K^3}\hat{\nabla}^2\left(\hat{\nabla}^2 u_A\right)+\frac{1}{K}\hat{\nabla}_A\bigg(\frac{u\cdot\nabla K}{K}\bigg)\\
&+\frac{1}{K}\bigg(\frac{\hat{\nabla}^2 u_A}{K}\bigg)\bigg(-2~(u\cdot K\cdot u)+4~\frac{u\cdot\nabla K}{K}+2~\lambda\frac{D}{K}-\frac{K}{D}\bigg)+\frac{1}{2~K}\bigg(\frac{\hat{\nabla}_A K}{K}\bigg)(u\cdot K\cdot u)
\end{split}
\end{equation}
\vspace{-0.2in}
\\And,
\begin{equation}\label{WAB}
\begin{split}
\widetilde{\cal W}_{AB}&=\frac{1}{2}K_{AB}-\frac{1}{2}\left(\hat{\nabla}_A u_B+\hat{\nabla}_B u_A\right)-\frac{1}{K}~K_{AB}\left(u\cdot K\cdot u\right)+\frac{1}{2~K}\left(\hat{\nabla}_A u_B+\hat{\nabla}_B u_A\right)\left(u\cdot K\cdot u\right)\\
&-\frac{1}{K}\left(\hat{\nabla}^F u_A\right)\left(\hat{\nabla}_F u_B\right)-\frac{1}{K}\bigg(\frac{\hat{\nabla}^2 u_A}{K}\bigg)\bigg(\frac{\hat{\nabla}^2 u_B}{K}\bigg)+\frac{\lambda}{K}~\Pi_{AB}\\
&+\frac{1}{2~K}\bigg[\hat{\nabla}_A\bigg(\frac{\hat{\nabla}^2 u_B}{K}\bigg)+\hat{\nabla}_B\bigg(\frac{\hat{\nabla}^2 u_A}{K}\bigg)+\hat{\nabla}_A\left(u^E K_{EB}\right)+\hat{\nabla}_B\left(u^E K_{EA}\right)-2~\hat{\nabla}_A\bigg(\frac{\hat{\nabla}_B K}{K}\bigg)\bigg]\\
\end{split}
\end{equation}
Now, we can calculate $\Delta$
\begin{equation}
\begin{split}
8\pi\Delta&=-\frac{1}{K}~8\pi\left(K^{AB}T^{(NT)}_{AB}\right)\\
&=-\frac{1}{2}\left(u\cdot K\cdot u\right)-\frac{1}{2K}~K^{AB}K_{AB}-\frac{1}{2K}\bigg(\frac{\hat{\nabla}^2 K}{K^2}-\lambda~\frac{D}{K}-\frac{K}{D}\bigg)\left(u\cdot K\cdot u\right)\\
&~~~~-\frac{2}{K}~u_A K^{AB}\bigg(\frac{1}{2}\frac{\hat{\nabla}_BK}{K}-\frac{\hat{\nabla}^2 u_B}{K}\bigg)+\frac{1}{K}~K^{AB}(\nabla_A u_B)
\end{split}
\end{equation}
So, the full stress tensor becomes
\begin{equation}
\boxed{8\pi T_{AB}={\cal S}_1~u_A u_B+{\cal V}_A ~u_B+{\cal V}_B ~u_A+\widetilde{\cal W}_{AB}+\tilde{\cal S}_2~\Pi_{AB}}
\end{equation}
Where, ${\cal S}_1$, ${\cal V}_A$, $\widetilde{\cal W}_{AB}$ are given respectively by \eqref{eq:S1}, \eqref{eq:VA}, \eqref{WAB} and $\widetilde{\cal S}_2$ is given by
\begin{equation}
\begin{split}
\widetilde{\cal S}_2&=-\frac{1}{2}\left(u\cdot K\cdot u\right)-\frac{1}{2K}~K^{AB}K_{AB}-\frac{1}{2K}\bigg(\frac{\hat{\nabla}^2 K}{K^2}-\lambda~\frac{D}{K}-\frac{K}{D}\bigg)\left(u\cdot K\cdot u\right)\\
&~~~~-\frac{2}{K}~u_A K^{AB}\bigg(\frac{1}{2}\frac{\hat{\nabla}_BK}{K}-\frac{\hat{\nabla}^2 u_B}{K}\bigg)+\frac{1}{K}~K^{AB}(\nabla_A u_B)
\end{split}
\end{equation}

\section{Conservation of the Membrane Stress Tensor}\label{sec:conservation}
The final expression of membrane stress tensor \eqref{eq:STRESS_TENSOR} is very large. It would be quite difficult to calculate the divergence of stress tensor by hand. We have written a $Mathematica$ code to calculate the divergence of the stress tensor, and verified that the divergence of the membrane stress tensor indeed gives the membrane equation. Specifically, we have checked the followings
\begin{itemize}
\item $u^A \hat{\nabla}^B T_{AB}$ gives scalar membrane equation ( second equation of 2.17 in \cite{secondorder} )
\item $P^A_C \hat{\nabla}^B T_{AB}$ gives vector membrane equation ( first equation of 2.17 in \cite{secondorder} )
\end{itemize}
Here, we want to make some comments about how we have done the large-$D$ calculation in \textit{Mathematica}. We choose the following background metric 
\begin{equation}
ds^2=-e^{2r} dt^2+dr^2+e^{2r} dx_a dx^a+e^{2r}  dx_\mu dx^\mu
\end{equation}
which is pure AdS metric written in a slightly different coordinates than usual Poincare patch coordinates$\big($ $r\rightarrow \log r$ will give usual Poincare patch metric $\big)$. Here, `$a$' runs over some finite $p$ dimension and $\mu$ runs over large $D-p-2$ dimension. $\psi$ and $u_A$ are only functions of $(t,r,x_a)$ but does not depend on $x_\mu$. We can effectively do our calculation in finite $p+2$ dimension. We will calculate the contribution that will come from the large $D-p-2$ dimension  by hand and will accordingly take into account. For example, if we want to calculate $\hat{\nabla}^B \hat{\nabla}_B u_A$ (where $A,B$ runs over full $D$ dimension), the first thing to note is that it has non zero component only along `$a$' direction and it is given by 
\begin{equation}
\hat{\nabla}^B \hat{\nabla}_B u_a=\hat{\nabla}^b \hat{\nabla}_b u_a+\frac{1}{2}\frac{D-p-2}{e^{2r}}\big(\hat{\nabla}^f e^{2r}\big)\big(\hat{\nabla}_f u_a\big)-\frac{D-p-2}{4~ e^{4r}}\big(\hat{\nabla}_a e^{2r}\big)\big[(u\cdot\partial)e^{2r}\big]
\end{equation}
Where $\nabla_b$ is covariant derivative with respect to finite $p+2$ dimensional metric. Similarly, we can calculate all the quantities appearing in the expression of the stress tensor.
\section{Conclusions}
In this note, we have calculated the membrane stress tensor up to order ${\cal O}\left(\frac{1}{D}\right)$ and showed that the conservation of this stress tensor gives the subleading order membrane equation.

Very briefly, our procedure is as follows : given the large-$D$ solution outside the membrane - linearize the solution - search for a regular solution inside the membrane region with the condition that the induced metric is continuous on both sides of the membrane - construct the Brown York stress tensor for inside and outside region - the difference of the Brown York stress tensor across the membrane is the membrane stress tensor.

As it turns out, the computation leading to the stress tensor at subsubleading orders is extremely tedious, though the final result is relatively compact and simple (presented in section 1.1). Still one might wonder what is the point of taking up such a calculation. The key motivation we have already mentioned in the introduction. It is about the finite $D$ completion of membrane stress tensor \citep{Dandekar:2017aiv}. Let us elaborate a little more on that.

Algebraically, the large $D$ expansion technique and membrane-gravity duality  are very similar to that of derivative expansion and fluid-gravity duality. It turns out that in fluid-gravity duality it is possible to write an  expression for the fluid stress tensor that works exactly for the complicated stationary solutions like rotating black holes. But that exact expression is nothing but a truncation of the fluid stress tensor at second order in derivative expansion.
Now in \citep{Dandekar:2017aiv} the authors have proposed a finite $D$ completion for large-$D$ stress tensor, which does not quite work and the  mismatch arises again at second order in terms of derivative expansion. All these facts and the experience with fluid gravity correspondence naturally lead to  the hope that a second order calculation in terms of $({1\over D})$ expansion would help in the final goal set by \citep{Dandekar:2017aiv}.

Though this second order membrane stress tensor is just a small step towards this final goal. We think, the following would be the next few steps, which might help to construct a finite $D$ completion of the membrane stress tensor (if it exists), by generating more data 
\begin{itemize}
\item A detailed matching with the hydrodynamic stress tensor dual to the same gravity system in the regime of overlap for these two perturbation techniques ( namely ${1\over D}$ expansion and derivative expansion (see \citep{prevwork,Bhattacharyya:2019mbz})). Now after computing the membrane   stress tensor, we could extend this matching  to include the effect of the gravitational radiation as well.
\item Recasting known rotating black hole solutions in arbitrary $D$, in the language of large $D$ expansion, capturing few  terms that could contribute in a stationary situation,  to all orders.

\item Finally, evaluating the second order membrane stress tensor on the rotating black holes, hoping some novel pattern or truncation would emerge out of this exercise, that will tell us in general how stationarity is encoded in this large-$D$ expansion technique.
\end{itemize}

We find all of the above projects are interesting, themselves. They will teach us a lot about how perturbation works in gravity and how they could be used to have  analytic control over  the otherwise difficult to handle dynamics of gravitating systems. We leave all these for future work.

%

\section*{Acknowledgements}
I am extremely grateful to Sayantani Bhattacharyya for suggesting me this problem and helping me through many difficulties at various stages of the project. I am also very grateful to her for going through the draft of this note and suggesting numerous improvements. I would like to thank Anirban Dinda, Suman Kundu and Milan Patra for collaboration at the initial stage. I would also like to thank Yogesh Dandekar and Suman Kundu for many illuminating discussions over the course of the project. I would like to acknowledge the hospitality of ICTS Bangalore, TIFR Mumbai and IISER Bhopal while this work was in progress. Finally, I would like to acknowledge my gratitude to the people of India for their steady and generous support to the research in basic sciences.

\appendix
\section{Calculation of integrals \eqref{funcn} at linear order}\label{app:integration}
\begin{equation}\label{integration1}
\begin{split}
t(R)&=-2\left(\frac{D}{K}\right)^2\int_R^\infty \frac{y~dy}{e^y-1}\\
&=-2\left(\frac{D}{K}\right)^2\bigg[-R~\text{Log}\left[1-e^{-R}\right]+\text{PolyLog}\left[2,e^{-R}\right]\bigg]
\end{split}
\end{equation}
Where $\text{PolyLog}[n,z]$ is defined as$$\text{PolyLog}[n,z]\equiv \text{Li}_n(z)=\sum_{k=1}^{\infty}\frac{z^k}{k^n}$$
We just want $e^{-R}$ term of the integration. Expand in $e^{-R}$ we get.
\begin{equation}
\begin{split}
t(R)&=-2\left(\frac{D}{K}\right)^2\left[R~e^{-R}+e^{-R}\right]+{\cal O}\left(e^{-2R}\right)\\
&=-2\left(\frac{D}{K}\right)^2 e^{-R}\left[R+1\right]
\end{split}
\end{equation}
\begin{equation}\label{t_R}
\boxed{
t(R)=-2\left(\frac{D}{K}\right)^2 e^{-R}\left[R+1\right]
+{\cal O}\left(e^{-2R}\right)}
\end{equation}

\begin{equation}\label{vec}
v(R)=2\left(\frac{D}{K}\right)^3\left[\int_R^\infty e^{-x}dx\int_0^x\frac{y~e^y}{e^y-1}dy-e^{-R}\int_0^\infty e^{-x}dx\int_0^x\frac{y~e^y}{e^y-1}dy\right]
\end{equation}
Now,
\begin{equation}\label{integration2}
\int_0^x\frac{y~e^y}{e^y-1}dy=\frac{\pi^2}{6}+\frac{x^2}{2}+x~\text{Log}\left[1-e^{-x}\right]-\text{PolyLog}\left[2,e^{-x}\right]
\end{equation}
\begin{equation}\label{integration3}
\begin{split}
&\Rightarrow \int_R^\infty e^{-x}dx\int_0^x\frac{y~e^y}{e^y-1}dy\\
&=\int_R^\infty e^{-x}\left(\frac{\pi^2}{6}+\frac{x^2}{2}+x~\text{Log}\left[1-e^{-x}\right]-\text{PolyLog}\left[2,e^{-x}\right]\right)dx\\
&=e^{-R}\left(\frac{\pi^2}{6}\right)+e^{-R}\left(\frac{R^2}{2}\right)-(1-e^{-R})R~\text{Log}\left[1-e^{-R}\right]+(1-e^{-R})~\text{PolyLog}\left[2,e^{-R}\right]
\end{split}
\end{equation}

\begin{equation}\label{vec2}
\Rightarrow \int_0^\infty e^{-x}dx\int_0^x\frac{y~e^y}{e^y-1}dy=\frac{\pi^2}{6}
\end{equation}
Substituting \eqref{integration3} and \eqref{vec2} in \eqref{vec} we get the final expression
\begin{equation}\label{vector}
v(R)=2\left(\frac{D}{K}\right)^3\left[e^{-R}\left(\frac{R^2}{2}\right)-(1-e^{-R})R~\text{Log}\left[1-e^{-R}\right]+(1-e^{-R})~\text{PolyLog}\left[2,e^{-R}\right]\right]
\end{equation}
Expanding as before in $e^{-R}$ we get
\begin{equation}\label{v_R}
\boxed{v(R)=2\left(\frac{D}{K}\right)^3\left(1+R+\frac{R^2}{2}\right)e^{-R}+{\cal O}\left(e^{-2R}\right)}
\end{equation}\\
The $f_1(R)$ integration is very straightforward
\begin{equation}
\begin{split}
f_1(R)&=2\left(\frac{D}{K}\right)^2\left[-\int_R^\infty x~e^{-x}dx+e^{-R}\int_0^\infty x~e^{-x}dx\right]\\
&=-2\left(\frac{D}{K}\right)^2R~e^{-R}
\end{split}
\end{equation}
\begin{equation}\label{f1_R}
\boxed{
f_1(R)=-2\left(\frac{D}{K}\right)^2R~e^{-R}+{\cal O}\left(e^{-2R}\right)}
\end{equation}\\
Calculation of $f_2(R)$ is a bit complicated
\begin{equation}
\begin{split}
&f_2(R)=\left(\frac{D}{K}\right)\Bigg[\int_R^{\infty}e^{-x}dx\int_0^x\frac{v(y)}{1-e^{-y}}dy-e^{-R}\int_0^{\infty}e^{-x}dx\int_0^x\frac{v(y)}{1-e^{-y}}dy\Bigg]\\
&~~~~~~~~-\left(\frac{D}{K}\right)^4\Bigg[\int_R^{\infty}e^{-x} dx\int_0^x\frac{y^2~ e^{-y}}{1-e^{-y}}dy-e^{-R}\int_0^{\infty}e^{-x} dx\int_0^x\frac{y^2 ~e^{-y}}{1-e^{-y}}dy\Bigg]\nonumber
\end{split}
\end{equation}
First we will calculate the second line of $f_2(R)$
\begin{equation}\label{integration4}
\int_0^x\frac{y^2 ~e^{-y}}{1-e^{-y}}dy=x^2~\text{Log}[1-e^{-x}]-2~x~\text{PolyLog}[2,e^{-x}]-2~\text{PolyLog}[3,e^{-x}]+2~\text{Zeta}[3]
\end{equation}
Where $\text{Zeta}[n]$ is the `Riemann Zeta function' given by $$\text{Zeta}[n]\equiv\zeta[n]=\sum_{k=1}^{\infty}\frac{1}{k^n}$$
Now, we need to do the following integration
\begin{equation}
\begin{split}
&~~~~\int_0^\infty e^{-x}dx\int_0^x\frac{y^2 ~e^{-y}}{1-e^{-y}}dy\\
&=\int_0^\infty e^{-x}\bigg[x^2~\text{Log}[1-e^{-x}]-2~x~\text{PolyLog}[2,e^{-x}]-2~\text{PolyLog}[3,e^{-x}]+2~\text{Zeta}[3]\bigg]dx\\
&=2\left(-1+\text{Zeta[3]}\right)\nonumber
\end{split}
\end{equation}
Now, we want to calculate the following integration
\begin{equation}
\begin{split}
&~~~~\int_R^\infty e^{-x}dx\int_0^x\frac{y^2 ~e^{-y}}{1-e^{-y}}dy\\
&=\int_R^\infty e^{-x}\bigg[x^2~\text{Log}[1-e^{-x}]-2~x~\text{PolyLog}[2,e^{-x}]-2~\text{PolyLog}[3,e^{-x}]+2~\text{Zeta}[3]\bigg]dx\nonumber
\end{split}
\end{equation}
We can expand the integrand in $e^{-x}$ and then can do the integration term by term. Doing the integration term by term, we get
\begin{equation}
\int_R^\infty e^{-x}dx\int_0^x\frac{y^2 ~e^{-y}}{1-e^{-y}}dy=2~e^{-R}~\text{Zeta}[3]+{\cal O}(e^{-2R})
\end{equation}
So, finally the second line of $f_2(R)$ becomes
\begin{equation}
\begin{split}
&-\left(\frac{D}{K}\right)^4\Bigg[\int_R^{\infty}e^{-x} dx\int_0^x\frac{y^2~ e^{-y}}{1-e^{-y}}dy-e^{-R}\int_0^{\infty}e^{-x} dx\int_0^x\frac{y^2 ~e^{-y}}{1-e^{-y}}dy\Bigg]\\
&=-2\left(\frac{D}{K}\right)^4 e^{-R}
\end{split}
\end{equation}
Now we will calculate the first line of $f_2(R)$
\begin{equation}
\left(\frac{D}{K}\right)\Bigg[\int_R^{\infty}e^{-x}dx\int_0^x\frac{v(y)}{1-e^{-y}}dy-e^{-R}\int_0^{\infty}e^{-x}dx\int_0^x\frac{v(y)}{1-e^{-y}}dy\Bigg]
\end{equation}
Using eq \eqref{vector} we get
\begin{equation}\label{integration5}
\begin{split}
\int_0^x\frac{v(y)}{1-e^{-y}}dy&=2\left(\frac{D}{K}\right)^3\int_0^x dy\left[\frac{y^2~e^{-y}}{2(1-e^{-y})}-y~\text{Log}\left[1-e^{-y}\right]+\text{PolyLog}\left[2,e^{-y}\right]\right]\\
&=2\left(\frac{D}{K}\right)^3\left[\frac{x^2}{2}~\text{Log}[1-e^{-x}]-2x\text{PolyLog}[2,e^{-x}]-3~\text{PolyLog}[3,e^{-x}]+3~\text{Zeta}[3]\right]
\end{split}
\end{equation}
Now we need to do the following integration
\begin{equation}
\begin{split}
&~~~~\int_0^\infty e^{-x}dx\int_0^x\frac{v(y)}{1-e^{-y}}dy\\
&=2\left(\frac{D}{K}\right)^3\int_0^\infty e^{-x}dx\left[\frac{x^2}{2}~\text{Log}[1-e^{-x}]-2~x~\text{PolyLog}[2,e^{-x}]-3~\text{PolyLog}[3,e^{-x}]+3~\text{Zeta}[3]\right]\\
&=2\left(\frac{D}{K}\right)^3~\text{Zeta}[3]\nonumber
\end{split}
\end{equation}
Now, we will calculate the following integration. Expanding the integrand in $e^{-x}$ and doing the integration term by term we get
\begin{equation}
\int_R^\infty e^{-x}dx\int_0^x\frac{v(y)}{1-e^{-y}}dy=2\left(\frac{D}{K}\right)^33 ~e^{-R}~\text{Zeta}[3]+{\cal O}(e^{-2R})
\end{equation}
So, finally the first line of $f_2(R)$ becomes
\begin{equation}
\begin{split}
&\left(\frac{D}{K}\right)\Bigg[\int_R^{\infty}e^{-x}dx\int_0^x\frac{v(y)}{1-e^{-y}}dy-e^{-R}\int_0^{\infty}e^{-x}dx\int_0^x\frac{v(y)}{1-e^{-y}}dy\Bigg]\\
&=4\left(\frac{D}{K}\right)^4e^{-R}~\text{Zeta}[3]+{\cal O}(e^{-2R})
\end{split}
\end{equation}
$f_2(R)$ becomes
\begin{equation}\label{f2_R}
\boxed{
f_2(R)=2\left(\frac{D}{K}\right)^4e^{-R}\left(2~\text{Zeta}[3]-1\right)+{\cal O}(e^{-2R})}
\end{equation}

\section{Some Details of Linearized Calculation }\label{Appendix:Gauge}
\subsection{Outside $(\psi>1)$}\label{app:outside}
From \eqref{eq:Gauge}
\begin{equation}
\begin{split}
&-n^A\sum_{m=0}^\infty (\psi-1)^m M^{(m)}_{AB}\\
&=\sum_{m=0}^\infty\bigg[-\frac{ND}{\psi}(\psi-1)^m\xi_B^{(m)}+m(\psi-1)^{m-1}N\xi_B^{(m)}+(\psi-1)^m(n\cdot\nabla)\xi_B^{(m)}\\
&~~~~-\frac{ND}{\psi}(\psi-1)^m ~n_B\left(n\cdot\xi^{(m)}\right)+N~m(\psi-1)^{m-1}~n_B\left(n\cdot\xi^{(m)}\right)+(\psi-1)^m ~n^A\nabla_B\xi_A^{(m)}\bigg]\\
&=\sum_{m=0}^\infty\bigg[-ND(\psi-1)^m\xi_B^{(m)}[1+(\psi-1)]^{-1}+m(\psi-1)^{m-1}N\xi_B^{(m)}+(\psi-1)^m(n\cdot\nabla)\xi_B^{(m)}\\
&-ND(\psi-1)^m n_B(n\cdot\xi^{(m)})[1+(\psi-1)]^{-1}+Nm(\psi-1)^{m-1}~n_B(n\cdot\xi^{(m)})+(\psi-1)^m n^A\nabla_B\xi_A^{(m)}\bigg]\nonumber
\end{split}
\end{equation}
Comparing coefficient of $(\psi-1)^0$ we get
\begin{equation}\label{eq1}
\begin{split}
n^A M_{AB}^{(0)}&=ND\left[\xi_B^{(0)}+n_B\left(n\cdot\xi^{(0)}\right)\right]-\left[(n\cdot\nabla)\xi_B^{(0)}+n^A\nabla_B\xi_A^{(0)}\right]-N\left[\xi_B^{(1)}+n_B\left(n\cdot\xi^{(1)}\right)\right]
\end{split}
\end{equation}
\vspace{-0.2in}
\\Comparing coefficient of $(\psi-1)^1$ we get
\begin{equation}\label{eq2}
\begin{split}
n^A M_{AB}^{(1)}&=ND\left[\xi_B^{(1)}-\xi_B^{(0)}\right]+ND\left[n_B\left(n\cdot\xi^{(1)}\right)-n_B\left(n\cdot\xi^{(0)}\right)\right]-\left[(n\cdot\nabla)\xi_B^{(1)}+n^A\nabla_B\xi_A^{(1)}\right]\\
&-2N\left[\xi_B^{(2)}+n_B\left(n\cdot\xi^{(2)}\right)\right]
\end{split}
\end{equation}
\vspace{-0.2in}
\\Comparing coefficient of $(\psi-1)^2$ we get
\begin{equation}\label{eq3}
\begin{split}
n^A M_{AB}^{(2)}&=ND\left[\xi_B^{(2)}-\xi_B^{(1)}+\xi_B^{(0)}+n_B\left(n\cdot\xi^{(2)}\right)-n_B\left(n\cdot\xi^{(1)}\right)+n_B\left(n\cdot\xi^{(0)}\right)\right]\\
&-\left[(n\cdot\nabla)\xi_B^{(2)}+n^A\nabla_B\xi_A^{(2)}\right]-3N\left[\xi_B^{(3)}+n_B\left(n\cdot\xi^{(3)}\right)\right]
\end{split}
\end{equation}
\vspace{-0.2in}
\\$M_{AB}$ is correct up to order ${\cal O}\left(\frac{1}{D}\right)^2$. So, we want $\xi_A$ to be correct up to order ${\cal O}\left(\frac{1}{D}\right)^3$. This implies we want $\xi_A^{(0)}$ to be correct up to order ${\cal O}\left(\frac{1}{D}\right)^3$, $\xi_A^{(1)}$ to be correct up to order ${\cal O}\left(\frac{1}{D}\right)^2$ and $\xi_A^{(2)}$ to be correct up to order ${\cal O}\left(\frac{1}{D}\right)$. Now, using the following expansion
\begin{equation}
\begin{split}
\xi_B^{(0)}&=\xi_B^{(0,0)}+\frac{1}{D}\xi_B^{(0,1)}+\frac{1}{D^2}\xi_B^{(0,2)}+\frac{1}{D^3}\xi_B^{(0,3)}+{\cal O}\left(\frac{1}{D}\right)^4\\
\xi_B^{(1)}&=\xi_B^{(1,0)}+\frac{1}{D}\xi_B^{(1,1)}+\frac{1}{D^2}\xi_B^{(1,2)}+{\cal O}\left(\frac{1}{D}\right)^3\\
\xi_B^{(2)}&=\xi_B^{(2,0)}+\frac{1}{D}\xi_B^{(2,1)}+{\cal O}\left(\frac{1}{D}\right)^2
\end{split}
\end{equation}
From \eqref{eq1} we get
\begin{equation}
\begin{split}
&ND\left[\xi_B^{(0,0)}+n_B\left(n\cdot\xi^{(0,0)}\right)\right]=0\\
\Rightarrow ~& ND\left[\left(n\cdot\xi^{(0,0)}\right)+\left(n\cdot\xi^{(0,0)}\right)\right]=0\\
\Rightarrow~&\left(n\cdot\xi^{(0,0)}\right)=0\\
\Rightarrow~&\xi_B^{(0,0)}=0
\end{split}
\end{equation}
From \eqref{eq2}, at leading order
\begin{equation}
\begin{split}
&ND\left[\xi_B^{(1,0)}-\xi_B^{(0,0)}+n_B\left(n\cdot\xi^{(1,0)}\right)-n_B\left(n\cdot\xi^{(0,0)}\right)\right]=0\\
\Rightarrow ~&ND\left[\xi_B^{(1,0)}+n_B\left(n\cdot\xi^{(1,0)}\right)\right]=0\\
\Rightarrow~&\xi_B^{(1,0)}=0
\end{split}
\end{equation}
Similarly, from \eqref{eq3}
\begin{equation}
\xi_B^{(2,0)}=0
\end{equation}
Now, we will calculate $\xi_B^{(2,1)}$. From \eqref{eq3} at ${\cal O}(1)$
\begin{equation}\label{eq:nM2}
n^A M_{AB}^{(2)}=N\left[\xi_B^{(2,1)}-\xi_B^{(1,1)}+\xi_B^{(0,1)}+n_B\left(n\cdot\xi^{(2,1)}\right)-n_B\left(n\cdot\xi^{(1,1)}\right)+n_B\left(n\cdot\xi^{(0,1)}\right)\right]
\end{equation}
From \eqref{eq2} at ${\cal O}(1)$
\begin{equation}\label{eq:nM1}
n^A M_{AB}^{(1)}=N\left[\xi_B^{(1,1)}-\xi_B^{(0,1)}+n_B\left(n\cdot\xi^{(1,1)}\right)-n_B\left(n\cdot\xi^{(0,1)}\right)\right]
\end{equation}
Adding \eqref{eq:nM1} and \eqref{eq:nM2} we get
\begin{equation}
\begin{split}
& n^A M_{AB}^{(2)}+n^A M_{AB}^{(1)}=N\left[\xi_B^{(2,1)}+n_B\left(n\cdot\xi^{(2,1)}\right)\right]\\
\Rightarrow~& n\cdot M^{(2)}\cdot n+n\cdot M^{(1)}\cdot n=2N\left(n\cdot\xi^{(2,1)}\right)\\
\Rightarrow~& n\cdot\xi^{(2,1)}=\frac{1}{2N}\left(n\cdot M^{(2)}\cdot n+n\cdot M^{(1)}\cdot n\right)
\end{split}
\end{equation}
Finally we get,
\begin{equation}\label{eq:xi21}
\xi_B^{(2,1)}=\frac{1}{N}\left[n^A M_{AB}^{(2)}+n^A M_{AB}^{(1)}-\frac{n_B}{2}\left(n\cdot M^{(2)}\cdot n+n\cdot M^{(1)}\cdot n\right)\right]
\end{equation}
Adding \eqref{eq1} and \eqref{eq2} we get,
\begin{equation}\label{eq:B.11}
\begin{split}
&~~~~n^A M_{AB}^{(1)}+n^A M_{AB}^{(0)}\\
&=ND\left[\xi_B^{(1)}+n_B\left(n\cdot\xi^{(1)}\right)\right]-\left[(n\cdot\nabla)\xi_B^{(1)}+(n\cdot\nabla)\xi_B^{(0)}\right]-\left[n^A\nabla_B\xi_A^{(1)}+n^A\nabla_B\xi_A^{(0)}\right]\\
&-N\left[\xi_B^{(1)}+n_B\left(n\cdot\xi^{(1)}\right)\right]-2N\left[\xi_B^{(2)}+n_B\left(n\cdot\xi^{(2)}\right)\right]
\end{split}
\end{equation}
From \eqref{eq:B.11}, at order ${\cal O}(1)$ we get
\begin{equation}\label{eq:xi11}
\begin{split}
&n^A M_{AB}^{(1)}+n^A M_{AB}^{(0)}=N\left[\xi_B^{(1,1)}+n_B\left(n\cdot\xi^{(1,1)}\right)\right]\\
\Rightarrow ~& n\cdot M^{(1)}\cdot n+n\cdot M^{(0)}\cdot n=2N\left(n\cdot\xi^{(1,1)}\right)\\
\Rightarrow ~&n\cdot\xi^{(1,1)}=\frac{1}{2N}\left(n\cdot M^{(1)}\cdot n+n\cdot M^{(0)}\cdot n\right)\\
\Rightarrow ~& \xi_B^{(1,1)}=\frac{1}{N}\left[n^A M_{AB}^{(1)}+n^A M_{AB}^{(0)}-\frac{n_B}{2}\left(n\cdot M^{(1)}\cdot n+n\cdot M^{(0)}\cdot n\right)\right]
\end{split}
\end{equation}
From \eqref{eq:B.11} at order ${\cal O}\left(\frac{1}{D}\right)$,
\begin{equation}\label{eq:xi12}
\begin{split}
&N\left[\xi_B^{(1,2)}+n_B\left(n\cdot\xi^{(1,2)}\right)\right]-\left[(n\cdot\nabla)\xi_B^{(1,1)}+(n\cdot\nabla)\xi_B^{(0,1)}\right]-\left[n^A\nabla_B\xi_A^{(1,1)}+n^A\nabla_B\xi_A^{(0,1)}\right]\\
&-N\left[\xi_B^{(1,1)}+n_B\left(n\cdot\xi^{(1,1)}\right)\right]-2N\left[\xi_B^{(2,1)}+n_B\left(n\cdot\xi^{(2,1)}\right)\right]=0\\
\Rightarrow~&\left(n\cdot\xi^{(1,2)}\right)=\frac{1}{N}\left[n^B(n\cdot\nabla)\xi_B^{(1,1)}+n^B(n\cdot\nabla)\xi_B^{(0,1)}\right]+2\left(n\cdot\xi^{(2,1)}\right)+\left(n\cdot\xi^{(1,1)}\right)\\
\Rightarrow~&\xi_B^{(1,2)}=\frac{1}{N}\left[(n\cdot\nabla)\xi_B^{(1,1)}+(n\cdot\nabla)\xi_B^{(0,1)}\right]+\frac{1}{N}\left[n^A\nabla_B\xi_A^{(1,1)}+n^A\nabla_B\xi_A^{(0,1)}\right]\\
&~~~~~~~+2~\xi_B^{(2,1)}+\xi_B^{(1,1)}-\frac{n_B}{N}\left[n^A(n\cdot\nabla)\xi_A^{(1,1)}+n^A(n\cdot\nabla)\xi_A^{(0,1)}\right]
\end{split}
\end{equation}
\vspace{-0.2in}
\\Now, we will calculate $\xi_A^{(0)}$. From \eqref{eq1}, at order ${\cal O}(1)$
\begin{equation}\label{eq:xi01}
\begin{split}
&n^A M_{AB}^{(0)}=N\left[\xi_B^{(0,1)}+n_B\left(n\cdot\xi^{(0,1)}\right)\right]\\
\Rightarrow~&n\cdot M\cdot n=2N\left(n\cdot\xi^{(0,1)}\right)\\
\Rightarrow~&\xi_B^{(0,1)}=\frac{1}{N}\left[n^A M_{AB}^{(0)}-\frac{n_B}{2}\left(n\cdot M^{(0)}\cdot n\right)\right]\\
\end{split}
\end{equation}
From \eqref{eq1} at order ${\cal O}\left(\frac{1}{D}\right)$
\begin{equation}\label{eq:xi02}
\begin{split}
&N\left[\xi_B^{(0,2)}+n_B\left(n\cdot\xi^{(0,2)}\right)\right]-\left[(n\cdot\nabla)\xi_B^{(0,1)}+n^A\nabla_B\xi_A^{(0,1)}\right]-N\left[\xi_B^{(1,1)}+n_B\left(n\cdot\xi^{(1,1)}\right)\right]=0\\
\Rightarrow~&2N\left(n\cdot\xi^{(0,2)}\right)=2~n^B(n\cdot\nabla)\xi_B^{(0,1)}+2N\left(n\cdot\xi^{(1,1)}\right)\\
\Rightarrow~&\xi_B^{(0,2)}=\frac{1}{N}\left[(n\cdot\nabla)\xi_B^{(0,1)}+n^A\nabla_B\xi_A^{(0,1)}\right]+\xi_B^{(1,1)}-\frac{n_B}{N}\left[n^A(n\cdot\nabla)\xi_A^{(0,1)}\right]
\end{split}
\end{equation}
From \eqref{eq1} at order ${\cal O}\left(\frac{1}{D}\right)^2$
\begin{equation}\label{eq:xi03}
\begin{split}
&N\left[\xi_B^{(0,3)}+n_B\left(n\cdot\xi^{(0,3)}\right)\right]-\left[(n\cdot\nabla)\xi_B^{(0,2)}+n^A\nabla_B\xi_A^{(0,2)}\right]-N\left[\xi_B^{(1,2)}+n_B\left(n\cdot\xi^{(1,2)}\right)\right]=0\\
\Rightarrow~&n\cdot\xi^{(0,3)}=\frac{1}{N}\left[n^B(n\cdot\nabla)\xi_B^{(0,2)}\right]+n\cdot\xi^{(1,2)}\\
\Rightarrow~&\xi_B^{(0,3)}=\frac{1}{N}\left[(n\cdot\nabla)\xi_B^{(0,2)}+n^A\nabla_B\xi_A^{(0,2)}\right]+\xi_B^{(1,2)}-\frac{n_B}{N}\left[n^A(n\cdot\nabla)\xi_A^{(0,2)}\right]
\end{split}
\end{equation}
Using, \eqref{eq:xi21} \eqref{eq:xi11}, \eqref{eq:xi12}, \eqref{eq:xi01}, \eqref{eq:xi02} and \eqref{eq:xi03} in \eqref{Mprime} we get 
\begin{equation}
\begin{split}
M^\prime_{AB}&=M_{AB}+\psi^D\nabla_A\left[\psi^{-D}\xi_B\right]+\psi^D\nabla_B\left[\psi^{-D}\xi_A\right]\\
&=M^{(0)}_{AB}+(\psi-1)M^{(1)}_{AB}+(\psi-1)^2M^{(2)}_{AB}+\underbrace{\nabla_A \xi_B-\left(\frac{ND}{\psi}\right)n_A\xi_B}_{{\cal L}_{AB}}+{\cal L}_{BA}+{\cal O}\left(\frac{1}{D}\right)^3\\
&=M^{(0)}_{AB}+(\psi-1)M^{(1)}_{AB}+(\psi-1)^2M^{(2)}_{AB}+\nabla_A\left[\xi_B^{(0)}+(\psi-1)\xi_B^{(1)}+(\psi-1)^2\xi_B^{(2)}\right]\\
&-ND[1+(\psi-1)]^{-1}n_A\left[\xi_B^{(0)}+(\psi-1)\xi_B^{(1)}+(\psi-1)^2\xi_B^{(2)}\right]+{\cal L}_{BA}\\\\
&=M^{(0)}_{AB}+(\psi-1)M^{(1)}_{AB}+(\psi-1)^2M^{(2)}_{AB}+\nabla_A\xi_B^{(0)}+Nn_A\xi_B^{(1)}+(\psi-1)\nabla_A\xi_B^{(1)}\\
&+(\psi-1)^2\nabla_A\xi_B^{(2)}+2N(\psi-1)n_A\xi_B^{(2)}-ND~n_A\xi_B^{(0)}-ND(\psi-1)n_A\xi_B^{(1)}\\
&-ND(\psi-1)^2n_A\xi_B^{(2)}+ND(\psi-1)n_A\xi_B^{(0)}+ND(\psi-1)^2n_A\xi_B^{(1)}-ND(\psi-1)^2n_A\xi_B^{(0)}+{\cal L}_{BA}\\\\
&=M^{(0)}_{AB}+\nabla_A\xi_B^{(0)}+Nn_A\xi_B^{(1)}-ND~n_A\xi_B^{(0)}\\
&+(\psi-1)\bigg[M^{(1)}_{AB}+\nabla_A\xi_B^{(1)}+2Nn_A\xi_B^{(2)}-ND~n_A\xi_B^{(1)}+ND~n_A\xi_B^{(0)}\bigg]\\
&+(\psi-1)^2\bigg[M^{(2)}_{AB}+\nabla_A\xi_B^{(2)}-ND~n_A\xi_B^{(2)}+ND~n_A\xi_B^{(1)}-ND~n_A\xi_B^{(0)}\bigg]+{\cal L}_{BA}\\
\end{split}
\end{equation}
Now writing the expression for ${\cal L}_{BA}$ we finally get
\begin{equation}\label{eq:B.18}
\begin{split}
M^\prime_{AB}&=\bigg[M^{(0)}_{AB}+\nabla_A\xi_B^{(0)}+Nn_A\xi_B^{(1)}-ND~n_A\xi_B^{(0)}+\nabla_B\xi_A^{(0)}+Nn_B\xi_A^{(1)}-ND~n_B\xi_A^{(0)}\bigg]\\
&+(\psi-1)\bigg[M^{(1)}_{AB}+\nabla_A\xi_B^{(1)}+2Nn_A\xi_B^{(2)}-ND~n_A\xi_B^{(1)}+ND~n_A\xi_B^{(0)}\\
&+\nabla_B\xi_A^{(1)}+2Nn_B\xi_A^{(2)}-ND~n_B\xi_A^{(1)}+ND~n_B\xi_A^{(0)}\bigg]\\
&+(\psi-1)^2\bigg[M^{(2)}_{AB}+\nabla_A\xi_B^{(2)}-ND~n_A\xi_B^{(2)}+ND~n_A\xi_B^{(1)}-ND~n_A\xi_B^{(0)}\\
&+\nabla_B\xi_A^{(2)}-ND~n_B\xi_A^{(2)}+ND~n_B\xi_A^{(1)}-ND~n_B\xi_A^{(0)}\bigg]
\end{split}
\end{equation}
Now, we will simplify \eqref{eq:B.18}. First, we will simplify the first square bracketed terms.
\begin{equation}\label{eq:B.19}
\begin{split}
&~~~~\nabla_A\xi_B^{(0)}-ND~n_A\xi_B^{(0)}+Nn_A\xi_B^{(1)}\\
&=\nabla_A\xi_B^{(0)}-Nn_A\left[\xi_B^{(0,1)}+\frac{1}{D}\xi_B^{(0,2)}+\frac{1}{D^2}\xi_B^{(0,3)}\right]+n_A~\frac{N}{D}\left[\xi_B^{(1,1)}+\frac{1}{D}\xi_B^{(1,2)}\right]+{\cal O}\left(\frac{1}{D}\right)^3\\
&=\nabla_A\xi_B^{(0)}-n_A\left[n^D M_{DB}^{(0)}-\frac{n_B}{2}\left(n\cdot M^{(0)}\cdot n\right)\right]-n_A~\frac{1}{D}\left[(n\cdot\nabla)\xi_B^{(0,1)}+n^D\nabla_B\xi_D^{(0,1)}\right]\\
&-\cancel{\frac{N}{D} n_A\xi_B^{(1,1)}}+n_An_B~\frac{1}{D}\left[n^D(n\cdot\nabla)\xi_D^{(0,1)}\right]-\frac{n_A}{D^2}\left[(n\cdot\nabla)\xi_B^{(0,2)}+n^D\nabla_B\xi_D^{(0,2)}\right]-\cancel{\frac{N}{D^2}n_A~\xi_B^{(1,2)}}\\
&+\frac{1}{D^2}n_A n_B\left[n^D(n\cdot\nabla)\xi_D^{(0,2)}\right]+\cancel{\frac{N}{D}n_A\xi_B^{(1,1)}}+\cancel{\frac{N}{D^2} n_A\xi_B^{(1,2)}}
\end{split}
\end{equation}
Using, \eqref{eq:B.19} and it's symmetric part the first square bracketed terms become
\begin{equation}\label{part1}
\begin{split}
&~~~~M^{(0)}_{AB}+\nabla_A\xi_B^{(0)}+Nn_A\xi_B^{(1)}-ND~n_A\xi_B^{(0)}+\nabla_B\xi_A^{(0)}+Nn_B\xi_A^{(1)}-ND~n_B\xi_A^{(0)}\\
&=\Pi^C_A\Pi^{C'}_B\left[M^{(0)}_{CC'}+\nabla_C\left(\frac{1}{D}\xi_{C'}^{(0,1)}+\frac{1}{D^2}\xi_{C'}^{(0,2)}\right)+\nabla_{C'}\left(\frac{1}{D}\xi_C^{(0,1)}+\frac{1}{D^2}\xi_C^{(0,2)}\right)\right]
\end{split}
\end{equation}
Now, we will simplify the second square bracketed term of \eqref{eq:B.18}
\begin{equation}
\begin{split}
&~~~~\nabla_A\xi_B^{(1)}+2Nn_A\xi_B^{(2)}-ND~n_A\xi_B^{(1)}+ND~n_A\xi_B^{(0)}\\
&=\nabla_A\xi_B^{(1)}+2Nn_A\xi_B^{(2)}-N~n_A\left[\xi_B^{(1,1)}+\frac{1}{D}\xi_B^{(1,2)}\right]+N~n_A\left[\xi_B^{(0,1)}+\frac{1}{D}\xi_B^{(0,2)}\right]+{\cal O}\left(\frac{1}{D}\right)^2\nonumber\\
\end{split}
\end{equation}
\begin{equation}\label{eq:B.20}
\begin{split}
&=\nabla_A\xi_B^{(1)}+\cancel{2Nn_A\xi_B^{(2)}}-n_A\left[n^D M_{DB}^{(1)}+\cancel{n^D M_{DB}^{(0)}}-\frac{n_B}{2}\left(n\cdot M^{(1)}\cdot n+\cancel{n\cdot M^{(0)}\cdot n}\right)\right]\\
&-\frac{n_A}{D}\left[(n\cdot\nabla)\xi_B^{(1,1)}+\cancel{(n\cdot\nabla)\xi_B^{(0,1)}}\right]-\frac{n_A}{D}\left[n^D\nabla_B\xi_D^{(1,1)}+\cancel{n^D\nabla_B\xi_D^{(0,1)}}\right]\\
&-\cancel{\frac{2}{D}~Nn_A\xi_B^{(2)}}-\cancel{\frac{N}{D} n_A ~\xi_B^{(1,1)}}+n_An_B\frac{1}{D}\left[n^D(n\cdot\nabla)\xi_D^{(1,1)}+\cancel{n^D(n\cdot\nabla)\xi_D^{(0,1)}}\right]\\
&+\frac{n_A}{D}\left[\cancel{n^D M_{DB}^{(0)}}-\cancel{\frac{n_B}{2}\left(n\cdot M^{(0)}\cdot n\right)}\right]+\frac{n_A}{D}\left[\cancel{(n\cdot\nabla)\xi_B^{(0,1)}}+\cancel{n^D\nabla_B\xi_D^{(0,1)}}\right]+\cancel{\frac{N}{D}n_A\xi_B^{(1,1)}}\\
&-\cancel{n_An_B\frac{1}{D}\left[n^D(n\cdot\nabla)\xi_D^{(0,1)}\right]}\\
&=\nabla_A\xi_B^{(1)}-n_A\left[n^D M_{DB}^{(1)}-\frac{n_B}{2}\left(n\cdot M^{(1)}\cdot n\right)\right]-\frac{n_A}{D}\left[(n\cdot\nabla)\xi_B^{(1,1)}\right]\\
&-\frac{n_A}{D}\left[n^D\nabla_B\xi_D^{(1,1)}\right]+n_An_B\frac{1}{D}\left[n^D(n\cdot\nabla)\xi_D^{(1,1)}\right]
\end{split}
\end{equation}
Adding $M^{(1)}_{AB}$, \eqref{eq:B.20} and it's symmetric part we get
\begin{equation}\label{part2}
\begin{split}
&M^{(1)}_{AB}+\nabla_A\xi_B^{(1)}+2Nn_A\xi_B^{(2)}-ND~n_A\xi_B^{(1)}+ND~n_A\xi_B^{(0)}\\
&+\nabla_B\xi_A^{(1)}+2Nn_B\xi_A^{(2)}-ND~n_B\xi_A^{(1)}+ND~n_B\xi_A^{(0)}+{\cal O}\left(\frac{1}{D}\right)^2\\
=~&\Pi^C_A\Pi^{C'}_B\left[M^{(1)}_{CC'}+\frac{1}{D}\left(\nabla_C\xi_{C'}^{(1,1)}+\nabla_{C'}\xi_C^{(1,1)}\right)\right]+{\cal O}\left(\frac{1}{D}\right)^2
\end{split}
\end{equation}
Finally, we will try to simplify the third square bracketed term of \eqref{Mprime}
\begin{equation}\label{eq:B.22}
\begin{split}
&~~~~\nabla_A\xi_B^{(2)}-ND~n_A\xi_B^{(2)}+ND~n_A\xi_B^{(1)}-ND~n_A\xi_B^{(0)}\\
&=-N~n_A\xi_B^{(2,1)}+N~n_A\xi_B^{(1,1)}-N~n_A\xi_B^{(0,1)}+{\cal O}\left(\frac{1}{D}\right)\\
&=-n_A\left[n^D M_{DB}^{(2)}+\cancel{n^D M_{DB}^{(1)}}-\frac{n_B}{2}\left(n\cdot M^{(2)}\cdot n+\cancel{n\cdot M^{(1)}\cdot n}\right)\right]\\
&+n_A\left[\cancel{n^D M_{DB}^{(1)}}+\cancel{n^D M_{DB}^{(0)}}-\frac{n_B}{2}\left(\cancel{n\cdot M^{(1)}\cdot n}+\cancel{n\cdot M^{(0)}\cdot n}\right)\right]\\
&-n_A\left[\cancel{n^D M_{DB}^{(0)}}-\cancel{\frac{n_B}{2}\left(n\cdot M^{(0)}\cdot n\right)}\right]\\
&=-n_A\left[n^D M_{DB}^{(2)}-\frac{n_B}{2}\left(n\cdot M^{(2)}\cdot n\right)\right]
\end{split}
\end{equation}
Using, \eqref{eq:B.22} and it's symmetric part the third square bracketed terms become
\begin{equation}\label{part3}
\begin{split}
&M^{(2)}_{AB}+\nabla_A\xi_B^{(2)}-ND~n_A\xi_B^{(2)}+ND~n_A\xi_B^{(1)}-ND~n_A\xi_B^{(0)}\\
&+\nabla_B\xi_A^{(2)}-ND~n_B\xi_A^{(2)}+ND~n_B\xi_A^{(1)}-ND~n_B\xi_A^{(0)}+{\cal O}\left(\frac{1}{D}\right)\\
=&~\Pi^C_A\Pi^{C'}_BM^{(2)}_{CC'}+{\cal O}\left(\frac{1}{D}\right)
\end{split}
\end{equation}
Finally, adding \eqref{part1}, \eqref{part2} and \eqref{part3} we get the final expression of $M^\prime_{AB}$\eqref{MPRIME}
\begin{equation}
\begin{split}
M^\prime_{AB}&=\Pi^C_A\Pi^{C'}_B\bigg[M^{(0)}_{CC'}+(\psi-1)M^{(1)}_{CC'}+(\psi-1)^2M^{(2)}_{CC'}\bigg]+\hat{\nabla}_A \xi_B^{(0)}+\hat{\nabla}_B \xi_A^{(0)}\\
&+(\psi-1)\left(\hat{\nabla}_A\xi_B^{(1)}+\hat{\nabla}_B\xi_A^{(1)}\right)+{\cal O}\left(\frac{1}{D}\right)^3
\end{split}
\end{equation}
Now we will calculate different terms in \eqref{MPRIME}. First we will calculate $\xi_A^{(0,1)}$
\begin{equation}\label{eq:xi_01}
\begin{split}
\xi_A^{(0,1)}&=\frac{1}{N}\left[n^B M_{BA}^{(0)}-\frac{n_A}{2}\left(n\cdot M^{(0)}\cdot n\right)\right]\\
&=\frac{1}{N}\bigg[O_A+\frac{2}{K^2}\bigg\{\left(\frac{D}{K}\right)^2\left(2~\text{Zeta}[3]-1\right)\mathfrak{s}_2O_A+\frac{D}{K}\mathfrak{v}_A\bigg\}\\
&-\frac{n_A}{2}\bigg\{1+\frac{2}{K^2}\left(\frac{D}{K}\right)^2\left(2~\text{Zeta}[3]-1\right)\mathfrak{s}_2\bigg\}\bigg]\\
&=\frac{1}{N}\left[\frac{n_A}{2}-u_A\right]+{\cal O}\left(\frac{1}{D}\right)^2
\end{split}
\end{equation}
Next, we will calculate $\xi_A^{(0,2)}$
\begin{equation}\label{eq:B.27}
\begin{split}
\xi_A^{(0,2)}&=\frac{1}{N}\left[(n\cdot\nabla)\xi_A^{(0,1)}+n^B\nabla_A\xi_B^{(0,1)}\right]-\frac{n_A}{N}\left[n^B(n\cdot\nabla)\xi_B^{(0,1)}\right]\\
&+\frac{1}{N}\left[n^B M_{BA}^{(1)}+n^B M_{BA}^{(0)}-\frac{n_A}{2}\left(n\cdot M^{(1)}\cdot n+n\cdot M^{(0)}\cdot n\right)\right]
\end{split}
\end{equation}
Now, we need to calculate different terms of \eqref{eq:B.27}
\begin{equation}\label{eq:part1}
\begin{split}
&\frac{1}{N}\left[n^B M_{BA}^{(1)}+n^B M_{BA}^{(0)}-\frac{n_A}{2}\left(n\cdot M^{(1)}\cdot n+n\cdot M^{(0)}\cdot n\right)\right]=\frac{1}{N}\left[\frac{n_A}{2}-u_A\right]+{\cal O}\left(\frac{1}{D}\right)^2\\
&\nabla_A\xi_B^{(0,1)}=\frac{1}{N}\left[\frac{\nabla_A n_B}{2}-\nabla_A u_B\right]-\frac{\nabla_A N}{N^2 D}\left[\frac{n_B}{2}-u_B\right]\\
&(n\cdot\nabla)\xi_B^{(0,1)}=\frac{1}{N}\left[\frac{(n\cdot\nabla)n_B}{2}-(n\cdot\nabla)u_B\right]-\frac{(n\cdot\nabla) N}{N^2}\left[\frac{n_B}{2}-u_B\right]\\
&n^B\nabla_A\xi_B^{(0,1)}=\frac{1}{N}\left[u^B\nabla_A n_B\right]-\frac{1}{2ND}\left(\frac{\nabla_A N}{N}\right)\\
&n^B(n\cdot\nabla)\xi_B^{(0,1)}=\frac{1}{N}\left[u^B(n\cdot\nabla)n_B\right]-\frac{1}{2ND}\left(\frac{(n\cdot\nabla) N}{N}\right)
\end{split}
\end{equation}
Using \eqref{eq:part1} in \eqref{eq:B.27} we get
\begin{equation}
\begin{split}
\xi_A^{(0,2)}
&=\frac{1}{N}\left[\frac{n_A}{2}-u_A\right]+\frac{1}{N^2}\bigg[\frac{(n\cdot\nabla)n_A}{2}-(n\cdot\nabla)u_A-\frac{(n\cdot\nabla) N}{N}\left(\frac{n_A}{2}-u_A\right)\\
&+u^B\nabla_A n_B-\frac{1}{2}\left(\frac{\nabla_A N}{N}\right)-n_A\left(u^B(n\cdot\nabla)n_B-\frac{1}{2}\frac{(n\cdot\nabla) N}{N}\right)\bigg]\nonumber\\
\end{split}
\end{equation}
\begin{equation}\label{eq:xi_02}
\begin{split}
&=\frac{1}{N}\left[\frac{n_A}{2}-u_A\right]+\frac{1}{N^2}\bigg[\frac{(n\cdot\nabla)n_A}{2}-(n\cdot\nabla)u_A+u_A\frac{(n\cdot\nabla) N}{N}+u^B{\cal K}_{AB}-\frac{1}{2}\frac{\nabla_A N}{N}\bigg]\\
&=\frac{1}{N}\left[\frac{n_A}{2}-u_A\right]-\frac{1}{N^2}\bigg[\frac{1}{2}n_A\frac{(n\cdot\nabla)N}{N}+(n\cdot\nabla)u_A-u_A\frac{(n\cdot\nabla) N}{N}-u^B{\cal K}_{AB}\bigg]+{\cal O}\left(\frac{1}{D}\right)
\end{split}
\end{equation}
Adding \eqref{eq:xi_01} and \eqref{eq:xi_02} we get the expression of $\xi_A^{(0)}$
\begin{equation}
\begin{split}
\xi_A^{(0)}&=\frac{1}{ND}\bigg[\frac{n_A}{2}-u_A\bigg]+\frac{1}{ND^2}\bigg[\frac{n_A}{2}-u_A\bigg]\\
&-\frac{1}{N^2 D^2}\bigg[\frac{n_A}{2}\bigg(\frac{n\cdot\nabla N}{N}\bigg)+(n\cdot\nabla)u_A-u_A\bigg(\frac{n\cdot\nabla N}{N}\bigg)-u^B K_{AB}\bigg]+{\cal O}\left(\frac{1}{D}\right)^3\nonumber
\end{split}
\end{equation}
Next, we will calculate $\xi_A^{(1,1)}$
\begin{equation}
\begin{split}
\xi_A^{(1,1)}&=\frac{1}{N}\left[n^B M_{BA}^{(1)}+n^B M_{BA}^{(0)}-\frac{n_A}{2}\left(n\cdot M^{(1)}\cdot n+n\cdot M^{(0)}\cdot n\right)\right]+{\cal O}\left(\frac{1}{D}\right)^2\\
&=\frac{1}{N}\left[\frac{n_A}{2}-u_A\right]+{\cal O}\left(\frac{1}{D}\right)^2
\end{split}
\end{equation}
So, expression of $\xi_A^{(1)}$ we get
\begin{equation}
\xi_A^{(1)}=\frac{1}{ND}\left[\frac{n_A}{2}-u_A\right]+{\cal O}\left(\frac{1}{D}\right)^2
\end{equation}
Now, we will calculate $\Pi^A_C\Pi^B_{C'}\left(\nabla_B \xi_A^{(0)}\right)$
\begin{equation}
\begin{split}
\Pi^A_C\Pi^B_{C'}\left(\nabla_B \xi_A^{(0)}\right)&=\frac{1}{ND}\Pi^A_C\Pi^B_{C'}u_A\left(\frac{\nabla_B N}{N}\right)+\frac{1}{ND}\Pi^A_C\Pi^B_{C'}\left[\frac{\nabla_B n_A}{2}-\nabla_B u_A\right]\\
&+\Pi^A_C\Pi^B_{C'}\bigg[(\nabla_B n_A)\left(\frac{1}{2ND^2}-\frac{1}{2N^2D^2}\frac{(n\cdot\nabla)N}{N}\right)-\nabla_B\left(\frac{u_A}{ND^2}\right)\\
&+\nabla_B\left\{\frac{1}{N^2D^2}\left(-(n\cdot\nabla)u_A+u_A\frac{(n\cdot\nabla)N}{N}+u^E K_{AE}\right)\right\}\bigg]+{\cal O}\left(\frac{1}{D}\right)^3
\end{split}
\end{equation}
Using the identity \eqref{eq:identity1} and \eqref{eq:identity2} we can write the above equation as
\begin{equation}
\begin{split}
&\Pi^A_C\Pi^B_{C'}\left(\nabla_B \xi_A^{(0)}\right)\\
&=\frac{1}{\psi K}\left(1-\frac{(n\cdot\nabla)N}{NK}+\frac{N}{\psi K}\right)\Pi^A_C\Pi^B_{C'}\left[u_A\left(\frac{\nabla_B K}{K}\right)+\frac{1}{2}\nabla_B n_A-\nabla_B u_A\right]\\
&+\frac{1}{\psi K^2}\Pi^A_C\Pi^B_{C'}\left[u_A\nabla_B\left(\frac{(n\cdot\nabla)K}{K}\right)-u_A\left(\frac{\nabla_B K}{K}\right)\frac{(n\cdot\nabla)K}{K}\right]\nonumber\\
\end{split}
\end{equation}
\begin{equation}\label{eq:ter1}
\begin{split}
&+\frac{1}{D^2}\Pi^A_C\Pi^B_{C'}\bigg[\frac{1}{2N}K_{BA}\left(1-\frac{1}{N}\frac{(n\cdot\nabla)N}{N}\right)+\frac{1}{N^2}u_A(\nabla_B N)-\frac{1}{N}(\nabla_B u_A)\\
&-\frac{2}{N^3}(\nabla_B N)\left(u^E K_{AE}-(n\cdot\nabla)u_A+u_A\frac{(n\cdot\nabla)N}{N}\right)\\
&+\frac{1}{N^2}\nabla_B\left(u^E K_{AE}-(n\cdot\nabla)u_A+u_A\frac{(n\cdot\nabla)N}{N}\right)\bigg]+{\cal O}\left(\frac{1}{D}\right)^3\\[10pt]
&=\frac{1}{\psi K}\left[u_C\left(\frac{\hat{\nabla}_{C'} K}{K}\right)+\frac{1}{2}K_{CC'}-\hat{\nabla}_{C'} u_C\right]-\frac{1}{K^2}\left(\frac{n\cdot\nabla K}{K}\right)\left[u_C\left(\frac{\hat{\nabla}_{C'} K}{K}\right)+\frac{1}{2}K_{CC'}-\hat{\nabla}_{C'} u_C\right]\\
&+\frac{1}{K^2}\left[u_C\hat{\nabla}_{C'}\left(\frac{(n\cdot\nabla)K}{K}\right)-u_C\left(\frac{\hat{\nabla}_{C'} K}{K}\right)\frac{(n\cdot\nabla)K}{K}\right]\\
&+\frac{1}{K^2}\bigg[-\frac{1}{2}K_{CC'}\left(\frac{(n\cdot\nabla)K}{K}\right)-2~\frac{\hat{\nabla}_{C'} K}{K}\left(u^E K_{CE}-\Pi_C^E(n\cdot\nabla)u_E+u_C\frac{(n\cdot\nabla)K}{K}\right)\\
&+\hat{\nabla}_{C'}\left(u^E K_{CE}-(n\cdot\nabla)u_C+u_C\frac{(n\cdot\nabla)K}{K}\right)\bigg]+{\cal O}\left(\frac{1}{D}\right)^3
\end{split}
\end{equation}
\vspace{-0.2in}
\\Now,
\vspace{-0.1in}
\begin{equation}\label{eq:ter3}
\Pi^A_C\Pi^B_{C'}\left(\nabla_B \xi_A^{(1)}\right)=\frac{1}{K}\left[u_C\left(\frac{\hat{\nabla}_{C'} K}{K}\right)+\frac{1}{2}K_{CC'}-\hat{\nabla}_{C'} u_C\right]+{\cal O}\left(\frac{1}{D}\right)^2
\end{equation}

\subsubsection*{Calculation of $h^{(0)}_{CC'}$}
From \eqref{eq:h0AB_def} we get
\begin{equation}
h^{(0)}_{AB}\big{|}_{\psi=1}=M'^{(0)}_{AB}\big{|}_{\psi=1}
\end{equation}
First we will write $\mathfrak{t}_{AB}$ and $\mathfrak{v}_A$ in a convenient way. From \eqref{structure},  $\mathfrak{t}_{AB}$ can be written as
\begin{equation}\label{eq:tAB}
\mathfrak{t}_{AB}={\cal Y}_{AB}+u_A {\cal X}_B+u_B {\cal X}_A+{\cal Z} u_A u_B
\end{equation}
Where,
\begin{equation}
\begin{split}
{\cal Y}_{AB}&=\frac{K}{D}K_{AB}-\frac{K}{2D}\left(\hat{\nabla}_A u_B+\hat{\nabla}_B u_A\right)-K^F_A K_{FB}+K^F_A~\hat{\nabla}_F u_B+K^F_B~\hat{\nabla}_F u_A-\big(\hat{\nabla}^F u_A\big)\big(\hat{\nabla}_F u_B\big)\\
&-\bigg(\frac{\hat{\nabla}^2 u_A}{K}\bigg)\bigg(\frac{\hat{\nabla}^2 u_B}{K}\bigg)+\bigg(\frac{\hat{\nabla}^2 u_A}{K}\bigg)\bigg(\frac{\hat{\nabla}_B K}{K}\bigg)+\bigg(\frac{\hat{\nabla}^2 u_B}{K}\bigg)\bigg(\frac{\hat{\nabla}_A K}{K}\bigg)-\bigg(\frac{\hat{\nabla}_A K}{K}\bigg)\bigg(\frac{\hat{\nabla}_B K}{K}\bigg)\\
{\cal X}_A&=\frac{K}{D}\left[u^C K_{CA}-\frac{1}{2}(u\cdot\hat{\nabla})u_A\right]-u^C K_{CE}K^E_A+u^C K_{EC}\left(\hat{\nabla}^E u_A\right)+\frac{(u\cdot\nabla)K}{K}\bigg[\frac{\hat{\nabla}^2 u_A}{K}-\frac{\hat{\nabla}_A K}{K}\bigg]\\
{\cal Z}&=\frac{K}{D}~u\cdot K\cdot u-u^C K^F_C K_{FD}u^D-\bigg(\frac{u\cdot\nabla K}{K}\bigg)^2
\end{split}
\end{equation}
From \eqref{structure},  $\mathfrak{v}_A$ can be written as
\begin{equation}\label{eq:vA}
\mathfrak{v}_A={\cal N}_A+{\cal J} u_A
\end{equation}
Where,
\begin{equation}
\begin{split}
{\cal N}_A&=\frac{K^2}{2D^2}\bigg[\frac{\hat{\nabla}^2 u_A}{K}-u^D K_{DA}\bigg]-\bigg[\frac{\hat{\nabla}_F K}{D}-\frac{K}{D}~ u^E K_{EF}\bigg]K^F_A+\bigg[\frac{\hat{\nabla}_F K}{D}-\frac{K}{D}~ u^E K_{EF}\bigg]\left(\hat{\nabla}^F u_A\right)\\
&+\bigg[\frac{(u\cdot\nabla)K}{D}-\frac{K}{D}~u\cdot K\cdot u\bigg]\frac{\hat{\nabla}^2 u_A}{K}-\bigg[\frac{(u\cdot\nabla)K}{D}-\frac{K}{D}~u\cdot K\cdot u\bigg]\frac{\hat{\nabla}_A K}{K}\\
{\cal J}&=-\frac{K^2}{2D^2}~u\cdot K\cdot u-u^B K_{BD}\bigg(\frac{\hat{\nabla}^D K}{D}-\frac{K}{D}~ u^E K^D_E\bigg)-\frac{u\cdot\nabla K}{K}\bigg(\frac{u\cdot\nabla K}{D}-\frac{K}{D}~u\cdot K\cdot u\bigg)
\end{split}
\end{equation}
Using \eqref{eq:tAB} and \eqref{eq:vA} we can write $h^{(0)}_{AB}\big{|}_{\psi=1}$as
\begin{equation}
h^{(0)}_{AB}=\widetilde{\cal S}^{(0)}~u_A u_B+u_A\widetilde{\cal H}^{(0)}_B+u_B\widetilde{\cal H}^{(0)}_A+{\cal W}^{(0)}_{AB}
\end{equation}
Where,
\begin{equation}
\begin{split}
\widetilde{\cal S}^{(0)}&=1-\frac{2}{K^2}\bigg[\cancel{\frac{K}{D}~u\cdot K\cdot u}-u\cdot K\cdot K\cdot u-\bigg(\frac{u\cdot\nabla K}{K}\bigg)^2\bigg]\\
&+\frac{2}{K^2}\left(2~\text{Zeta}[3]-1\right)\Bigg[-\frac{{K}}{D}\left(\frac{u\cdot{\nabla}{K}}{K}-u\cdot {K}\cdot u\right)- 2~\lambda- (u\cdot {K}\cdot {K}\cdot u)+2 \left(\frac{{\nabla}_A{K}}{K}\right)u^B {K}^A_B\\
&-\left(\frac{u\cdot{\nabla}{K}}{K}\right)^2+2\left(\frac{u\cdot{\nabla}{K}}{K}\right)(u\cdot {K}\cdot u)-\left(\frac{\hat{\nabla}^D K}{K}\right)\left(\frac{\hat{\nabla}_D K}{K}\right)-(u\cdot {K}\cdot u)^2+\lambda\Bigg]\\
&-\frac{2}{K^2}\bigg[\cancel{-\frac{K}{D}~u\cdot K\cdot u}-2~u^B K_{BD}\bigg(\frac{\hat{\nabla}^D K}{K}- u^E K^D_E\bigg)-2~\frac{u\cdot\nabla K}{K}\bigg(\frac{u\cdot\nabla K}{K}-u\cdot K\cdot u\bigg)\bigg]\\[10pt]
&=1-\frac{2}{K^2}\bigg[u\cdot K\cdot K\cdot u-3\bigg(\frac{(u\cdot\nabla)K}{K}\bigg)^2-2~u^B K_{BD}\bigg(\frac{\hat{\nabla}^D K}{K}\bigg)+2~u\cdot K\cdot u~\bigg(\frac{u\cdot\nabla K}{K}\bigg)\bigg]\\
&+\frac{2}{K^2}\left(2~\text{Zeta}[3]-1\right)\bigg[-\frac{K}{D}\bigg(\frac{(u\cdot\nabla)K}{K}-u\cdot K\cdot u\bigg)-\lambda-u\cdot K\cdot K\cdot u+2\left(\frac{\nabla_A K}{K}\right)u^B K^A_B\\
&-\bigg(\frac{u\cdot\nabla K}{K}\bigg)^2+2~\frac{(u\cdot\nabla)K}{K}(u\cdot K\cdot u)-\bigg(\frac{\hat{\nabla}^D K}{K}\bigg)\bigg(\frac{\hat{\nabla}_D K}{K}\bigg)-(u\cdot K\cdot u)^2\bigg]
\end{split}
\end{equation}
\begin{equation}
\begin{split}
\widetilde{\cal H}^{(0)}_A&=\frac{1}{K}\frac{\hat{\nabla}_A K}{K}-\frac{2}{K^2}\bigg[\frac{K}{D}\left(u^C K_{CA}-\frac{1}{2}(u\cdot\hat{\nabla})u_A\right)-u^C K_{CE}K^E_A+u^C K_{EC}\left(\hat{\nabla}^E u_A\right)\\
&+\frac{u\cdot\nabla K}{K}\bigg(\frac{\hat{\nabla}^2 u_A}{K}-\frac{\hat{\nabla}_A K}{K}\bigg)\bigg]-\frac{2}{K^2}\frac{D}{K}\bigg[\frac{K^2}{2D^2}\bigg(\frac{\hat{\nabla}^2 u_A}{K}-u^D K_{DA}\bigg)-\bigg(\frac{\hat{\nabla}_F K}{D}-\frac{K}{D} u^E K_{EF}\bigg)K^F_A\\
&+\bigg(\frac{\hat{\nabla}_F K}{D}-\frac{K}{D}~ u^E K_{EF}\bigg)\hat{\nabla}^F u_A+\bigg(\frac{u\cdot\nabla K}{D}-\frac{K}{D}~u\cdot K\cdot u\bigg)\frac{\hat{\nabla}^2 u_A}{K}\\
&-\bigg(\frac{u\cdot\nabla K}{D}-\frac{K}{D}~u\cdot K\cdot u\bigg)\frac{\hat{\nabla}_A K}{K}\bigg]-\frac{4}{K^2}\left(\frac{n\cdot\nabla K}{K}\right)\left(\frac{\hat{\nabla}_A K}{K}\right)+\frac{2}{K^2}\hat{\nabla}_A\left(\frac{n\cdot\nabla K}{K}\right)
\end{split}
\end{equation}
Using the following two identity
\begin{equation}\label{eq:ident2}
\begin{split}
\hat{\nabla}_A\left(\frac{n\cdot\nabla K}{K}\right)&=\hat{\nabla}_A\bigg(\frac{\hat{\nabla}^2 K}{K^2}\bigg)+\lambda \frac{D}{K}\bigg(\frac{\hat{\nabla}_A K}{K}\bigg)-\frac{\hat{\nabla}_A K}{D}\\
\left(u\cdot\hat{\nabla}\right)u_A&=\frac{\hat{\nabla}^2 u_A}{K}-\frac{\hat{\nabla}_A K}{K}+u^D K_{DA}+u_A\left(-\frac{(u\cdot\nabla)K}{K}+u\cdot K\cdot u\right)
\end{split}
\end{equation}
we get,
\begin{equation}
\begin{split}
\widetilde{\cal H}^{(0)}_A&=\frac{1}{K}\frac{\hat{\nabla}_A K}{K}+u^C K_{CA}\left[\cancel{\frac{2}{K^2}\frac{K}{2D}}-\cancel{\frac{2}{K^2}\frac{K}{D}}+\cancel{\frac{2}{K^2}\frac{K}{2D}}\right]+u^C K_{EC}K^E_A\left[\cancel{\frac{2}{K^2}}-\cancel{\frac{2}{K^2}}\right]+\frac{2}{K^2}\hat{\nabla}_A\bigg(\frac{\hat{\nabla}^2 K}{K^2}\bigg)\\
&+u^C K_{EC}\left(\hat{\nabla}^E u_A\right)\left[\cancel{-\frac{2}{K^2}}+\cancel{\frac{2}{K^2}}\right]+\frac{2}{K^2}\frac{D}{K}K^F_A\bigg(\frac{\hat{\nabla}_F K}{D}\bigg)-\frac{2}{K^2}\frac{D}{K}\left(\hat{\nabla}^F u_A\right)\bigg(\frac{\hat{\nabla}_F K}{D}\bigg)\\
&+\frac{\hat{\nabla}^2 u_A}{K}\bigg[\cancel{\frac{2}{K^2}\frac{K}{2D}}-\frac{2}{K^2}\left(\frac{u\cdot\nabla K}{K}\right)-\cancel{\frac{2}{K^2}\frac{K}{2D}}-\frac{2}{K^2}\left(\frac{u\cdot\nabla K}{K}-u\cdot K\cdot u\right)\bigg]\\
&+\frac{\hat{\nabla}_A K}{K}\bigg[-\frac{2}{K^2}\frac{K}{2D}+\frac{2}{K^2}\frac{(u\cdot\nabla)K}{K}+\frac{2}{K^2}\left(\frac{u\cdot\nabla K}{K}-u\cdot K\cdot u\right)+\lambda\frac{2}{K^2}\frac{D}{K}-\frac{2}{K^2}\frac{K}{D}\\
&-\frac{4}{K^2}\left(2~\frac{u\cdot\nabla K}{K}-u\cdot K\cdot u-\frac{K}{D}\right)\bigg]+\frac{2}{K^2}\frac{K}{2D}u_A\left(-\frac{u\cdot\nabla K}{K}+u\cdot K\cdot u\right)\\[8pt]
&=\frac{1}{K}\bigg(\frac{\hat{\nabla}_A K}{K}\bigg)+\frac{2}{K^2}\hat{\nabla}_A\bigg(\frac{\hat{\nabla}^2 K}{K^2}\bigg)+\frac{2}{K^2}K^F_A\bigg(\frac{\hat{\nabla}_F K}{K}\bigg)-\frac{2}{K^2}\left(\hat{\nabla}^F u_A\right)\bigg(\frac{\hat{\nabla}_F K}{K}\bigg)\\
&+\frac{2}{K^2}\bigg(\frac{\hat{\nabla}^2 u_A}{K}\bigg)\bigg[u\cdot K\cdot u-2~\frac{(u\cdot\nabla)K}{K}\bigg]+\frac{2}{K^2}\bigg(\frac{\hat{\nabla}_A K}{K}\bigg)\bigg[u\cdot K\cdot u-2~\frac{(u\cdot\nabla)K}{K}+\lambda~\frac{D}{K}+\frac{K}{2D}\bigg]\\
&+\frac{2}{K^2}\frac{K}{2D}u_A\left(-\frac{u\cdot\nabla K}{K}+u\cdot K\cdot u\right)
\end{split}
\end{equation}
\begin{equation}
\begin{split}
{\cal W}^{(0)}_{AB}&=\frac{1}{K}\left[K_{AB}-\hat{\nabla}_A u_B-\hat{\nabla}_B u_A\right]-\frac{2}{K^2}\bigg[\frac{K}{D} K_{AB}-\frac{K}{2D}\left(\hat{\nabla}_A u_B+\hat{\nabla}_B u_A\right)-K^F_A K_{FB}+K^F_A\hat{\nabla}_F u_B\\
&+K^F_B\hat{\nabla}_F u_A-\left(\hat{\nabla}^F u_A\right)\left(\hat{\nabla}_F u_B\right)-\frac{\hat{\nabla}^2 u_A}{K}\frac{\hat{\nabla}^2 u_B}{K}+\cancel{\frac{\hat{\nabla}^2 u_A}{K}\frac{\hat{\nabla}_B K}{K}}+\cancel{\frac{\hat{\nabla}^2 u_B}{K}\frac{\hat{\nabla}_A K}{K}}-\frac{\hat{\nabla}_A K}{K}\frac{\hat{\nabla}_B K}{K}\bigg]\\
&-\frac{2}{K^2}\bigg[\frac{n\cdot\nabla K}{K}\left(K_{AB}-\hat{\nabla}_A u_B-\hat{\nabla}_B u_A\right)\bigg]\\
&+\frac{1}{K^2}\bigg[\hat{\nabla}_A\bigg(u^E K_{BE}-\frac{\hat{\nabla}^2 u_B}{K}\bigg)+\hat{\nabla}_B\bigg(u^E K_{AE}-\frac{\hat{\nabla}^2 u_A}{K}\bigg)+2K_{AB}\frac{u\cdot\nabla K}{K}\bigg]\\
&-\frac{2}{K^2}\bigg[\bigg(\frac{\hat{\nabla}_B K}{K}\bigg)u^E K_{AE}-\cancel{\frac{\hat{\nabla}_B K}{K}\frac{\hat{\nabla}^2 u_A}{K}}+\bigg(\frac{\hat{\nabla}_A K}{K}\bigg)u^E K_{BE}-\cancel{\frac{\hat{\nabla}_A K}{K}\frac{\hat{\nabla}^2 u_B}{K}}\bigg]\nonumber\\[10pt]
\end{split}
\end{equation}
\begin{equation}
\begin{split}
&=\frac{1}{K}\left[K_{AB}-\hat{\nabla}_A u_B-\hat{\nabla}_B u_A\right]\\
&-\frac{2}{K^2}~K_{AB}\left[\frac{(u\cdot\nabla)K}{K}-u\cdot K\cdot u\right]-\frac{2}{K^2}\left(\hat{\nabla}_A u_B+\hat{\nabla}_B u_A\right)\left[\frac{K}{2D}-2~\frac{(u\cdot\nabla)K}{K}+u\cdot K\cdot u\right]\\
&+\frac{2}{K^2}K^F_A K_{FB}-\frac{2}{K^2}\left(K^F_A\hat{\nabla}_F u_B+K^F_B\hat{\nabla}_F u_A\right)+\frac{2}{K^2}\left(\hat{\nabla}^F u_A\right)\left(\hat{\nabla}_F u_B\right)+\frac{2}{K^2}\frac{\hat{\nabla}^2 u_A}{K}~\frac{\hat{\nabla}^2 u_B}{K}\\
&+\frac{2}{K^2}\bigg(\frac{\hat{\nabla}_A K}{K}\bigg)\bigg(\frac{\hat{\nabla}_B K}{K}\bigg)-\frac{2}{K^2}\bigg[\bigg(\frac{\hat{\nabla}_A K}{K}\bigg)u^EK_{EB}+\bigg(\frac{\hat{\nabla}_B K}{K}\bigg)u^EK_{EA}\bigg]\\
&+\frac{1}{K^2}\left[\hat{\nabla}_A\left(u^E K_{EB}\right)+\hat{\nabla}_B\left(u^E K_{EA}\right)\right]-\frac{1}{K^2}\bigg[\hat{\nabla}_A\bigg(\frac{\hat{\nabla}^2 u_B}{K}\bigg)+\hat{\nabla}_B\bigg(\frac{\hat{\nabla}^2 u_A}{K}\bigg)\bigg]
\end{split}
\end{equation}
\subsubsection*{Calculation of $h^{(1)}_{CC'}$}
From \eqref{eq:h1AB_def} $h^{(1)}_{AB}$ on $\psi=1$ is given by
\begin{equation}
\begin{split}
h^{(1)}_{AB}&={M^\prime}^{(1)}_{AB}+C^{(0)}_{AB}\\
&=C^{(0)}_{AB}-\frac{2D}{K^2}\left[\mathfrak{t}_{AB}+\mathfrak{s}_1~ u_A u_B+\frac{D}{K}\big(~ \mathfrak{v}_A u_B + \mathfrak{v}_B u_A\big)\right]\\
&+\frac{1}{K}\left[u_A\frac{\hat{\nabla}_B K}{K}+u_B\frac{\hat{\nabla}_A K}{K}+K_{AB}-\hat{\nabla}_B u_A-\hat{\nabla}_A u_B\right]+{\cal O}\left(\frac{1}{D}\right)^2
\end{split}
\end{equation}
From \eqref{eq:C0AB_def} 
\begin{equation}
\begin{split}
C^{(0)}_{CC'}&=\frac{1}{N}\Pi^A_C\Pi^B_{C'}(n.\nabla){M^\prime}^{(0)}_{AB}\\
&=\frac{1}{N}\left[u_C \Pi^E_{C'}(n\cdot\nabla)u_E+u_{C'} \Pi^E_C(n\cdot\nabla)u_E\right]\\
&-\frac{1}{N K}\frac{(n\cdot\nabla)N}{N}\bigg[u_C\frac{\hat{\nabla}_{C'} K}{K}+u_{C'}\frac{\hat{\nabla}_C K}{K}+K_{CC'}-\left(\hat{\nabla}_C u_{C'}+\hat{\nabla}_{C'} u_C\right)\bigg]\\
&+\frac{1}{NK}\Pi^E_C\Pi^F_{C'}(n\cdot\nabla)\bigg[u_E\Pi^B_F\frac{\nabla_B K}{K}+u_F\Pi^B_E\frac{\nabla_B K}{K}+K_{EF}-\Pi^A_E\Pi^B_F\left(\nabla_A u_B+\nabla_B u_A\right)\hspace{-0.05in}\bigg]+{\cal O}\left(\frac{1}{D}\right)^2
\end{split}
\end{equation}
To simplify the above expression we will use the following identity. We will not give the derivations of these identities. The derivations are quite straightforward
\begin{equation}\label{eq:ident1}
\begin{split}
\Pi^E_C\Pi^F_{C'}(n\cdot\nabla)K_{EF}&=-\frac{\hat{\nabla}_C K}{K}\frac{\hat{\nabla}_{C'} K}{K}-\lambda~ \Pi_{CC'}+\hat{\nabla}_C\bigg(\frac{\hat{\nabla}_{C'}K}{K}\bigg)-K^E_C K_{EC'}\\
\Pi^E_C\Pi^F_{C'}(n\cdot\nabla)\left[u_E \Pi^B_F\frac{\nabla_B K}{K}\right]&=\frac{\hat{\nabla}^2 u_C}{K}\frac{\hat{\nabla}_{C'} K}{K}-u_C\left(\frac{n\cdot\nabla K}{K}\right)\frac{\hat{\nabla}_{C'} K}{K}+u_C\bigg[\frac{1}{K^2}\hat{\nabla}_{C'}\left(\hat{\nabla}^2 K\right)\\
&-\frac{\hat{\nabla}_{C'} K}{K}\bigg(2\frac{\hat{\nabla}^2 K}{K^2}-\lambda~\frac{D}{K}-\frac{K}{D}\bigg)-2~\frac{K}{D}\left(\frac{\hat{\nabla}_{C'}K}{K}\right)-K^D_{C'}\bigg(\frac{\nabla_D K}{K}\bigg)\bigg]\\
\end{split}
\end{equation}
\begin{equation}\label{eq:eq:B.54}
\begin{split}
\Pi^E_C\Pi^F_{C'}(n\cdot\nabla)\left[\Pi^A_E\Pi^B_F\nabla_A u_B\right]&=-\frac{\hat{\nabla}_C K}{K}\frac{\hat{\nabla}^2 u_{C'}}{K}+\frac{\hat{\nabla}_{C'} K}{K}u^B K_{BC}-K_{CC'}\frac{u\cdot\nabla K}{K}\\
&+\hat{\nabla}_C\bigg(\frac{\hat{\nabla}^2 u_{C'}}{K}\bigg)-K^D_C\left(\hat{\nabla}_D u_{C'}\right)
\end{split}
\end{equation}
Using \eqref{eq:ident1} and \eqref{eq:eq:B.54} we can write $C^{(0)}_{CC'}$ as
\begin{equation}\label{eq:C0AB}
C^{(0)}_{CC'}=u_C ~\tau_{C'}+u_{C'} ~\tau_C+\Xi_{CC'}
\end{equation}
Where,
\begin{equation}
\begin{split}
\tau_C&=\frac{1}{N}\Pi^E_C(n\cdot\nabla)u_E+\frac{1}{NK}\bigg[-\frac{\hat{\nabla}_C K}{K}\bigg(2~\frac{(n\cdot\nabla)K}{K}+2~\frac{\hat{\nabla}^2 K}{K^2}-\lambda~\frac{D}{K}\bigg)+\frac{1}{K^2}\hat{\nabla}_C\left(\hat{\nabla}^2 K\right)\\
&-2~\frac{K}{D}\bigg(\frac{\hat{\nabla}_C K}{K}\bigg)-K^D_C\bigg(\frac{\nabla_D K}{K}\bigg)\bigg]\\
\Xi_{CC'}&=-\frac{1}{NK}\frac{(n\cdot\nabla)N}{N}\left[K_{CC'}-\hat{\nabla}_C u_{C'}-\hat{\nabla}_{C'}u_C\right]+\frac{1}{NK}\bigg[2~\frac{\hat{\nabla}^2 u_C}{K}\frac{\hat{\nabla}_{C'}K}{K}+2~\frac{\hat{\nabla}^2 u_{C'}}{K}\frac{\hat{\nabla}_C K}{K}\\
&-\frac{\hat{\nabla}_C K}{K}\frac{\hat{\nabla}_{C'} K}{K}-\lambda~\Pi_{CC'}+\hat{\nabla}_C\bigg(\frac{\hat{\nabla}_{C'}K}{K}\bigg)-K^E_C K_{EC'}-\frac{\hat{\nabla}_{C'}K}{K} u^B K_{BC}-\frac{\hat{\nabla}_CK}{K} u^B K_{BC'}\\
&+2~K_{CC'}\frac{(u\cdot\nabla)K}{K}+K^D_C\left(\hat{\nabla}_D u_{C'}\right)+K^D_{C'}\left(\hat{\nabla}_D u_C\right)-\hat{\nabla}_C\bigg(\frac{\hat{\nabla}^2 u_{C'}}{K}\bigg)-\hat{\nabla}_{C'}\bigg(\frac{\hat{\nabla}^2 u_C}{K}\bigg)\bigg]
\end{split}
\end{equation}
Using \eqref{eq:tAB}, \eqref{eq:vA} and \eqref{eq:C0AB} we can write $h^{(1)}_{AB}$ as
\begin{equation}
h^{(1)}_{AB}=\Phi~u_A u_B+u_A~\Omega_B+u_B~\Omega_A+{\cal W}^{(1)}_{AB}
\end{equation}
Where,
\begin{equation}
\begin{split}
\Phi&=-2~\frac{D}{K^2}\bigg[\frac{K}{D}~u\cdot K\cdot u-u^C K^F_C K_{FD}u^D-\bigg(\frac{u\cdot\nabla K}{K}\bigg)^2\bigg]-2~\frac{D}{K^2}\bigg[\lambda+\left(\frac{u\cdot{\nabla}K}{K}\right)^2\\
&+\frac{\hat{\nabla}_A {K}}{K}\bigg(4~u^B {K}^A_B-2\left[(u\cdot\hat{\nabla})u^A\right]-\frac{\hat{\nabla}^A {K}}{K}\bigg)-(\hat{\nabla}_A u_B)(\hat{\nabla}^A u^B)-[(u\cdot\hat{\nabla})u_A][(u\cdot\hat{\nabla})u^A]\\
&-(u\cdot {K}\cdot u)^2+2\left[(u\cdot\hat{\nabla})u^A\right](u^B {K}_{BA})-3~(u\cdot {K}\cdot {K}\cdot u)-\frac{K}{D}\left(\frac{u\cdot{\nabla}{K}}{K}-u\cdot {K}\cdot u\right)\bigg]\\
&-4~\frac{D^2}{K^3}\bigg[-\frac{K^2}{2D^2}~u\cdot K\cdot u-u^B K_{BD}\bigg(\frac{\hat{\nabla}^D K}{D}-\frac{K}{D}~u^E K^D_E\bigg)-\frac{u\cdot\nabla K}{K}\bigg(\frac{u\cdot\nabla K}{D}-\frac{K}{D} u\cdot K\cdot u\bigg)\bigg]\nonumber\\[10pt]
\end{split}
\end{equation}
\begin{equation}
\begin{split}
&=-2~\frac{D}{K^2}\bigg[-2~u\cdot K\cdot K\cdot u+\lambda+\frac{\hat{\nabla}_A {K}}{K}\bigg(2~u^B {K}^A_B-\frac{\hat{\nabla}^A {K}}{K}\bigg)-(\hat{\nabla}_A u_B)(\hat{\nabla}^A u^B)\\
&-(u\cdot {K}\cdot u)^2-[(u\cdot\hat{\nabla})u_A]\bigg((u\cdot\hat{\nabla})u^A+2~\frac{\hat{\nabla}^A K}{K}-2~u_B K^{BA}\bigg)+\frac{K}{D}~u\cdot K\cdot u\\
&-\frac{u\cdot\nabla K}{K}\bigg(\frac{K}{D}+2~\frac{u\cdot\nabla K}{K}-2~u\cdot K\cdot u\bigg)\bigg]
\end{split}
\end{equation}
Using 2nd identity of \eqref{eq:ident2} we can write the above equation as
\begin{equation}
\begin{split}
\Phi
&=-2~\frac{D}{K^2}\bigg[-2~u\cdot K\cdot K\cdot u+\lambda+\frac{\hat{\nabla}_A {K}}{K}\bigg(2~u^B {K}^A_B-\frac{\hat{\nabla}^A {K}}{K}\bigg)-(\hat{\nabla}_A u_B)(\hat{\nabla}^A u^B)\\
&-(u\cdot {K}\cdot u)^2-\bigg(\frac{\hat{\nabla}^2 u_A}{K}-\frac{\hat{\nabla}_A K}{K}+u^E K_{EA}\bigg)\bigg(\frac{\hat{\nabla}^2 u^A}{K}+\frac{\hat{\nabla}^A K}{K}-u_D K^{DA}\bigg)+\frac{K}{D}~u\cdot K\cdot u\\
&+\bigg(\frac{u\cdot\nabla K}{K}-u\cdot K\cdot u\bigg)\bigg(\frac{u\cdot\nabla K}{K}-u\cdot K\cdot u\bigg)-\frac{u\cdot\nabla K}{K}\bigg(\frac{K}{D}+2~\frac{u\cdot\nabla K}{K}-2~u\cdot K\cdot u\bigg)\bigg]\\[8pt]
&=-2~\frac{D}{K^2}\bigg[\lambda-u\cdot K\cdot K\cdot u-(\hat{\nabla}_A u_B)(\hat{\nabla}^A u^B)-\frac{\hat{\nabla}^2 u_A}{K}\frac{\hat{\nabla}^2 u^A}{K}-\bigg(\frac{u\cdot\nabla K}{K}\bigg)^2\\
&-\frac{K}{D}~\frac{u\cdot\nabla K}{K}+\frac{K}{D}~u\cdot K\cdot u\bigg]\\
\end{split}
\end{equation}
\begin{equation}
\begin{split}
\Omega_A&=-2~\frac{D}{K^2}\bigg[\frac{K}{D}\left(u^C K_{CA}-\frac{1}{2}(u\cdot\hat{\nabla})u_A\right)-u^C K_{CE}K^E_A+u^C K_{EC}\left(\hat{\nabla}^E u_A\right)\\
&+\frac{(u\cdot\nabla)K}{K}\bigg(\frac{\hat{\nabla}^2 u_A}{K}-\frac{\hat{\nabla}_A K}{K}\bigg)\bigg]-2~\frac{D}{K^2}\bigg[\frac{K}{2D}\bigg(\frac{\hat{\nabla}^2 u_A}{K}-u^D K_{DA}\bigg)-\bigg(\frac{\hat{\nabla}_F K}{K}- u^E K_{EF}\bigg)K^F_A\\
&+\bigg(\frac{\hat{\nabla}_F K}{K}-u^E K_{EF}\bigg)\left(\hat{\nabla}^F u_A\right)+\bigg(\frac{u\cdot\nabla K}{K}-u\cdot K\cdot u\bigg)\frac{\hat{\nabla}^2 u_A}{K}-\bigg(\frac{u\cdot\nabla K}{K}-u\cdot K\cdot u\bigg)\frac{\hat{\nabla}_A K}{K}\bigg]\\
&+\frac{1}{K}\frac{\hat{\nabla}_A K}{K}+\frac{1}{N}\Pi^E_A(n\cdot\nabla)u_E+\frac{D}{K^2}\bigg[-\frac{\hat{\nabla}_A K}{K}\bigg(2~\frac{(n\cdot\nabla)K}{K}+2~\frac{\hat{\nabla}^2 K}{K^2}-\lambda~\frac{D}{K}\bigg)\\
&+\frac{1}{K^2}\hat{\nabla}_A\left(\hat{\nabla}^2 K\right)-2~\frac{K}{D}\bigg(\frac{\hat{\nabla}_A K}{K}\bigg)-K^D_A\bigg(\frac{\nabla_D K}{K}\bigg)\bigg]\\
\end{split}
\end{equation}
To simplify the above expression we will use the following identity
\begin{equation}\label{eq:ident3}
\begin{split}
\frac{1}{N}\Pi^E_A(n\cdot\nabla)u_E&=\frac{D}{K}\bigg(\frac{\hat{\nabla}^2 u_A}{K}\bigg)+\frac{D}{K^2}\bigg[-u^B K_{BD}K^D_A+\frac{1}{K^2}\hat{\nabla}^2\hat{\nabla}^2 u_A-\frac{\hat{\nabla}^B K}{K}~\hat{\nabla}_B u_A-\frac{u\cdot\nabla K}{K}\frac{\hat{\nabla}_A K}{K}\\
&+\frac{\hat{\nabla}^2 u_A}{K}\left(-8~\frac{u\cdot\nabla K}{K}+4~u\cdot K\cdot u-2~\lambda~\frac{D}{K}+2~\frac{K}{D}\right)\bigg]\\
&+u_A ~\frac{D}{K^2}\bigg[-(\hat{\nabla}_D u_E)(\hat{\nabla}^D u^E)-u\cdot K\cdot K\cdot u-\frac{\hat{\nabla}^2 u_E}{K}~\frac{\hat{\nabla}^2 u^E}{K}-\bigg(\frac{u\cdot\nabla K}{K}\bigg)^2\bigg]
\end{split}
\end{equation}
To prove the above identity we have used subsidiary condition $P^A_B(O\cdot\nabla)O_A=0$ and the second order membrane equation(  2.17 in \cite{secondorder} ). Using \eqref{eq:ident3} we get
\begin{equation}
\begin{split}
\Omega_A&=\frac{D}{K}~\frac{\hat{\nabla}^2 u_A}{K}+\frac{D}{K^2}\frac{K}{D}\bigg(\frac{\hat{\nabla}_A K}{K}\bigg)+\frac{D}{K^2}\bigg[-u^B K_{BD}K^D_A+\frac{1}{K^2}\hat{\nabla}^2\hat{\nabla}^2 u_A-\frac{\hat{\nabla}^B K}{K}~\hat{\nabla}_B u_A\\
&-\frac{u\cdot\nabla K}{K}\frac{\hat{\nabla}_A K}{K}+\frac{\hat{\nabla}^2 u_A}{K}\left(-8~\frac{u\cdot\nabla K}{K}+4~u\cdot K\cdot u-2~\lambda~\frac{D}{K}+2~\frac{K}{D}\right)\bigg]\\
&+\frac{D}{K^2}\frac{\hat{\nabla}_A K}{K}\bigg(-2~\frac{n\cdot\nabla K}{K}-2~\frac{\hat{\nabla}^2 K}{K^2}+\lambda~\frac{D}{K}\bigg)+\frac{D}{K^2}\frac{1}{K^2}\hat{\nabla}_A\left(\hat{\nabla}^2 K\right)-2~\frac{D}{K^2}~\frac{K}{D}\bigg(\frac{\hat{\nabla}_A K}{K}\bigg)\\
&-\frac{D}{K^2}K^D_A\bigg(\frac{\hat{\nabla}_D K}{K}\bigg)+\frac{D}{K^2}\bigg[-\cancel{2~\frac{K}{D}u^C K_{CA}}+\frac{K}{D}\bigg(\frac{\hat{\nabla}^2 u_A}{K}-\frac{\hat{\nabla}_A K}{K}+\cancel{u^D K_{DA}}\bigg)+2~u^C K_{CE}K^E_A\\
&-\cancel{2~u^C K_{EC}\left(\hat{\nabla}^E u_A\right)}-2~\frac{(u\cdot\nabla)K}{K}\bigg(\frac{\hat{\nabla}^2 u_A}{K}-\frac{\hat{\nabla}_A K}{K}\bigg)-\frac{K}{D}\bigg(\frac{\hat{\nabla}^2 u_A}{K}-\cancel{u^D K_{DA}}\bigg)\\
&+2~K^F_A\bigg(\frac{\hat{\nabla}_F K}{K}- u^E K_{EF}\bigg)-2~(\hat{\nabla}^F u_A)\bigg(\frac{\hat{\nabla}_F K}{K}-\cancel{u^E K_{EF}}\bigg)-2~\frac{\hat{\nabla}^2 u_A}{K}\bigg(\frac{u\cdot\nabla K}{K}-u\cdot K\cdot u\bigg)\\
&+2~\frac{\hat{\nabla}_A K}{K}\bigg(\frac{u\cdot\nabla K}{K}-u\cdot K\cdot u\bigg)\bigg]+u_A ~\frac{D}{K^2}\bigg[-(\hat{\nabla}_D u_E)(\hat{\nabla}^D u^E)-u\cdot K\cdot K\cdot u\\
&-\frac{\hat{\nabla}^2 u_E}{K}~\frac{\hat{\nabla}^2 u^E}{K}-\bigg(\frac{u\cdot\nabla K}{K}\bigg)^2+\frac{K}{D}\bigg(-\frac{u\cdot\nabla K}{K}+u\cdot K\cdot u\bigg)\bigg]\\[10pt]
&=\frac{D}{K}\bigg(\frac{\hat{\nabla}^2 u_A}{K}\bigg)+\frac{D}{K^2}\bigg(\frac{\hat{\nabla}_A K}{K}\bigg)\bigg[-5~\frac{(u\cdot\nabla)K}{K}+2~u\cdot K\cdot u-\lambda~\frac{D}{K}\bigg]\\
&+\frac{D}{K^2}\bigg(\frac{\hat{\nabla}^2 u_A}{K}\bigg)\bigg[-12~\frac{(u\cdot\nabla)K}{K}+6~u\cdot K\cdot u-2~\lambda~\frac{D}{K}+2~\frac{K}{D}\bigg]\\
&+\frac{D}{K^2}\bigg[-u^B K_{BD}K^D_A+\frac{1}{K^2}\hat{\nabla}^2\left(\hat{\nabla}^2 u_A\right)-3\bigg(\frac{\hat{\nabla}^B K}{K}\bigg)\hat{\nabla}_B u_A+\frac{1}{K^2}\hat{\nabla}_A\left(\hat{\nabla}^2 K\right)+K^D_A\bigg(\frac{\hat{\nabla}_D K}{K}\bigg)\bigg]\\
&+u_A ~\frac{D}{K^2}\bigg[-(\hat{\nabla}_D u_E)(\hat{\nabla}^D u^E)-u\cdot K\cdot K\cdot u-\frac{\hat{\nabla}^2 u_E}{K}~\frac{\hat{\nabla}^2 u^E}{K}-\bigg(\frac{u\cdot\nabla K}{K}\bigg)^2\\
&+\frac{K}{D}\bigg(-\frac{u\cdot\nabla K}{K}+u\cdot K\cdot u\bigg)\bigg]
\end{split}
\end{equation}

\begin{equation}
\begin{split}
{\cal W}^{(1)}_{AB}&=-2\frac{D}{K^2}\bigg[\frac{K}{D}K_{AB}-\frac{K}{2D}(\hat{\nabla}_A u_B+\hat{\nabla}_B u_A)-K^F_A K_{FB}+K^F_A~\hat{\nabla}_F u_B+K^F_B~\hat{\nabla}_F u_A-\hat{\nabla}^F u_A \hat{\nabla}_F u_B\\
&-\frac{\hat{\nabla}^2 u_A}{K}\frac{\hat{\nabla}^2 u_B}{K}+\cancel{\frac{\hat{\nabla}^2 u_A}{K}\frac{\hat{\nabla}_B K}{K}}+\cancel{\frac{\hat{\nabla}^2 u_B}{K}\frac{\hat{\nabla}_A K}{K}}-\frac{\hat{\nabla}_A K}{K}\frac{\hat{\nabla}_B K}{K}\bigg]+\cancel{\frac{1}{K}\left[K_{AB}-\hat{\nabla}_B u_A-\hat{\nabla}_A u_B\right]}\\
&-\frac{1}{NK}\bigg[\cancel{N}+\frac{(n\cdot\nabla)K}{K}\bigg]\left[K_{AB}-\hat{\nabla}_A u_{B}-\hat{\nabla}_{B}u_A\right]+\frac{D}{K^2}\bigg[\cancel{2~\frac{\hat{\nabla}^2 u_A}{K}\frac{\hat{\nabla}_{B}K}{K}}+\cancel{2~\frac{\hat{\nabla}^2 u_{B}}{K}\frac{\hat{\nabla}_A K}{K}}\\
&-\frac{\hat{\nabla}_A K}{K}\frac{\hat{\nabla}_{B} K}{K}-\lambda~\Pi_{AB}+\hat{\nabla}_A\bigg(\frac{\hat{\nabla}_{B}K}{K}\bigg)-K^E_A K_{EB}-\frac{\hat{\nabla}_{B}K}{K} u^E K_{EA}-\frac{\hat{\nabla}_AK}{K} u^E K_{EB}\\
&+2~K_{AB}\frac{(u\cdot\nabla)K}{K}+K^D_A\left(\hat{\nabla}_D u_{B}\right)+K^D_{B}\left(\hat{\nabla}_D u_A\right)-\hat{\nabla}_A\bigg(\frac{\hat{\nabla}^2 u_{B}}{K}\bigg)-\hat{\nabla}_{B}\bigg(\frac{\hat{\nabla}^2 u_A}{K}\bigg)\bigg]\\[10pt]
&=\frac{D}{K^2}\left[u\cdot K\cdot u-\frac{K}{D}\right]K_{AB}+\frac{D}{K^2}\bigg[\frac{\hat{\nabla}^2 K}{K^2}-\lambda~\frac{D}{K}\bigg]\left(\hat{\nabla}_A u_B+\hat{\nabla}_B u_A\right)+\frac{D}{K^2}~K^F_A K_{FB}\\
&-\frac{D}{K^2}~\lambda~ \Pi_{AB}-\frac{D}{K^2}\left(K^F_A~\hat{\nabla}_F u_B+K^F_B~\hat{\nabla}_F u_A\right)+2~\frac{D}{K^2}\left(\hat{\nabla}^F u_A\right)\left(\hat{\nabla}_F u_B\right)\\
&+2~\frac{D}{K^2}\bigg(\frac{\hat{\nabla}^2 u_A}{K}\bigg)\bigg(\frac{\hat{\nabla}^2 u_B}{K}\bigg)+\frac{D}{K^2}\frac{1}{K}\hat{\nabla}_A\left(\hat{\nabla}_B K\right)-\frac{D}{K^2}\bigg[\bigg(\frac{\hat{\nabla}_A K}{K}\bigg)u^E K_{EB}+\bigg(\frac{\hat{\nabla}_B K}{K}\bigg)u^E K_{EA}\bigg]\\
&-\frac{D}{K^2}\frac{1}{K}\left[\hat{\nabla}_A\left(\hat{\nabla}^2 u_B\right)+\hat{\nabla}_B\left(\hat{\nabla}^2 u_A\right)\right]+\frac{D}{K^2}\bigg[\bigg(\frac{\hat{\nabla}_A K}{K}\bigg)\bigg(\frac{\hat{\nabla}^2 u_B}{K}\bigg)+\bigg(\frac{\hat{\nabla}_B K}{K}\bigg)\bigg(\frac{\hat{\nabla}^2 u_A}{K}\bigg)\bigg]
\end{split}
\end{equation}

\subsection{Inside$(\psi<1)$}\label{app:inside}
From \eqref{eq:Einstein_in}
\begin{equation}\label{eq:B.58}
\begin{split}
\Rightarrow ~~&\Pi^A_C\Pi^B_{C'}\bigg[\underbrace{\nabla_A\nabla_E\tilde{\mathfrak{h}}^E_B+\nabla_B\nabla_E\tilde{\mathfrak{h}}^E_A}_{\text{Part-1}}\underbrace{-\nabla^2 \tilde{\mathfrak{h}}_{AB}}_{\text{Part-2}}\underbrace{-\nabla_B\nabla_A\tilde{\mathfrak{h}}}_{\text{Part-3}}\\
&\underbrace{+2~\bar{R}_{EABC}\tilde{\mathfrak{h}}^{EC}+\bar{R}_{AC}\tilde{\mathfrak{h}}^C_B+\bar{R}_{BC}\tilde{\mathfrak{h}}^C_A-2(D-1)\lambda\tilde{\mathfrak{h}}_{AB}}_{\text{Part-4}}\bigg]=0
\end{split}
\end{equation}
Now, we will simplify the above equation
\begin{equation}
\begin{split}
\text{Part-1}&=\Pi^A_C\Pi^B_{C'}\left[\nabla_A\nabla_E\tilde{\mathfrak{h}}^E_B+\nabla_B\nabla_E\tilde{\mathfrak{h}}^E_A\right]\\
&=\Pi^A_C\Pi^B_{C'}\sum_{m=0}^\infty\left[\nabla_A\left((\psi-1)^m\nabla_E[\tilde{h}^{(m)}]^E_B\right)+\nabla_B\left((\psi-1)^m\nabla_E[\tilde{h}^{(m)}]^E_A\right)\right]\\
&=\Pi^A_C\Pi^B_{C'}\sum_{m=0}^\infty(\psi-1)^m\left[\nabla_A\nabla_E[\tilde{h}^{(m)}]^E_B+\nabla_B\nabla_E[\tilde{h}^{(m)}]^E_A\right]\\
\end{split}
\end{equation}
\begin{equation}
\begin{split}
\text{Part-2}&=-\Pi^A_C\Pi^B_{C'}\nabla^2\tilde{\mathfrak{h}}_{AB}\\
&=-\Pi^A_C\Pi^B_{C'}\nabla^D\sum_{m=0}^\infty\left[m(\psi-1)^{m-1}Nn_D\tilde{h}_{AB}^{(m)}+(\psi-1)^m\nabla_D\tilde{h}^{(m)}_{AB}\right]\\
&=-\Pi^A_C\Pi^B_{C'}\sum_{m=0}^\infty\bigg[m(m-1)(\psi-1)^{m-2}N^2\tilde{h}^{(m)}_{AB}+m(\psi-1)^{m-1}[(n\cdot\nabla)N]\tilde{h}^{(m)}_{AB}\\
&+m(\psi-1)^{m-1}NK\tilde{h}^{(m)}_{AB}+2m(\psi-1)^{m-1}N(n\cdot\nabla)\tilde{h}^{(m)}_{AB}+(\psi-1)^m\nabla^2\tilde{h}^{(m)}_{AB}\bigg]
\end{split}
\end{equation}
\begin{equation}
\begin{split}
\text{Part-3}&=-\Pi^{A}_{C}\Pi^{B}_{C'}\left[\nabla_B\nabla_A\tilde{\mathfrak{h}}\right]\\
&=-\Pi^{A}_{C}\Pi^{B}_{C'}\nabla_B\sum_{m=0}^\infty\left[m(\psi-1)^{m-1}Nn_A\tilde{h}^{(m)}+(\psi-1)^m\nabla_A\tilde{h}^{(m)}\right]\\
&=-\Pi^{A}_{C}\Pi^{B}_{C'}\sum_{m=0}^\infty\left[m(\psi-1)^{m-1}N(\nabla_Bn_A)\tilde{h}^{(m)}+(\psi-1)^m\nabla_B\nabla_A\tilde{h}^{(m)}\right]\\
\end{split}
\end{equation}
\begin{equation}
\begin{split}
\text{Part-4}&=\Pi^{A}_{C}\Pi^{B}_{C'}\left[2~\bar{R}_{EABC}\tilde{\mathfrak{h}}^{EC}+\bar{R}_{AC}\tilde{\mathfrak{h}}^C_B+\bar{R}_{BC}\tilde{\mathfrak{h}}^C_A-2(D-1)\lambda\tilde{\mathfrak{h}}_{AB}\right]\\
&=\Pi^{A}_{C}\Pi^{B}_{C'}\left[2\lambda\left(g_{EB}g_{AC}-g_{EC}g_{AB}\right)\tilde{\mathfrak{h}}^{EC}+2\lambda(D-1)\tilde{\mathfrak{h}}_{AB}-2\lambda(D-1)\lambda\tilde{\mathfrak{h}}_{AB}\right]\\
&=\Pi^{A}_{C}\Pi^{B}_{C'}~2\lambda\sum_{m=0}^\infty(\psi-1)^m\left[\tilde{h}^{(m)}_{AB}-\tilde{h}^{(m)}g_{AB}\right]
\end{split}
\end{equation}
Collecting the coefficient of $(\psi-1)^0$ of \eqref{eq:B.58}
\begin{equation}\label{Inside}
\begin{split}
&\Pi^{A}_{C}\Pi^{B}_{C'}\bigg[\nabla_A\nabla_E\left[\tilde{h}^{(0)}\right]^E_B  +\nabla_{B}\nabla_{E}\left[\tilde{h}^{(0)}\right]^E_A-2N^{2}\tilde{h}^{(2)}_{AB}-{[(n\cdot\nabla)N]}\tilde{h}^{(1)}_{AB}-N{ K}\tilde{h}^{(1)}_{AB}\\
&-2N(n\cdot\nabla)\tilde{h}^{(1)}_{AB}-\nabla^{2}\tilde{h}^{(0)}_{AB}-N~{ K}_{AB}~\tilde{h}^{(1)}-\nabla_B\nabla_{A}\tilde{h}^{(0)}+2~\lambda ~\tilde{h}^{(0)}_{AB}-2~\lambda ~\tilde{h}^{(0)}~g_{AB}\bigg]=0\\
\end{split}
\end{equation}
Using \eqref{eq:h1t_def}, the leading order(${\cal O}(D)$) terms of \eqref{Inside}
\begin{equation}\label{eq:H11lead}
\begin{split}
&\Pi^{A}_{C}\Pi^{B}_{C'}\bigg[\nabla_A\nabla_E\big[\tilde{h}^{(0)}\big]^E_B  +\nabla_{B}\nabla_{E}\big[\tilde{h}^{(0)}\big]^E_A-N{ K}\tilde{h}^{(1,1)}_{CC'}-\nabla^{2}\tilde{h}^{(0)}_{AB}-NK_{AB}\tilde{h}^{(1,1)}\bigg]=0\\
\end{split}
\end{equation}
In the last equation, we have used the fact that $\tilde{h}^{(0)}$ can nowhere be ${\cal O}(D)$. Taking trace of \eqref{eq:H11lead}
\begin{equation}
\begin{split}
&\Pi^{AB}\bigg[2~\nabla_A\nabla_E\big[\tilde{h}^{(0)}\big]^E_B-\nabla^{2}\tilde{h}^{(0)}_{AB}\bigg]-2NK\tilde{h}^{(1,1)}=0\\
\Rightarrow~&\tilde{h}^{(1,1)}=\frac{1}{2NK}\Pi^{AB}\bigg[2~\nabla_A\nabla_E\big[\tilde{h}^{(0)}\big]^E_B-\nabla^{2}\tilde{h}^{(0)}_{AB}\bigg]
\end{split}
\end{equation}
Now, from \eqref{eq:H11lead}
\begin{equation}\label{eq:ht11_expr}
\begin{split}
&\tilde{h}^{(1,1)}_{CC'}=\Pi^{A}_{C}\Pi^{B}_{C'}\frac{1}{NK}\bigg[\nabla_A\nabla_E\big[\tilde{h}^{(0)}\big]^E_B  +\nabla_{B}\nabla_{E}\big[\tilde{h}^{(0)}\big]^E_A-\nabla^{2}\tilde{h}^{(0)}_{AB}\bigg]-\frac{1}{K}K_{CC'}\tilde{h}^{(1,1)}\\
\end{split}
\end{equation}
From, subleading order(${\cal O}(1)$) of \eqref{Inside}
\begin{equation}\label{eq:subleading}
\begin{split}
&\Pi^{A}_{C}\Pi^{B}_{C'}\bigg[-2N^{2}\tilde{h}^{(2)}_{AB}-{[(n\cdot\nabla)N]}\tilde{h}^{(1,1)}_{AB}-\frac{NK}{D}\tilde{h}^{(1,2)}_{AB}-2N(n\cdot\nabla)\tilde{h}^{(1,1)}_{AB}-\frac{N}{D}K_{AB}\tilde{h}^{(1,2)}\\
&-\nabla_B\nabla_{A}\tilde{h}^{(0)}+2~\lambda \tilde{h}^{(0)}_{AB}-2~\lambda \tilde{h}^{(0)}g_{AB}\bigg]={\cal O}\left(\frac{1}{D}\right)\\
\end{split}
\end{equation}
Taking trace,
\begin{equation}
\tilde{h}^{(1,2)}=\frac{D}{2NK}\bigg[-2N^{2}\tilde{h}^{(2)}-{[(n\cdot\nabla)N]}\tilde{h}^{(1,1)}-2\lambda \tilde{h}^{(0)}(D-2)-\Pi^{AB}\bigg\{2N(n\cdot\nabla)\tilde{h}^{(1,1)}_{AB}+\nabla_B\nabla_{A}\tilde{h}^{(0)}\bigg\}\bigg]+{\cal O}\left(1\right)
\end{equation}
Now, from \eqref{eq:subleading}
\begin{equation}\label{eq:ht12_expr}
\begin{split}
\tilde{h}^{(1,2)}_{CC'}&=\frac{D}{NK}\Pi^{A}_{C}\Pi^{B}_{C'}\bigg[-2N^{2}\tilde{h}^{(2)}_{AB}-{[(n\cdot\nabla)N]}\tilde{h}^{(1,1)}_{AB}-2N(n\cdot\nabla)\tilde{h}^{(1,1)}_{AB}-\frac{N}{D}{ K}_{AB}\tilde{h}^{(1,2)}\\
&-\nabla_B\nabla_{A}\tilde{h}^{(0)}+2~\lambda \tilde{h}^{(0)}_{AB}-2~\lambda \tilde{h}^{(0)}g_{AB}\bigg]+{\cal O}\left(\frac{1}{D}\right)
\end{split}
\end{equation}
Collecting coefficients of $(\psi-1)$ of \eqref{eq:B.58} at order(${\cal O}(D)$)
\begin{equation}\label{eq:htildesub}
\begin{split}
&\Pi^{A}_{C}\Pi^{B}_{C'}\bigg[\nabla_A\nabla_E\big[\tilde{h}^{(1,1)}\big]^E_B  +\nabla_{B}\nabla_{E}\big[\tilde{h}^{(1,1)}\big]^E_A-2N{ K}\tilde{h}^{(2)}_{AB}-\nabla^2\tilde{h}^{(1,1)}_{AB}-2NK_{AB}\tilde{h}^{(2)}\\
&-\nabla_B\nabla_A \tilde{h}^{(1,1)}-2\lambda \tilde{h}^{(1,1)}g_{AB}\bigg]={\cal O}(1)
\end{split}
\end{equation}
Taking trace,
\begin{equation}
\begin{split}
&\tilde{h}^{(2)}=\Pi^{AB}\frac{1}{4NK}\bigg[2~\nabla_A\nabla_E\big[\tilde{h}^{(1,1)}\big]^E_B  -\nabla^2\tilde{h}^{(1,1)}_{AB}-\nabla_B\nabla_A \tilde{h}^{(1,1)}-2\lambda \tilde{h}^{(1,1)}g_{AB}\bigg]+{\cal O}(1)\\
\end{split}
\end{equation}
From \eqref{eq:htildesub}
\begin{equation}\label{eq:ht2_expr}
\begin{split}
\tilde{h}^{(2)}_{CC'}=\Pi^{A}_{C}\Pi^{B}_{C'}\frac{1}{2NK}&\bigg[\nabla_A\nabla_E\big[\tilde{h}^{(1,1)}\big]^E_B  +\nabla_{B}\nabla_{E}\big[\tilde{h}^{(1,1)}\big]^E_A-\nabla^{2}\tilde{h}^{(1,1)}_{AB}-2NK_{AB}\tilde{h}^{(2)}-\nabla_B\nabla_A \tilde{h}^{(1,1)}\\
&-2\lambda \tilde{h}^{(1,1)}g_{AB}\bigg]+{\cal O}\left(\frac{1}{D}\right)
\end{split}
\end{equation}

\subsubsection*{Calculation of $\tilde{h}^{(1,1)}_{CC'}$}
From, \eqref{eq:ht11_expr}
\begin{equation}
\begin{split}
&\tilde{h}^{(1,1)}_{CC'}=\underbrace{\Pi^{A}_{C}\Pi^{B}_{C'}\frac{1}{NK}\bigg[\nabla_A\nabla_E\big[\tilde{h}^{(0)}\big]^E_B  +\nabla_{B}\nabla_{E}\big[\tilde{h}^{(0)}\big]^E_A\bigg]}_{\tilde{h}^{(1,1)}_{CC'}{\large{|}}_{\text{part-1}}}\underbrace{-\Pi^{A}_{C}\Pi^{B}_{C'}\frac{1}{NK}\nabla^{2}\tilde{h}^{(0)}_{AB}}_{\tilde{h}^{(1,1)}_{CC'}{\large{|}}_{\text{part-2}}}\underbrace{-\frac{1}{K}K_{CC'}\tilde{h}^{(1,1)}}_{\tilde{h}^{(1,1)}_{CC'}{\large{|}}_{\text{part-3}}}\\
\end{split}
\end{equation}
\vspace{-0.2in}
\begin{equation}
\tilde{h}^{(1,1)}_{CC'}{\large{|}}_{\text{part-1}}=\Pi^{A}_{C}\Pi^{B}_{C'}\frac{1}{NK}\bigg[\nabla_A\nabla_E\big[h^{(0)}\big]^E_B  +\nabla_{B}\nabla_{E}\big[h^{(0)}\big]^E_A\bigg]+{\cal O}\left(\frac{1}{D}\right)^2
\end{equation}
We want to calculate the above expression on $\psi=1$. But to calculate $\tilde{h}^{(1,1)}_{CC'}{\large{|}}_{\text{part-1}}$ on 
$\psi=1$ we need the $(\psi-1)$ dependent terms of $h^{(0)}_{AB}$. From \eqref{eq:h0AB_def}
\begin{equation}
\begin{split}
\big[h^{(0)}\big]^E_B&=\big[M'^{(0)}\big]^E_B-(\psi-1)\big[C^{(0)}\big]^E_B+{\cal O}(\psi-1)^2\\
\Rightarrow~\nabla_E\big[h^{(0)}\big]^E_B&=\nabla_E\big[M'^{(0)}\big]^E_B+{\cal O}(\psi-1)
\end{split}
\end{equation}
Now,
\begin{equation}
\big[M'^{(0)}\big]^E_B=u^E u_B+\frac{1}{K}\bigg[u^E\Pi^C_B\bigg(\frac{\nabla_C K}{K}\bigg)+u_B\Pi^{CE}\bigg(\frac{\nabla_C K}{K}\bigg)+K^E_B-\Pi^{CE}\Pi^{C'}_B(\nabla_C u_{C'}+\nabla_{C'}u_C)\bigg]+{\cal O}\left(\frac{1}{D}\right)^2
\end{equation}
After a bit of simplification divergence of the above equation becomes
\begin{equation}
\begin{split}
\nabla_E\big[M'^{(0)}\big]^E_B&=u_B~(\nabla\cdot u)+(u\cdot\nabla)u_B\\
&+\frac{1}{K}\bigg[u_B\frac{\hat{\nabla}^2 K}{K}+\hat{\nabla}_B K-n_B~K^{AC}K_{AC}-\hat{\nabla}^2 u_B-Ku^C K_{CB}-\lambda~ D u_B\bigg]+{\cal O}\left(\frac{1}{D}\right)
\end{split}
\end{equation}
In the derivation of the above equation we have used the following identities
\begin{equation}
\begin{split}
\nabla^2  K&=\hat{\nabla}^2  K +  K(n\cdot \nabla) K+{\cal O}(D)\\
\hat{\nabla}^2 u_A&=\Pi^D_A\left[\nabla^2u_D-K(n\cdot \nabla)u_D\right]+{\cal O}(1)\\
\nabla^A K_{AB}&=\hat{\nabla}_B K-n_B K^{AC} K_{AC}+{\cal O}(1)
\end{split}
\end{equation}
Now,
\begin{equation}\label{eq:B.72}
\begin{split}
\nabla_E\big[h^{(0)}\big]^E_B&=u_B~(\nabla\cdot u)+u_B\frac{\hat{\nabla}^2 K}{K^2}-n_B~\frac{K}{D}-\lambda~ \frac{D}{K} u_B-n_B(u\cdot K\cdot u)\\
&+\bigg[-\frac{\hat{\nabla}^2 u_B}{K}+\frac{\hat{\nabla}_B K}{K}-u^E K_{EB}+(u\cdot\hat{\nabla})u_B\bigg]+{\cal O}\left(\frac{1}{D}\right)\\
&=-2~u_B\frac{u\cdot\nabla K}{K}+u_B\frac{\hat{\nabla}^2 K}{K^2}-n_B~\frac{K}{D}-\lambda~ \frac{D}{K} u_B-n_B(u\cdot K\cdot u)+u_B(u\cdot K\cdot u)\\
&=-n_B~\frac{K}{D}-n_B(u\cdot K\cdot u)
\end{split}
\end{equation}
In the last line we have used the divergence of leading order membrane equation
\begin{equation}
\frac{\hat{\nabla}^2 K}{K^2}=2~\frac{u\cdot\nabla K}{K}-u\cdot K\cdot u+\lambda~ \frac{D}{K}+{\cal O}\left(\frac{1}{D}\right)
\end{equation}\vspace{-0.2in}
\\From \eqref{eq:B.72}
\vspace{-0.2in}
\begin{equation}
\Pi^{A}_{C}\Pi^{B}_{C'}\frac{1}{NK}\nabla_A\nabla_E\big[h^{(0)}\big]^E_B=-\frac{1}{NK}\left[\frac{K}{D}+u\cdot K\cdot u\right]K_{CC'}
\end{equation}
So, finally we get
\vspace{-0.2in}
\begin{equation}\label{eq:h11part-1}
\tilde{h}^{(1,1)}_{CC'}{\large{|}}_{\text{part-1}}=-\frac{2}{NK}\left[\frac{K}{D}+u\cdot K\cdot u\right]K_{CC'}+{\cal O}\left(\frac{1}{D}\right)^2
\end{equation}
Now, we will calculate 
\vspace{-0.2in}
\begin{equation}
\tilde{h}^{(1,1)}_{CC'}{\large{|}}_{\text{part-2}}=-\Pi^{A}_{C}\Pi^{B}_{C'}\frac{1}{NK}\nabla^{2}h^{(0)}_{AB}+{\cal O}\left(\frac{1}{D}\right)^2
\end{equation}
We want to calculate the above expression on $\psi=1$. But to calculate $\tilde{h}^{(1,1)}_{CC'}{\large{|}}_{\text{part-2}}$ on 
$\psi=1$ we need the $(\psi-1)$ and $(\psi-1)^2$ dependent terms of $h^{(0)}_{AB}$.
\begin{equation}
\tilde{h}^{(1,1)}_{CC'}{\large{|}}_{\text{part-2}}=\underbrace{-\Pi^{A}_{C}\Pi^{B}_{C'}\frac{1}{NK}\nabla^{2}M'^{(0)}_{AB}}_{\text{Term-1}}+\underbrace{\Pi^{A}_{C}\Pi^{B}_{C'}\frac{1}{NK}\nabla^{2}\left[(\psi-1) C^{(0)}_{AB}\right]}_{\text{Term-2}}+\underbrace{\Pi^{A}_{C}\Pi^{B}_{C'}\frac{1}{NK}\nabla^{2}\left[(\psi-1)^2 E^{(0)}_{AB}\right]}_{\text{Term-3}}
\end{equation}
\begin{equation}
\begin{split}
\text{Term-3}&=\Pi^{A}_{C}\Pi^{B}_{C'}\frac{1}{NK}\nabla^{2}\left[(\psi-1)^2 E^{(0)}_{AB}\right]\\
&=\frac{2N}{K}E^{(0)}_{CC'}
\end{split}
\end{equation}
From \eqref{eq:E0AB_def}
\begin{equation}\label{eq:B.79}
\begin{split}
E^{(0)}_{CC'}&=-\frac{1}{2N}\Pi^{A}_{C}\Pi^{B}_{C'}(n.\nabla)C^{(0)}_{AB}\\
&=\frac{1}{2N}\frac{(n\cdot\nabla) N}{N^2}\bigg[u_C~\frac{\hat{\nabla}^2 u_{C'}}{K}+u_{C'}~\frac{\hat{\nabla}^2 u_C}{K}\bigg]-\frac{1}{2N^2}\bigg[2~\frac{\hat{\nabla}^2 u_C}{K}\frac{\hat{\nabla}^2 u_{C'}}{K}+u_C\bigg(\frac{\hat{\nabla} _{C'} K}{K}\bigg)\frac{(u\cdot\nabla)K}{K}\\
&+u_{C'}\bigg(\frac{\hat{\nabla} _C K}{K}\bigg)\frac{(u\cdot\nabla)K}{K}+u_C\Pi^B_{C'}(n\cdot\nabla)(n\cdot\nabla)u_B+u_{C'}\Pi^B_C(n\cdot\nabla)(n\cdot\nabla)u_B\bigg]+{\cal O}\left(\frac{1}{D}\right)
\end{split}
\end{equation}
Using \eqref{eq:B.79} we can write Term-3 as
\begin{equation}\label{eq:Term-3}
\text{Term-3}={\cal A}^{(3)}_{CC'}+u_C{\cal B}^{(3)}_{C'}+u_{C'}{\cal B}^{(3)}_C
\end{equation}
Where,
\begin{equation}
\begin{split}
{\cal A}_{CC'}^{(3)}&=-2~\frac{D}{K^2}\bigg(\frac{\hat{\nabla}^2 u_C}{K}\bigg)\bigg(\frac{\hat{\nabla}^2 u_{C'}}{K}\bigg)\\
{\cal B}^{(3)}_C&=\frac{D}{K^2}\bigg[\frac{\hat{\nabla}^2 K}{K^2}-\lambda\frac{D}{K}\bigg]\frac{\hat{\nabla}^2 u_C}{K}-\frac{D}{K^2}\bigg[\frac{\hat{\nabla}_C K}{K}\frac{(u\cdot\nabla)K}{K}+\Pi^B_C(n\cdot\nabla)(n\cdot\nabla)u_B\bigg]
\end{split}
\end{equation}
Now,
\begin{equation}
\begin{split}
\text{Term-2}&=\Pi^{A}_{C}\Pi^{B}_{C'}\frac{1}{NK}\nabla^{2}\left[(\psi-1) C^{(0)}_{AB}\right]+{\cal O}\left(\frac{1}{D}\right)^2\\
&=C^{(0)}_{CC'}+\frac{1}{K}\frac{(n\cdot\nabla)N}{N}C^{(0)}_{CC'}+\frac{2}{K}\Pi^A_C\Pi^B_{C'}(n\cdot\nabla)C^{(0)}_{AB}\\
\end{split}
\end{equation}
Using \eqref{eq:C0AB} and \eqref{eq:B.79} we can write Term-2 as
\begin{equation}\label{eq:Term-2}
\text{Term-2}={\cal A}^{(2)}_{CC'}+u_C{\cal B}^{(2)}_{C'}+u_{C'}{\cal B}^{(2)}_C
\end{equation}
Where,
\begin{equation}
\begin{split}
{\cal B}^{(2)}_C&=\frac{1}{N}\Pi^E_C(n\cdot\nabla)u_E+\frac{1}{NK}\bigg[-\frac{\hat{\nabla}_C K}{K}\bigg(2~\frac{(n\cdot\nabla)K}{K}+2~\frac{\hat{\nabla}^2 K}{K^2}-\lambda~\frac{D}{K}\bigg)+\frac{1}{K^2}\hat{\nabla}_C\left(\hat{\nabla}^2 K\right)\\
&-2~\frac{K}{D}\bigg(\frac{\hat{\nabla}_C K}{K}\bigg)-K^D_C\bigg(\frac{\nabla_D K}{K}\bigg)\bigg]+\frac{1}{NK}\frac{(n\cdot\nabla)N}{N}\left[\Pi^E_C(n\cdot\nabla)u_E\right]\\
&+\frac{2}{K}\bigg[-\frac{1}{N}\left(\frac{n\cdot\nabla N}{N}\right)\frac{\hat{\nabla}^2 u_C}{K}+\frac{1}{N}\bigg(\frac{\hat{\nabla}_C K}{K}\bigg)\frac{u\cdot\nabla K}{K}+\frac{1}{N}\Pi^B_C(n\cdot\nabla)(n\cdot\nabla)u_B\bigg]\\
{\cal A}^{(2)}_{CC'}&=-\frac{1}{NK}\frac{(n\cdot\nabla)N}{N}\left[K_{CC'}-\hat{\nabla}_C u_{C'}-\hat{\nabla}_{C'}u_C\right]+\frac{1}{NK}\bigg[2~\frac{\hat{\nabla}^2 u_C}{K}\frac{\hat{\nabla}_{C'}K}{K}+2~\frac{\hat{\nabla}^2 u_{C'}}{K}\frac{\hat{\nabla}_C K}{K}\\
&-\frac{\hat{\nabla}_C K}{K}\frac{\hat{\nabla}_{C'} K}{K}-\lambda~\Pi_{CC'}+\hat{\nabla}_C\bigg(\frac{\hat{\nabla}_{C'}K}{K}\bigg)-K^E_C K_{EC'}-\frac{\hat{\nabla}_{C'}K}{K} u^B K_{BC}-\frac{\hat{\nabla}_CK}{K} u^B K_{BC'}\\
&+2~K_{CC'}\frac{(u\cdot\nabla)K}{K}+K^D_C\left(\hat{\nabla}_D u_{C'}\right)+K^D_{C'}\left(\hat{\nabla}_D u_C\right)-\hat{\nabla}_C\bigg(\frac{\hat{\nabla}^2 u_{C'}}{K}\bigg)-\hat{\nabla}_{C'}\bigg(\frac{\hat{\nabla}^2 u_C}{K}\bigg)\bigg]\\
&+\frac{2}{NK}\bigg[2~\frac{\hat{\nabla}^2 u_C}{K}\frac{\hat{\nabla}^2 u_{C'}}{K}\bigg]
\end{split}
\end{equation}
\begin{equation}
\text{Term-1}=-\Pi^{A}_{C}\Pi^{B}_{C'}\frac{1}{NK}\nabla^{2}M'^{(0)}_{AB}+{\cal O}\left(\frac{1}{D}\right)^2
\end{equation}
Here,
\vspace{-0.1in}
\begin{equation}
M'^{(0)}_{AB}=u_A u_B+\frac{1}{ND}\bigg[u_A\Pi^E_B\bigg(\frac{\nabla_E K}{K}\bigg)+u_B\Pi_A^E\bigg(\frac{\nabla_E K}{K}\bigg)+K_{AB}-\Pi^E_A\Pi^F_B(\nabla_E u_F+\nabla_Fu_E)\bigg]+{\cal O}\left(\frac{1}{D}\right)^2
\end{equation}
\begin{equation}\label{eq:B.87}
\begin{split}
&~~~~\Pi^{A}_{C}\Pi^{B}_{C'}\nabla^{2}M'^{(0)}_{AB}\\
&=u_C\Pi^B_{C'}\nabla^2 u_B+u_{C'}\Pi^B_C\nabla^2 u_B+2~\Pi^A_C\Pi^B_{C'}(\nabla^D u_A)(\nabla_D u_B)\\
&-\frac{1}{ND}\bigg(\frac{\nabla^2 N}{N}\bigg)\bigg[u_C~\Pi^E_{C'}\frac{\nabla_E K}{K}+u_{C'}~\Pi_C^E\frac{\nabla_E K}{K}+K_{CC'}-\Pi^E_C\Pi^F_{C'}(\nabla_E u_F+\nabla_Fu_E)\bigg]\\
&+\frac{1}{ND}\bigg[\Pi^A_C(\nabla^2 u_A)\Pi^E_{C'}\frac{\nabla_E K}{K}+u_C\Pi^B_{C'}\nabla^2\left(\Pi^E_B\frac{\nabla_E K}{K}\right)+\Pi^A_{C'}(\nabla^2 u_A)\Pi^E_C\frac{\nabla_E K}{K}\\
&+u_{C'}\Pi^B_C\nabla^2\bigg(\Pi^E_B\frac{\nabla_E K}{K}\bigg)+\Pi^A_C\Pi^B_{C'}\nabla^2 K_{AB}-\Pi^A_C\Pi^B_{C'}\nabla^2\Big\{\Pi^E_A\Pi^F_B(\nabla_E u_F+\nabla_F u_E)\Big\}\bigg]+{\cal O}\left(\frac{1}{D}\right)
\end{split}
\end{equation}
We will use the following identities to simplify \eqref{eq:B.87}. we are just stating the identities without proof, proofs are quite straightforward.
\begin{equation}\label{eq:ident4}
\begin{split}
\frac{1}{N}&=\frac{D}{K}\bigg[1-\frac{1}{K}\bigg(\frac{\hat{\nabla}^2 K}{K^2}-\lambda~\frac{D}{K}-\frac{K}{D}\bigg)\bigg]+{\cal O}\left(\frac{1}{D}\right)^2\\
\Pi^D_B\nabla^2 u_D&=\hat{\nabla}^2 u_B+K\Pi^D_B(n\cdot\nabla)u_D-K^F_B u^D K_{FD}+\Pi^D_B(n\cdot\nabla)(n\cdot\nabla)u_D\\
&-\frac{\hat{\nabla}^F K}{K}\big(\hat{\nabla}_F u_B\big)+{\cal O}\left(\frac{1}{D}\right)\\
\Pi^A_C\Pi^B_{C'}(\nabla^2 K_{AB})&=-2~\big(\hat{\nabla}_C K\big)\bigg(\frac{\hat{\nabla}_{C'} K}{K}\bigg)-2~\lambda K \Pi_{CC'}+\lambda(D-1)K_{CC'}\\
&+2\hat{\nabla}_C\big(\hat{\nabla}_{C'} K\big)-K_{CC'}\frac{K^2}{D}\\
\Pi^B_{C'}\nabla^2\bigg(\Pi^E_B\frac{\nabla_E K}{K}\bigg)&=-\big(\hat{\nabla}_{C'}K\big)\bigg[4~\frac{\hat{\nabla}^2 K}{K^2}-3~\lambda\frac{D}{K}\bigg]+\frac{2}{K}\hat{\nabla}_{C'}\big(\hat{\nabla}^2 K\big)\\
\Pi^A_C\Pi^B_{C'}\nabla^2\big[\Pi^E_A\Pi^F_B\nabla_E u_F\big]&=-2~\big(\hat{\nabla}_C K\big)\bigg(\frac{\hat{\nabla}^2 u_{C'}}{K}\bigg)+2~\big(\hat{\nabla}_{C'} K\big)u^FK_{FC}+\lambda(D-1)\big(\hat{\nabla}_C u_{C'}\big)\\
&+2~\hat{\nabla}_C\big(\hat{\nabla}^2 u_{C'}\big)-2~K_{CC'}(u\cdot\nabla)K\\
\Pi^A_C\Pi^B_{C'}\big(\nabla^D u_A\big)\big(\nabla_D u_A\big)&=\big(\hat{\nabla}^D u_C\big)\big(\hat{\nabla}_D u_{C'}\big)+\bigg(\frac{\hat{\nabla}^2 u_C}{K}\bigg)\bigg(\frac{\hat{\nabla}^2 u_{C'}}{K}\bigg)\\
\end{split}
\end{equation}\vspace{-0.2in}
\\Using \eqref{eq:ident4}, we can write Term-1 as
\begin{equation}\label{eq:Term-1}
\text{Term-1}={\cal A}^{(1)}_{CC'}+u_C{\cal B}^{(1)}_{C'}+u_{C'}{\cal B}^{(1)}_C
\end{equation}
Where,
\begin{equation}
\begin{split}
{\cal B}^{(1)}_C&=-\frac{D}{K^2}\bigg[1-\frac{1}{K}\bigg(\frac{\hat{\nabla}^2 K}{K^2}-\lambda~\frac{D}{K}-\frac{K}{D}\bigg)\bigg]\left[\hat{\nabla}^2 u_C+K\Pi^D_C(n\cdot\nabla)u_D\right]\\
&-\frac{D}{K^2}\bigg[-K^F_C K_{FD}u^D+\Pi^D_C(n\cdot\nabla)(n\cdot\nabla)u_D-\frac{\hat{\nabla}^F K}{K}\big(\hat{\nabla}_F u_C\big)\bigg]-2~\frac{D}{K^4}\hat{\nabla}_C\big(\hat{\nabla}^2 K\big)\\
&+\frac{1}{K}\bigg[1+\frac{D}{K}\bigg(2~\frac{\hat{\nabla}^2 K}{K^2}-\lambda~\frac{D}{K}-\frac{K}{D}\bigg)\bigg]\frac{\hat{\nabla}_C K}{K}+\frac{D}{K^2}\bigg[4~\frac{\hat{\nabla}^2 K}{K^2}-3~\lambda\frac{D}{K}\bigg]\frac{\hat{\nabla}_C K}{K}
\end{split}
\end{equation}
\begin{equation}
\begin{split}
{\cal A}^{(1)}_{CC'}&=-2~\frac{D}{K^2}\big(\hat{\nabla}^D u_C\big)\big(\hat{\nabla}_D u_{C'}\big)-2~\frac{D}{K^2}\frac{\hat{\nabla}^2 u_C}{K}\frac{\hat{\nabla}^2 u_{C'}}{K}\\
&+\frac{1}{K}\bigg[1+\frac{D}{K}\bigg(2~\frac{\hat{\nabla}^2 K}{K^2}-\lambda\frac{D}{K}-\frac{K}{D}\bigg)\bigg]\left[K_{CC'}-\hat{\nabla}_C u_{C'}-\hat{\nabla}_{C'}u_C\right]\\
&-\frac{D}{K^2}\bigg[4~\frac{\hat{\nabla}^2 u_C}{K}\frac{\hat{\nabla}_{C'} K}{K}+4~\frac{\hat{\nabla}^2 u_{C'}}{K}\frac{\hat{\nabla}_C K}{K}-2~\frac{\hat{\nabla}_C K}{K}\frac{\hat{\nabla}_{C'} K}{K}-2~\lambda\Pi_{CC'}+\lambda~\frac{D}{K}K_{CC'}\nonumber\\
\end{split}
\end{equation}
\begin{equation}
\begin{split}
&+\frac{2}{K}\hat{\nabla}_C\big(\hat{\nabla}_{C'}K\big)-\frac{K}{D}K_{CC'}-2~\frac{\hat{\nabla}_{C'} K}{K} u^F K_{FC}-2~\frac{\hat{\nabla}_C K}{K} u^F K_{FC'}-\lambda~\frac{D}{K}\left(\hat{\nabla}_C u_{C'}+\hat{\nabla}_{C'} u_C\right)\\
&-\frac{2}{K}\hat{\nabla}_C\big(\hat{\nabla}^2 u_{C'}\big)-\frac{2}{K}\hat{\nabla}_{C'}\big(\hat{\nabla}^2 u_C\big)+4~K_{CC'}\frac{u\cdot\nabla K}{K}\bigg]
\end{split}
\end{equation}
Adding, \eqref{eq:Term-1}, \eqref{eq:Term-2} and \eqref{eq:Term-3} we get final expression of $\tilde{h}^{(1,1)}_{CC'}{\large{|}}_{\text{part-2}}$
\begin{equation}\label{eq:h11part-2}
\tilde{h}^{(1,1)}_{CC'}{\large{|}}_{\text{part-2}}=\left({\cal A}^{(1)}_{CC'}+{\cal A}^{(2)}_{CC'}+{\cal A}^{(3)}_{CC'}\right)+u_C\left({\cal B}^{(1)}_{C'}+{\cal B}^{(2)}_{C'}+{\cal B}^{(3)}_{C'}\right)+u_{C'}\left({\cal B}^{(1)}_C+{\cal B}^{(2)}_C+{\cal B}^{(3)}_C\right)
\end{equation}
\begin{equation}\label{eq:h11part-3}
\begin{split}
\tilde{h}^{(1,1)}_{CC'}{\large{|}}_{\text{part-3}}&=-\frac{1}{K}K_{CC'}\tilde{h}^{(1,1)}+{\cal O}\left(\frac{1}{D}\right)^2\\
&=-\frac{1}{K}K_{CC'}\frac{1}{2}\Pi^{AB}\left[\tilde{h}^{(1,1)}_{AB}{\large{|}}_{\text{part-1}}+\tilde{h}^{(1,1)}_{AB}{\large{|}}_{\text{part-2}}\right]+{\cal O}\left(\frac{1}{D}\right)^2\\
&=-\frac{1}{2K}K_{CC'}\bigg[-\frac{2}{N}\left(\frac{K}{D}+u\cdot K\cdot u\right)\bigg]-\frac{1}{2K}K_{CC'}\left[\Pi^{AB}\tilde{h}^{(1,1)}_{AB}{\large{|}}_{\text{part-2}}\right]\\
&=\frac{D}{K^2}K_{CC'}\left(\frac{K}{D}+u\cdot K\cdot u\right)-\frac{1}{2K}K_{CC'}\left[\Pi^{AB}\left({\cal A}^{(1)}_{AB}+{\cal A}^{(2)}_{AB}\right)\right]\\
&=\frac{D}{K^2}K_{CC'}\left(\frac{K}{D}+u\cdot K\cdot u\right)+{\cal O}\left(\frac{1}{D}\right)^2\\
\end{split}
\end{equation}
In the derivation of \eqref{eq:h11part-3} we have used the following identity
\begin{equation}
\frac{1}{K}\Pi^{AB}\hat{\nabla}_A\bigg(\frac{\hat{\nabla}^2 u_B}{K}\bigg)=\frac{u\cdot\nabla K}{K}+{\cal O}\left(\frac{1}{D}\right)
\end{equation}
Adding \eqref{eq:h11part-1},\eqref{eq:h11part-2} and \eqref{eq:h11part-3} we get the final expression of $\tilde{h}^{(1,1)}_{CC'}$ as given in \eqref{eq:htilde11}.

\subsection*{Calculation of $\tilde{h}^{(2)}_{CC'}$}
From \eqref{eq:ht2_expr}, the non-vanishing terms of $\tilde{h}^{(2)}_{CC'}$ are the following
\begin{equation}\label{eq:ht2}
\begin{split}
\tilde{h}^{(2)}_{CC'}=\underbrace{\Pi^{A}_{C}\Pi^{B}_{C'}\frac{1}{2NK}\bigg[\nabla_A\nabla_E\big[\tilde{h}^{(1,1)}\big]^E_B  +\nabla_{B}\nabla_{E}\big[\tilde{h}^{(1,1)}\big]^E_A\bigg]}_{\tilde{h}^{(2)}_{CC'}\big{|}_{\text{Part-1}}}\underbrace{-\Pi^{A}_{C}\Pi^{B}_{C'}\frac{1}{2NK}\left[\nabla^{2}\tilde{h}^{(1,1)}_{AB}\right]}_{\tilde{h}^{(2)}_{CC'}\big{|}_{\text{Part-2}}}+{\cal O}\left(\frac{1}{D}\right)
\end{split}
\end{equation}
\vspace{-0.2in}
\\For the calculation of $\tilde{h}^{(2)}_{CC'}$ we need $(\psi-1)$ dependent terms of $\tilde{h}^{(1,1)}_{CC'}$. The expression of 
$\tilde{h}^{(1,1)}_{CC'}$ upto the relevant order is given by
\begin{equation}
\begin{split}
\tilde{h}^{(1,1)}_{CC'}&=-\frac{1}{NK}\left[u_C\Pi^B_{C'}\left(\nabla^2 u_B\right)+u_{C'}\Pi^B_C\left(\nabla^2 u_B\right)\right]+\frac{1}{N}\left[u_C\Pi^B_{C'}(n\cdot\nabla) u_B+u_{C'}\Pi^B_C(n\cdot\nabla) u_B\right]\\
&-(\psi-1)\frac{1}{N}\left[1+\frac{D}{K}\left(\frac{\nabla^2 K}{K^2}\right)-\frac{(n\cdot\nabla)N}{N^2}\right]\bigg[u_C\bigg(\frac{\hat{\nabla}^2 u_{C'}}{K}\bigg)+u_{C'}\bigg(\frac{\hat{\nabla}^2 u_C}{K}\bigg)\bigg]\\
&+(\psi-1)\frac{1}{N^2}\bigg[2\bigg(\frac{\hat{\nabla}^2 u_C}{K}\bigg)\bigg(\frac{\hat{\nabla}^2 u_{C'}}{K}\bigg)+u_C\bigg(\frac{\hat{\nabla}^2\hat{\nabla}^2 u_{C'}}{K^2}\bigg)+u_{C'}\bigg(\frac{\hat{\nabla}^2\hat{\nabla}^2 u_C}{K^2}\bigg)\\
&-u_C\bigg(\frac{\hat{\nabla}^2 K}{K^2}\bigg)\bigg(\frac{\hat{\nabla}^2 u_{C'}}{K}\bigg)-u_{C'}\bigg(\frac{\hat{\nabla}^2 K}{K^2}\bigg)\bigg(\frac{\hat{\nabla}^2 u_C}{K}\bigg)\bigg]\\
&-(\psi-1)\frac{D}{NK}\bigg[\left(\hat{\nabla}_C u_{C'}+\hat{\nabla}_{C'} u_C\right)\left(\frac{u\cdot\nabla K}{K}\right)+u_C\hat{\nabla}_{C'}\left(\frac{u\cdot\nabla K}{K}\right)+u_{C'}\hat{\nabla}_C\bigg(\frac{u\cdot\nabla K}{K}\bigg)\bigg]
\end{split}
\end{equation}
\vspace{-0.1in}
\begin{equation}
\begin{split}
\nabla_E\big[\tilde{h}^{(1,1)}\big]^E_B&=-\frac{1}{N}\left[-u_B~Kn^C\bigg(\frac{\nabla^2 u_C}{K}\bigg)\right]+\frac{1}{N}\left[-Ku_B~ n^C(n\cdot\nabla)u_C\right]\\
&=\frac{K}{N}u_B~\frac{n^C\nabla^2 u_C}{K}-\frac{K}{N}u_B\left[n^C(n\cdot\nabla)u_C\right]\\
&=-\frac{K}{N}u_B\left(\frac{u\cdot\nabla K}{K}\right)
\end{split}
\end{equation}
From, \eqref{eq:ht2}
\begin{equation}\label{eq:ht2-part1}
\tilde{h}^{(2)}_{CC'}\big{|}_{\text{Part-1}}=-\frac{D}{2NK}\bigg[\left(\hat{\nabla}_C u_{C'}+\hat{\nabla}_{C'} u_C\right)\frac{u\cdot\nabla K}{K}+u_C\hat{\nabla}_{C'}\left(\frac{u\cdot\nabla K}{K}\right)+u_{C'}\hat{\nabla}_C\left(\frac{u\cdot\nabla K}{K}\right)\bigg]
\end{equation}
From, \eqref{eq:ht2}
\begin{equation}\label{eq:ht2-part2}
\begin{split}
\tilde{h}^{(2)}_{CC'}\big{|}_{\text{Part-2}}&=-\frac{1}{2NK}\Pi^{A}_{C}\Pi^{B}_{C'}\left[\nabla^{2}\tilde{h}^{(1,1)}_{AB}\right]\\
&=\underbrace{-\frac{1}{2NK}\Pi^{A}_{C}\Pi^{B}_{C'}\nabla^2\bigg[-\frac{1}{NK}\left(u_A\Pi^E_B\nabla^2 u_E+u_B\Pi^E_A\nabla^2 u_E\right)\bigg]}_{\text{Term-1}}\\
&\underbrace{-\frac{1}{2NK}\Pi^{A}_{C}\Pi^{B}_{C'}\nabla^2\bigg[\frac{1}{N}\left\{u_A\Pi^E_B(n\cdot\nabla)u_E+u_B\Pi^E_A(n\cdot\nabla)u_E\right\}\bigg]}_{\text{Term-2}}\\
&+\frac{1}{2N}\bigg[1+\frac{D}{K}~\frac{\nabla^2 K}{K^2}-\frac{(n\cdot\nabla)N}{N^2}\bigg]\bigg[u_C\frac{\hat{\nabla}^2 u_{C'}}{K}+u_{C'}\frac{\hat{\nabla}^2 u_C}{K}\bigg]\\
&-\frac{1}{2N^2}\bigg[2\frac{\hat{\nabla}^2 u_C}{K}\frac{\hat{\nabla}^2 u_{C'}}{K}+u_C\frac{\hat{\nabla}^2\hat{\nabla}^2 u_{C'}}{K^2}+u_{C'}\frac{\hat{\nabla}^2\hat{\nabla}^2 u_C}{K^2}-u_C\frac{\hat{\nabla}^2 K}{K^2}\frac{\hat{\nabla}^2 u_{C'}}{K}-u_{C'}\frac{\hat{\nabla}^2 K}{K^2}\frac{\hat{\nabla}^2 u_C}{K}\bigg]\\
&+\frac{D}{2NK}\bigg[\left(\hat{\nabla}_C u_{C'}+\hat{\nabla}_{C'} u_C\right)\left(\frac{u\cdot\nabla K}{K}\right)+u_C\hat{\nabla}_{C'}\left(\frac{u\cdot\nabla K}{K}\right)+u_{C'}\hat{\nabla}_C\bigg(\frac{u\cdot\nabla K}{K}\bigg)\bigg]
\end{split}
\end{equation}
\vspace{-0.1in}
\begin{equation}
\begin{split}
\text{Term-1}&=-\frac{1}{2NK}\Pi^{A}_{C}\Pi^{B}_{C'}\nabla^2\bigg[-\frac{1}{NK}\left(u_A\Pi^E_B\nabla^2 u_E+u_B\Pi^E_A\nabla^2 u_E\right)\bigg]\\
&=-\frac{1}{2NK}\bigg[1+\frac{2D}{K}~\frac{\nabla^2 K}{K^2}\bigg]\left[u_C\Pi^E_{C'}\nabla^2 u_E+u_{C'}\Pi^E_C\nabla^2 u_E\right]\\
&+\frac{1}{2NK}\frac{D}{K^2}\bigg[2~\Pi^E_C\big(\nabla^2 u_E\big)\Pi^F_{C'}\big(\nabla^2 u_F\big)+u_C\Pi^E_{C'}\nabla^2\big(\Pi^F_E\nabla^2 u_F\big)+u_{C'}\Pi^E_C\nabla^2\big(\Pi^F_E\nabla^2 u_F\big)\bigg]
\end{split}
\end{equation}
Using the identity,
\begin{equation}
\begin{split}
\Pi^B_F\nabla^2\big(\Pi^C_B\nabla^2 u_C\big)&=2~\hat{\nabla}^2\left(\hat{\nabla}^2 u_F\right)+K^2\bigg(\frac{\hat{\nabla}_F K}{K}\bigg)\bigg(\frac{u\cdot\nabla K}{K}\bigg)-\lambda(D-1)K\bigg(\frac{\hat{\nabla}^2 u_F}{K}\bigg)\\
&+K\hat{\nabla}^2\bigg(\frac{\hat{\nabla}^2 u_F}{K}\bigg)+K^2\Pi^E_F(n\cdot\nabla)(n\cdot\nabla)u_E-2K\big(\hat{\nabla}^E K\big)\big(\hat{\nabla}_E u_F\big)-3\frac{K^3}{D}\bigg(\frac{\hat{\nabla}^2 u_F}{K}\bigg)
\end{split}
\end{equation}
\vspace{-0.2in}
\\we get
\begin{equation}\label{eq:term1}
\begin{split}
\text{Term-1}&=\frac{4D}{NK}\frac{\hat{\nabla}^2 u_C}{K}\frac{\hat{\nabla}^2 u_{C'}}{K}-\frac{1}{2N}\bigg[1+\frac{2D}{K}\bigg(\frac{\hat{\nabla}^2 K}{K^2}+\frac{n\cdot\nabla K}{K}\bigg)\bigg]\bigg[2u_C\bigg(\frac{\hat{\nabla}^2 u_{C'}}{K}\bigg)+2u_{C'}\bigg(\frac{\hat{\nabla}^2 u_C}{K}\bigg)\bigg]\\
&+\frac{D}{2NK}u_C\bigg[-\lambda\frac{D}{K}\frac{\hat{\nabla}^2 u_{C'}}{K}+3~\frac{\hat{\nabla}^2\hat{\nabla}^2 u_{C'}}{K^2}-\frac{\hat{\nabla}^2 K}{K^2}\frac{\hat{\nabla}^2 u_{C'}}{K}+\frac{\hat{\nabla}_{C'} K}{K}\frac{u\cdot\nabla K}{K}\\
&+\Pi^E_{C'}(n\cdot\nabla)(n\cdot\nabla)u_E-2~\frac{\hat{\nabla}^E K}{K}\big(\hat{\nabla}_E u_{C'}\big)-3~\frac{K}{D}\frac{\hat{\nabla}^2 u_{C'}}{K}\bigg]\\
&+\frac{D}{2NK}u_{C'}\bigg[-\lambda\frac{D}{K}\frac{\hat{\nabla}^2 u_C}{K}+3~\frac{\hat{\nabla}^2\hat{\nabla}^2 u_C}{K^2}-\frac{\hat{\nabla}^2 K}{K^2}\frac{\hat{\nabla}^2 u_C}{K}+\frac{\hat{\nabla}_C K}{K}\frac{u\cdot\nabla K}{K}\\
&+\Pi^E_C(n\cdot\nabla)(n\cdot\nabla)u_E-2~\frac{\hat{\nabla}^E K}{K}\big(\hat{\nabla}_E u_C\big)-3~\frac{K}{D}\frac{\hat{\nabla}^2 u_C}{K}\bigg]
\end{split}
\end{equation}
\begin{equation}
\begin{split}
\text{Term-2}&=-\frac{1}{2NK}\Pi^{A}_{C}\Pi^{B}_{C'}\nabla^2\bigg[\frac{1}{N}\left\{u_A\Pi^E_B(n\cdot\nabla)u_E+u_B\Pi^E_A(n\cdot\nabla)u_E\right\}\bigg]\\
&=\frac{1}{2NK}\frac{\nabla^2 N}{N^2}\big[u_C\Pi^E_{C'}(n\cdot\nabla)u_E+u_{C'}\Pi^E_C(n\cdot\nabla)u_E\big]\\
&-\frac{1}{2N^2 K}\bigg[\Pi^A_C\big(\nabla^2 u_A\big)\Pi^E_{C'}(n\cdot\nabla)u_E+\Pi^A_{C'}\big(\nabla^2 u_A\big)\Pi^E_C(n\cdot\nabla)u_E\\
&+u_C\Pi^B_{C'}\nabla^2\{\Pi^E_B(n\cdot\nabla)u_E\}+u_{C'}\Pi^B_C\nabla^2\{\Pi^E_B(n\cdot\nabla)u_E\}\bigg]\\
&=\frac{1}{2N}\bigg[1+\frac{D}{K}\bigg(\frac{\nabla^2 K}{K^2}\bigg)\bigg]\bigg[u_C\bigg(\frac{\hat{\nabla}^2 u_{C'}}{K}\bigg)+u_{C'}\bigg(\frac{\hat{\nabla}^2 u_C}{K}\bigg)\bigg]-\frac{2}{N^2}\frac{\hat{\nabla}^2 u_C}{K}\frac{\hat{\nabla}^2 u_{C'}}{K}\\
&-\frac{1}{2N^2}u_C\bigg[\frac{\hat{\nabla}^2\hat{\nabla}^2 u_{C'}}{K^2}-\frac{\hat{\nabla}^2 K}{K^2}\frac{\hat{\nabla}^2 u_{C'}}{K}+\frac{\hat{\nabla}_{C'} K}{K}\frac{u\cdot\nabla K}{K}+\Pi^E_{C'}(n\cdot\nabla)(n\cdot\nabla)u_E\bigg]\nonumber\\
\end{split}
\end{equation}
\begin{equation}\label{eq:term2}
\begin{split}
&-\frac{1}{2N^2}u_{C'}\bigg[\frac{\hat{\nabla}^2\hat{\nabla}^2 u_C}{K^2}-\frac{\hat{\nabla}^2 K}{K^2}\frac{\hat{\nabla}^2 u_C}{K}+\frac{\hat{\nabla}_C K}{K}\frac{u\cdot\nabla K}{K}+\Pi^E_C(n\cdot\nabla)(n\cdot\nabla)u_E\bigg]
\end{split}
\end{equation}
Adding \eqref{eq:ht2-part1} and \eqref{eq:ht2-part2} we get the final expression of $\tilde{h}^{(2)}_{CC'}$ as given in \eqref{eq:htilde2} after using \eqref{eq:term1} and \eqref{eq:term2}


\subsection*{Calculation of $\tilde{h}^{(1,2)}_{CC'}$}
From \eqref{eq:ht12_expr}, the non-vanishing terms of $\tilde{h}^{(1,2)}_{CC'}$ are the followings
\begin{equation}
\begin{split}
\tilde{h}^{(1,2)}_{CC'}&=\underbrace{-2~\frac{D}{K}\Pi^{A}_{C}\Pi^{B}_{C'}(n\cdot\nabla)\tilde{h}^{(1,1)}_{AB}}_{\tilde{h}^{(1,2)}_{CC'}\big{|}_{\text{Part-1}}}\underbrace{-\frac{D}{NK}\Pi^{A}_{C}\Pi^{B}_{C'}\left[2N^{2}\tilde{h}^{(2)}_{AB}+\{(n\cdot\nabla)N\}\tilde{h}^{(1,1)}_{AB}-2\lambda \tilde{h}^{(0)}_{AB}\right]}_{\tilde{h}^{(1,2)}_{CC'}\big{|}_{\text{Part-2}}}+{\cal O}\left(\frac{1}{D}\right)
\end{split}
\end{equation}
\begin{equation}\label{eq:ht12-part1}
\begin{split}
\tilde{h}^{(1,2)}_{CC'}\big{|}_{\text{Part-1}}&=-2~\frac{D}{K}\Pi^{A}_{C}\Pi^{B}_{C'}(n\cdot\nabla)\tilde{h}^{(1,1)}_{AB}\\
&=\underbrace{2~\frac{D}{K}\Pi^{A}_{C}\Pi^{B}_{C'}(n\cdot\nabla)\left[\frac{1}{NK}\big\{u_A\Pi^E_B\left(\nabla^2 u_E\right)+u_B\Pi^E_A\left(\nabla^2 u_E\right)\big\}\right]}_{\text{term-1}}\\
&\underbrace{-2~\frac{D}{K}\Pi^{A}_{C}\Pi^{B}_{C'}(n\cdot\nabla)\left[\frac{1}{N}\big\{u_A\Pi^E_B(n\cdot\nabla) u_E+u_B\Pi^E_A(n\cdot\nabla) u_E\big\}\right]}_{\text{term-2}}\\
&+2~\frac{D}{K}\left[1+\frac{D}{K}\left(\frac{\nabla^2 K}{K^2}\right)-\frac{(n\cdot\nabla)N}{N^2}\right]\bigg[u_C\bigg(\frac{\hat{\nabla}^2 u_{C'}}{K}\bigg)+u_{C'}\bigg(\frac{\hat{\nabla}^2 u_C}{K}\bigg)\bigg]\\
&-2~\frac{D}{NK}\bigg[2\bigg(\frac{\hat{\nabla}^2 u_C}{K}\bigg)\bigg(\frac{\hat{\nabla}^2 u_{C'}}{K}\bigg)+u_C\bigg(\frac{\hat{\nabla}^2\hat{\nabla}^2 u_{C'}}{K^2}\bigg)+u_{C'}\bigg(\frac{\hat{\nabla}^2\hat{\nabla}^2 u_C}{K^2}\bigg)\\
&-u_C\bigg(\frac{\hat{\nabla}^2 K}{K^2}\bigg)\bigg(\frac{\hat{\nabla}^2 u_{C'}}{K}\bigg)-u_{C'}\bigg(\frac{\hat{\nabla}^2 K}{K^2}\bigg)\bigg(\frac{\hat{\nabla}^2 u_C}{K}\bigg)\bigg]\\
&+2~\frac{D^2}{K^2}\bigg[\left(\hat{\nabla}_C u_{C'}+\hat{\nabla}_{C'} u_C\right)\left(\frac{u\cdot\nabla K}{K}\right)+u_C\hat{\nabla}_{C'}\left(\frac{u\cdot\nabla K}{K}\right)+u_{C'}\hat{\nabla}_C\bigg(\frac{u\cdot\nabla K}{K}\bigg)\bigg]
\end{split}
\end{equation}
\begin{equation}
\begin{split}
\text{term-1}&=\frac{2D}{K}\Pi^{A}_{C}\Pi^{B}_{C'}(n\cdot\nabla)\left[\frac{1}{NK}\big\{u_A\Pi^E_B\left(\nabla^2 u_E\right)+u_B\Pi^E_A\left(\nabla^2 u_E\right)\big\}\right]\\
&=-\frac{2D}{NK^2}\left[\frac{(n\cdot\nabla)N}{N}+\frac{(n\cdot\nabla)K}{K}\right]\left[u_C\Pi^E_{C'}\nabla^2 u_E+u_{C'}\Pi^E_C\nabla^2 u_E\right]\\
&+\frac{2D}{NK^2}\bigg[\big\{\Pi^A_C(n\cdot\nabla)u_A\big\}\Pi^E_{C'}\nabla^2 u_E+\big\{\Pi^A_{C'}(n\cdot\nabla)u_A\big\}\Pi^E_C\nabla^2 u_E+u_C\Pi^B_{C'}(n\cdot\nabla)\big(\Pi^E_B\nabla^2 u_E\big)\\
&+u_{C'}\Pi^B_C(n\cdot\nabla)\big(\Pi^E_B\nabla^2 u_E\big)\bigg]\nonumber\\
\end{split}
\end{equation}
\begin{equation}\label{eq:term-1}
\begin{split}
&=-\frac{2D}{NK}\left[N+2~\frac{(n\cdot\nabla)K}{K}\right]\left[2~u_C\frac{\hat{\nabla}^2 u_{C'}}{K}+2~u_{C'}\frac{\hat{\nabla}^2 u_C}{K}\right]\\
&+\frac{2D}{NK}\bigg[4~\frac{\hat{\nabla}^2 u_C}{K}\frac{\hat{\nabla}^2 u_{C'}}{K}-\frac{1}{K}u_C\Pi^B_{C'}\big\{(n\cdot\nabla)n_B\big\}n^E\nabla^2 u_E+\frac{1}{K}u_C\Pi^E_{C'}(n\cdot\nabla)\big(\nabla^2 u_E\big)\\
&-\frac{1}{K}u_{C'}\Pi^B_C\big\{(n\cdot\nabla)n_B\big\}n^E\nabla^2 u_E+\frac{1}{K}u_{C'}\Pi^E_C(n\cdot\nabla)\big(\nabla^2 u_E\big)\bigg]\\
\end{split}
\end{equation}\vspace{-0.2in}
\\Using the identity
\begin{equation}
\begin{split}
\Pi^C_F(n\cdot\nabla)\big(\nabla^2 u_C\big)&=-\lambda~D\frac{\hat{\nabla}^2 u_F}{K}+\frac{\hat{\nabla}^2\hat{\nabla}^2 u_F}{K}-K\frac{\hat{\nabla}^2 K}{K^2}\frac{\hat{\nabla}^2 u_F}{K}-K\frac{\hat{\nabla}_F K}{K}\frac{(u\cdot\nabla) K}{K}\\
&+K\Pi^E_F(n\cdot\nabla)(n\cdot\nabla)u_E-2\big(\hat{\nabla}^E K\big)\big(\hat{\nabla}_E u_F\big)-3~\frac{K^2}{D}\frac{\hat{\nabla}^2 u_F}{K}
\end{split}
\end{equation}\vspace{-0.2in}
\\we get
\begin{equation}
\begin{split}
\text{term-1}&=-\frac{2D}{NK}\left[N+2~\frac{(n\cdot\nabla)K}{K}\right]\left[2~u_C\frac{\hat{\nabla}^2 u_{C'}}{K}+2~u_{C'}\frac{\hat{\nabla}^2 u_C}{K}\right]\\
&+\frac{2D}{NK}\bigg[4~\frac{\hat{\nabla}^2 u_C}{K}\frac{\hat{\nabla}^2 u_{C'}}{K}+2~u_C\frac{\hat{\nabla}_{C'}K}{K}\frac{(u\cdot\nabla)K}{K}+2~u_{C'}\frac{\hat{\nabla}_C K}{K}\frac{(u\cdot\nabla)K}{K}\bigg]\\
&+\frac{2D}{NK}u_C\bigg[-\lambda~\frac{D}{K}\frac{\hat{\nabla}^2 u_{C'}}{K}+\frac{\hat{\nabla}^2\hat{\nabla}^2 u_{C'}}{K^2}-\frac{\hat{\nabla}^2 K}{K^2}\frac{\hat{\nabla}^2 u_{C'}}{K}-\frac{\hat{\nabla}_{C'} K}{K}\frac{(u\cdot\nabla) K}{K}\\
&+\Pi^E_{C'}(n\cdot\nabla)(n\cdot\nabla)u_E-2\bigg(\frac{\hat{\nabla}^E K}{K}\bigg)\big(\hat{\nabla}_E u_{C'}\big)-3~\frac{K}{D}\frac{\hat{\nabla}^2 u_{C'}}{K}\bigg]\\
&+\frac{2D}{NK}u_{C'}\bigg[-\lambda~\frac{D}{K}\frac{\hat{\nabla}^2 u_C}{K}+\frac{\hat{\nabla}^2\hat{\nabla}^2 u_C}{K^2}-\frac{\hat{\nabla}^2 K}{K^2}\frac{\hat{\nabla}^2 u_C}{K}-\frac{\hat{\nabla}_C K}{K}\frac{(u\cdot\nabla) K}{K}\\
&+\Pi^E_C(n\cdot\nabla)(n\cdot\nabla)u_E-2\bigg(\frac{\hat{\nabla}^E K}{K}\bigg)\big(\hat{\nabla}_E u_C\big)-3~\frac{K}{D}\frac{\hat{\nabla}^2 u_C}{K}\bigg]
\end{split}
\end{equation}
\begin{equation}\label{eq:term-2}
\begin{split}
\text{term-2}&=-\frac{2D}{K}\Pi^{A}_{C}\Pi^{B}_{C'}(n\cdot\nabla)\left[\frac{1}{N}\big\{u_A\Pi^E_B(n\cdot\nabla) u_E+u_B\Pi^E_A(n\cdot\nabla) u_E\big\}\right]\\
&=-\frac{2D}{K}\Pi^A_C\Pi^B_{C'}\left(-\frac{n\cdot\nabla N}{N^2}\right)\left[u_A\Pi^E_B(n\cdot\nabla) u_E+u_B\Pi^E_A(n\cdot\nabla) u_E\right]\\
&-\frac{2D}{NK}\Pi^A_C\Pi^B_{C'}\bigg[\big\{(n\cdot\nabla)u_A\big\}\Pi^E_B(n\cdot\nabla)u_E-u_A\big\{(n\cdot\nabla)n_B\big\}n^E(n\cdot\nabla)u_E\\
&+u_A~\Pi^E_B(n\cdot\nabla)(n\cdot\nabla)u_E+\big\{(n\cdot\nabla)u_B\big\}\Pi^E_A(n\cdot\nabla)u_E-u_B\big\{(n\cdot\nabla)n_A\big\}n^E(n\cdot\nabla)u_E\\
&+u_B~\Pi^E_A(n\cdot\nabla)(n\cdot\nabla)u_E\bigg]\nonumber\\
\end{split}
\end{equation}
\begin{equation}\label{eq:term-2}
\begin{split}
&=\frac{2D}{NK}\left[N+\frac{n\cdot\nabla K}{K}\right]\left[u_C\frac{\hat{\nabla}^2 u_{C'}}{K}+u_{C'}\frac{\hat{\nabla}^2 u_C}{K}\right]-\frac{2D}{NK}\bigg[2~\frac{\hat{\nabla}^2 u_C}{K}\frac{\hat{\nabla}^2u_{C'}}{K}+u_C\frac{\hat{\nabla}_{C'}K}{K}\frac{(u\cdot\nabla)K}{K}\\
&+u_{C'}\frac{\hat{\nabla}_C K}{K}\frac{(u\cdot\nabla)K}{K}+u_C\Pi^E_{C'}(n\cdot\nabla)(n\cdot\nabla)u_E+u_{C'}\Pi^E_C(n\cdot\nabla)(n\cdot\nabla)u_E\bigg]
\end{split}
\end{equation}
\begin{equation}\label{eq:ht12-part2}
\begin{split}
\tilde{h}^{(1,2)}_{CC'}\big{|}_{\text{Part-2}}&=-\frac{D}{NK}\Pi^{A}_{C}\Pi^{B}_{C'}\left[2N^{2}\tilde{h}^{(2)}_{AB}+\{(n\cdot\nabla)N\}\tilde{h}^{(1,1)}_{AB}-2\lambda \tilde{h}^{(0)}_{AB}\right]\\
&=-2~\tilde{h}^{(2)}_{CC'}-\frac{D}{K}\left[\frac{n\cdot\nabla K}{K}+N\right]\tilde{h}^{(1,1)}_{AB}+2~\lambda\left(\frac{D}{K}\right)^2 u_C u_{C'}\\[8pt]
&=2~\lambda\frac{D^2}{K^2}u_C u_{C'}-2~\frac{D^2}{K^2}\frac{\hat{\nabla}^2 u_C}{K}~\frac{\hat{\nabla}^2 u_{C'}}{K}+\frac{D^2}{K^2}\bigg[\frac{\hat{\nabla}^2 K}{K^2}-\lambda\frac{D}{K}\bigg]\bigg[u_C\frac{\hat{\nabla}^2 u_{C'}}{K}+u_{C'}\frac{\hat{\nabla}^2 u_C}{K}\bigg]\\
&-2u_C\bigg[\frac{D}{K}\bigg\{-\frac{1}{2}-2~\frac{D}{K}\frac{\hat{\nabla}^2 K}{K^2}+\lambda~\frac{D^2}{K^2}\bigg\}\frac{\hat{\nabla}^2 u_{C'}}{K}+\frac{D^2}{2K^2}\bigg\{\frac{\hat{\nabla}^2\hat{\nabla}^2 u_{C'}}{K^2}-2\frac{\hat{\nabla}^E K}{K}\big(\hat{\nabla}_E u_{C'}\big)\bigg\}\bigg]\\
&-2u_{C'}\bigg[\frac{D}{K}\bigg\{-\frac{1}{2}-2~\frac{D}{K}\frac{\hat{\nabla}^2 K}{K^2}+\lambda~\frac{D^2}{K^2}\bigg\}\frac{\hat{\nabla}^2 u_C}{K}+\frac{D^2}{2K^2}\bigg\{\frac{\hat{\nabla}^2\hat{\nabla}^2 u_C}{K^2}-2\frac{\hat{\nabla}^E K}{K}\big(\hat{\nabla}_E u_C\big)\bigg\}\bigg]
\end{split}
\end{equation}
Adding \eqref{eq:ht12-part1} and \eqref{eq:ht12-part2} we get the final expression of $\tilde{h}^{(1,2)}_{CC'}$ as given in \eqref{eq:htilde12} after using \eqref{eq:term-1} and \eqref{eq:term-2}

\section{Some Details of Stress Tensor Calculation}\label{app:stress}
\subsection*{Outside($\psi>1$)}
\begin{equation}\label{Outside}
G_{AB}^{\text{(out)}}=g_{AB}+\psi^{-D}\mathfrak{h}_{AB}
\end{equation}
Inverse of \eqref{Outside} at linear order is
\begin{equation}
G^{AB}_{\text{(out)}}=g^{AB}-\psi^{-D}\mathfrak{h}^{AB}+{\cal O}(h)^2 \quad\quad \text{here, }\mathfrak{h}^{AB}=g^{AC}g^{BD}\mathfrak{h}_{CD}
\end{equation}
Using, the gauge condition $n^A\mathfrak{h}_{AB}=0$, we get
\begin{equation}
n_A^{(out)}=n_A
\end{equation}
Now,
\begin{equation}
\begin{split}
\mathfrak{p}_{AB}^{\text{(out)}}&=G_{AB}^{\text{(out)}}-n_A^{\text{(out)}}n_B^{\text{(out)}}\\
&=g_{AB}+\psi^{-D}\mathfrak{h}_{AB}-n_A n_B\\
&=\Pi_{AB}+\psi^{-D}\mathfrak{h}_{AB}
\end{split}
\end{equation}
\begin{equation}
\begin{split}
\left[\mathfrak{p}^{\text{(out)}}\right]^A_B&=\delta^A_B-n^A n_B=\Pi^A_B
\end{split}
\end{equation}
Now, from \eqref{eq:KABout_def}
\begin{equation}
\begin{split}
K^{\text{(out)}}_{AB}&=\left[\mathfrak{p}^{\text{(out)}}\right]^C_A \left[\mathfrak{p}^{\text{(out)}}\right]^{C'}_B \left(\tilde{\nabla}_C n_{C'}\right)_{\psi=1}\\
&=\Pi^C_A\Pi^{C'}_B
\left(\partial_C n_{C'}-\tilde{\Gamma}^E_{CC'}n_E\right)\bigg{|}_{\psi=1}
\end{split}
\end{equation}
Where,
\begin{equation}
\tilde{\Gamma}^E_{CC'}=\Gamma^E_{CC'}+\delta\tilde{\Gamma}^E_{CC'}
\end{equation}
Here, $\Gamma^E_{CC'}$ is Christoffel symbol with respect to $g_{AB}$ and $\delta\tilde{\Gamma}^E_{CC'}$ is defined as
\begin{equation}
\delta\tilde{\Gamma}^E_{CC'}=\frac{1}{2}[G^{\text{(out)}}]^{EF}\left[\nabla_C(\psi^{-D}{\mathfrak{h}}_{C'F})+\nabla_{C'}(\psi^{-D}{\mathfrak{h}}_{CF})-\nabla_F(\psi^{-D}{\mathfrak{h}}_{CC'})\right]
\end{equation}
Here, $\nabla_C$ is covarint derivative with respect to $g_{AB}$
\begin{equation}
K^{\text{(out)}}_{AB}=K_{AB}-\Pi^C_A\Pi^{C'}_B n_E\delta\tilde{\Gamma}^E_{CC'}\bigg{|}_{\psi=1}
\end{equation}
Now,
\begin{equation}
\begin{split}
&~~~~-\Pi^C_A\Pi^{C'}_B n_E\delta\tilde{\Gamma}^E_{CC'}\bigg{|}_{\psi=1}\\
&=-\frac{1}{2}\Pi^C_A\Pi^{C'}_B n^F\left[\nabla_C(\psi^{-D}{\mathfrak{h}}_{C'F})+\nabla_{C'}(\psi^{-D}{\mathfrak{h}}_{CF})-\nabla_F(\psi^{-D}{\mathfrak{h}}_{CC'})\right]\bigg{|}_{\psi=1}\\
&=-\frac{1}{2}\Pi^C_A\Pi^{C'}_B n^F\left[\psi^{-D}\nabla_C{\mathfrak{h}}_{C'F}+\psi^{-D}\nabla_{C'}{\mathfrak{h}}_{CF}+ND\psi^{-D-1}n_F{\mathfrak{h}}_{CC'}-\psi^{-D}\nabla_F{\mathfrak{h}}_{CC'}\right]\\
&=-\frac{1}{2}\Pi^C_A\Pi^{C'}_B\left[-{\mathfrak{h}}_{C'F}(\nabla_C n^F)-{\mathfrak{h}}_{CF}(\nabla_{C'}n^F)+ND{\mathfrak{h}}_{CC'}-(n\cdot\nabla){\mathfrak{h}}_{CC'}\right]\\
&=-\frac{1}{2}\Pi^C_A\Pi^{C'}_B\left[-h^{(0)}_{C'F}\left(\nabla_C n^F\right)-h^{(0)}_{CF}\left(\nabla_{C'} n^F\right)+ND h^{(0)}_{CC'}-Nh^{(1)}_{CC'}-(n\cdot\nabla)h^{(0)}_{CC'}\right]\bigg{|}_{\psi=1}\\
&=-\frac{1}{2}\Pi^C_A\Pi^{C'}_B\left[-h^{(0)}_{C'F}K^F_C-h^{(0)}_{CF}K^F_{C'}+ND h^{(0)}_{CC'}-Nh^{(1)}_{CC'}\right]
\end{split}
\end{equation}
Finally, we get
\begin{equation}\label{eq:KABoutexpr}
K_{AB}^{\text{(out)}}=K_{AB}-\frac{ND}{2}h^{(0)}_{AB}+\frac{N}{2}h^{(1)}_{AB}+\frac{1}{2}\left(h^{(0)}_{BD}K^D_A+h^{(0)}_{AD}K^D_B\right)
\end{equation}
Trace of $K_{AB}^{\text{(out)}}$
\begin{equation}\label{eq:Kexpr}
\begin{split}
K^{\text{(out)}}&=\left(g^{AB}-\psi^{-D}\mathfrak{h}^{AB}\right)K_{AB}^{\text{(out)}}\bigg{|}_{\psi=1}\\
&=K-\frac{ND}{2}h^{(0)}+\frac{N}{2}h^{(1)}+\frac{1}{2}g^{AB}\left(h^{(0)}_{BD}K^D_A+h^{(0)}_{AD}K^D_B\right)-h^{(0)}_{AB}K^{AB}\\
&=K-\frac{ND}{2}h^{(0)}+\frac{N}{2}h^{(1)}
\end{split}
\end{equation}
\subsection*{Inside($\psi<1$)}
As, in the previous subsection
\begin{equation}
n_A^{\text{(in)}}=n_A,~~~\mathfrak{p}_{AB}^{\text{(in)}}=\Pi_{AB}+\tilde{\mathfrak{h}}_{AB}~~\text{and,}~~~\left[\mathfrak{p}^{\text{(in)}}\right]^A_B=\Pi^A_B
\end{equation} 
Now, from \eqref{eq:KABin_def}
\begin{equation}
\begin{split}
K^{(in)}_{AB}&=\left[\mathfrak{p}^{(in)}\right]^C_A \left[\mathfrak{p}^{(in)}\right]^{C'}_B \left(\dot{\nabla}_C n_{C'}^{(in)}\right)_{\psi=1}\\
&=\Pi^C_A\Pi^{C'}_B
\left(\partial_C n_{C'}-\hat{\Gamma}^E_{CC'}n_E\right)\bigg{|}_{\psi=1}
\end{split}
\end{equation}
Where,
\begin{equation}
\hat{\Gamma}^E_{CC'}=\Gamma^E_{CC'}+\delta\hat{\Gamma}^E_{CC'}
\end{equation}
Here, $\Gamma^E_{CC'}$ is Christoffel symbol with respect to $g_{AB}$ and $\delta\hat{\Gamma}^E_{CC'}$ is defined as
\begin{equation}
\delta\hat{\Gamma}^E_{CC'}=\frac{1}{2}[G^{\text{(in)}}]^{EF}\left(\nabla_C\tilde{\mathfrak{h}}_{C'F}+\nabla_{C'}\tilde{\mathfrak{h}}_{CF}-\nabla_F\tilde{\mathfrak{h}}_{CC'}\right)
\end{equation}
Here, $\nabla_C$ is covarint derivative with respect to $g_{AB}$
Now,
\begin{equation}
K^{\text{(in)}}_{AB}=K_{AB}-\Pi^C_A\Pi^{C'}_B n_E\delta\hat{\Gamma}^E_{CC'}\bigg{|}_{\psi=1}
\end{equation}
Now,
\begin{equation}
\begin{split}
-\Pi^C_A\Pi^{C'}_B n_E\delta\hat{\Gamma}^E_{CC'}\bigg{|}_{\psi=1}&=-\frac{1}{2}\Pi^C_A\Pi^{C'}_B n^F\left(\nabla_C\tilde{\mathfrak{h}}_{C'F}+\nabla_{C'}\tilde{\mathfrak{h}}_{CF}-\nabla_F\tilde{\mathfrak{h}}_{CC'}\right)\\
&=\frac{1}{2}\Pi^C_A\Pi^{C'}_B\left[\tilde{\mathfrak{h}}_{C'F}(\nabla_C n^F)+\tilde{\mathfrak{h}}_{CF}(\nabla_{C'} n^F)+(n\cdot\nabla)\sum_{m=0}^\infty(\psi-1)^m\tilde{h}^{(m)}_{CC'}\right]_{\psi=1}\\
&=\frac{1}{2}\Pi^C_A\Pi^{C'}_B\left[\tilde{h}^{(0)}_{C'F}K^F_C+\tilde{h}^{(0)}_{CF}K^F_{C'}+N\tilde{h}^{(1)}_{CC'}\right]_{\psi=1}\\
&=\frac{1}{2}\tilde{h}^{(0)}_{BF}K^F_A+\frac{1}{2}\tilde{h}^{(0)}_{AF}K^F_{B}+\frac{1}{2}N\tilde{h}^{(1)}_{AB}\\
\end{split}
\end{equation}
So, we get
\begin{equation}
K^{\text{(in)}}_{AB}=K_{AB}+\frac{1}{2}\left(\tilde{h}^{(0)}_{BF}K^F_A+\tilde{h}^{(0)}_{AF}K^F_{B}+N\tilde{h}^{(1)}_{AB}\right)
\end{equation}
Stress of extrinsic curvature is given by
\begin{equation}
\begin{split}
K^{\text{(in)}}&=\left(g^{AB}-\tilde{\mathfrak{h}}^{AB}\right)K^{\text{(in)}}_{AB}\bigg{|}_{\psi=1}\\
&=\left(g^{AB}-[\tilde{h}^{(0)}]^{AB}\right)K^{\text{(in)}}_{AB}\\
&=K+\frac{1}{2}\left(\tilde{h}^{(0)}_{AF}K^{FA}+\tilde{h}^{(0)}_{AF}K^{FA}+N\tilde{h}^{(1)}\right)-[\tilde{h}^{(0)}]^{AB}K_{AB}\\
&=K+\frac{N}{2}\tilde{h}^{(1)}
\end{split}
\end{equation}
\section{Important Identities}
In this appendix we will mention the identities we have used in this note. The identities have been calculated on $\psi=1$ hypersurface. We are not giving the derivations simply due to the fact that the derivations are very lengthy but nevertheless the derivations are quite straightforward.
\subsection*{Identity-1:}
\begin{equation}\label{eq:identity1}
\frac{\hat{\nabla}_B N}{N}=\frac{\hat{\nabla}_B K}{K}+\frac{1}{K}\hat{\nabla}_B\left(\frac{n\cdot\nabla K}{K}\right)-\frac{1}{K}\left(\frac{\hat{\nabla}_B K}{K}\right)\left(\frac{n\cdot\nabla K}{K}\right)+{\cal O}\left(\frac{1}{D}\right)^2
\end{equation}
\subsection*{Identity-2:}
\begin{equation}\label{eq:identity2}
\frac{(n\cdot\nabla)N}{N}=\frac{K}{D}+\frac{(n\cdot\nabla)K}{K}+\frac{1}{D}\frac{(n\cdot\nabla)K}{K}+\frac{(n\cdot\nabla)(n\cdot\nabla)K}{K^2}-\frac{2}{K}\bigg(\frac{ n\cdot\nabla K}{K}\bigg)^2+{\cal O}\left(\frac{1}{D}\right)^2
\end{equation}
\vspace{-0.2in}
\subsection*{Identity-3:}
\begin{equation}\label{eq:identity3}
ND=K+\frac{(n\cdot\nabla)K}{K}+\frac{(n\cdot\nabla)(n\cdot\nabla)K}{K^2}-\frac{2}{K}\bigg(\frac{n\cdot\nabla K}{K}\bigg)^2+{\cal O}\left(\frac{1}{D}\right)^2
\end{equation}
\subsection*{Identity-4:}
\begin{equation}
\begin{split}
\frac{(n\cdot\nabla)K}{K}&=\frac{\hat{\nabla}^2 K}{K^2}-\frac{1}{K}K_{AB}K^{AB}-\frac{\lambda(D-1)}{K}+\frac{1}{K^4}\hat{\nabla}^2\big(\hat{\nabla}^2 K\big)-\frac{2}{K}\bigg(\frac{\hat{\nabla}^2 K}{K^2}\bigg)\bigg(\frac{\hat{\nabla}^2 K}{K^2}\bigg)\\
&+\lambda~\frac{D}{K^2}\bigg(\frac{\hat{\nabla}^2 K}{K^2}\bigg)-\frac{1}{D}\bigg(\frac{\hat{\nabla}^2 K}{K^2}\bigg)-\frac{1}{K}\bigg(\frac{\hat{\nabla}^2 K}{K^2}-\lambda~\frac{D}{K}-\frac{K}{D}\bigg)\bigg(\frac{\hat{\nabla}^2 K}{K^2}\bigg)\\
&-\frac{2}{K}\bigg(\frac{\hat{\nabla}^E K}{K}\bigg)\bigg(\frac{\hat{\nabla}_E K}{K}\bigg)+{\cal O}\left(\frac{1}{D}\right)^2
\end{split}
\end{equation}
\vspace{-0.2in}
\subsection*{Identity-5:}
\begin{equation}
\begin{split}
\frac{(n\cdot\nabla)(n\cdot\nabla)K}{2K^2}&=\frac{1}{K}\bigg[-\frac{3}{2}\bigg(\frac{\hat{\nabla}^2 K}{K^2}\bigg)\bigg(\frac{\hat{\nabla}^2 K}{K^2}\bigg)+\lambda~\frac{D}{K}\bigg(\frac{\hat{\nabla}^2 K}{K^2}\bigg)+\frac{1}{2K^3}\hat{\nabla}^2\big(\hat{\nabla}^2 K\big)\\
&-\bigg(\frac{\hat{\nabla}^E K}{K}\bigg)\bigg(\frac{\hat{\nabla}_E K}{K}\bigg)-2~\frac{K}{D}\bigg(\frac{\hat{\nabla}^2 K}{K^2}\bigg)+\lambda+\frac{K^2}{D^2}\bigg]+{\cal O}\left(\frac{1}{D}\right)^2
\end{split}
\end{equation}

\section{Notations}\label{app:notation}
In this appendix, we shall summarize the notations we have used in this note.
\begin{table}[ht]
\caption{Notations} 
\vspace{0.2cm}
\centering 
\begin{tabular}{|c| c|} 
\hline 
\hline
\vspace{-0.2cm}
& \\
Background space-time indices & Capital Latin ($A,B,C,D$)\\ [1ex]
\hline
\vspace{-0.2cm}
& \\
Indices on the membrane & Small Greek ($\alpha,\beta,\mu,\nu$)\\ [1ex]
\hline
\vspace{-0.3cm}
& \\
Background metric & $g_{AB}$\\ [1ex]
\hline
\vspace{-0.3cm}
& \\
Induced metric on the membrane as embedded in $g_{AB}$ & $g^{(ind)}_{\mu\nu}$\\ [1ex]
\hline
\vspace{-0.3cm}
& \\
~~~~~~~~Full non-linear metric outside the membrane ~~& ${\cal G}_{AB}$ \\
 as read off from \citep{secondorder} & \\[1ex]
\hline
\vspace{-0.3cm}
& \\
Linearized metric outside the membrane & $G^{(out)}_{AB}=g_{AB}+{\psi}^{-D}\mathfrak{h}_{AB}$\\ [1ex]
\hline
\vspace{-0.3cm}
& \\
Linearized metric inside the membrane & $G^{(in)}_{AB}=g_{AB}+\mathfrak{\tilde{h}}_{AB}$\\ [1ex]
\hline
\vspace{-0.3cm}
& \\
Projector on the membrane as embedded in $g_{AB}$&$\Pi_{AB}=g_{AB}-n_A~n_B$\\ [1ex]
\hline
\vspace{-0.3cm}
& \\
Projector perpendicular to both the normal of the&$P_{AB}=g_{AB}-n_A~n_B+u_A u_B$\\
membrane as embedded in $g_{AB}$ and the velocity&\\ [1ex]
\hline
\vspace{-0.3cm}
& \\
Projector on the membrane as embedded in $G_{AB}^{(out)}$ &$\mathfrak{p}^{(out)}_{AB}=G_{AB}^{(out)}-n_A^{(out)}n_B^{(out)}$\\ [1ex]
\hline
\vspace{-0.3cm}
& \\
Projector on the membrane as embedded in $G_{AB}^{(in)}$ &$\mathfrak{p}^{(in)}_{AB}=G_{AB}^{(in)}-n_A^{(in)}n_B^{(in)}$\\ [1ex]
\hline
\vspace{-0.3cm}
& \\
Covariant derivative w.r.t. $g_{AB}$& $\nabla_A$\\
[1ex]
\hline
\vspace{-0.3cm}
&\\
Covariant derivative w.r.t.  $g^{(ind)}_{\mu\nu}$& $\bar\nabla_\mu$\\
[1ex]
\hline
\vspace{-0.3cm}
&\\
Covariant derivative w.r.t.  $G_{AB}^{(out)}$& $\tilde{\nabla}_A$\\
[1ex]
\hline
\vspace{-0.3cm}
&\\
Covariant derivative w.r.t.  $G_{AB}^{(in)}$& $\dot{\nabla}_A$\\
[1ex]
\hline
\vspace{-0.3cm}
&\\
~~~~~Covariant derivative w.r.t. $g_{AB}$ projected ~~& ~~~~~~~~~~~~~~~~$\hat\nabla_A$~~~~~~~~~~~~~~~ \\
 along the membrane & ~See equation \eqref{eq:hat_def} for definition\\
[1ex]
\hline
\vspace{-0.3cm}
&\\
Extrinsic curvature of the membrane& $K^{(out)}_{AB}$\\
 when embedded in  $G_{AB}^{(out)}$&\\
[1ex]
\hline
\vspace{-0.3cm}
&\\
Extrinsic curvature of the membrane & $K^{(in)}_{AB}$\\
when embedded in $G_{AB}^{(in)}$&\\
[1ex]
\hline
\vspace{-0.3cm}
&\\
Extrinsic curvature of the membrane & $K_{AB}$\\
when embedded in $g_{AB}$&\\
[1ex]
\hline
\hline
\end{tabular}
\end{table}
\FloatBarrier

\bibliographystyle{JHEP}
\bibliography{larged}

\end{document}